\documentclass[acmcsur]{acmtrans2m}

\title{Some Facets of Complexity Theory and Cryptography:
       A Five-Lectures Tutorial}

\author{
J\"{O}RG ROTHE \\ 
Heinrich-Heine-Universit\"at D\"usseldorf
}

\usepackage{eepic,epic}
\usepackage{latexsym}
\usepackage{amsfonts}
\usepackage{amsmath}
\usepackage{amssymb}

\usepackage[latin1]{inputenc}    
\usepackage{psfig}

\newlength{\filength}
\newsavebox{\gcbox}
\sbox{\gcbox}{\framebox[\filength]{\rule{0ex}{2ex}}}

\newtheorem{theorem}{Theorem}[section]

\newcommand{\qedblob}{\mbox{\rule[-1.5pt]{5pt}{10.5pt}}}
\def\literalqed{{\ \nolinebreak\hfill\mbox{\qedblob\quad}}}

\def\qed{\literalqed}

\newtheorem{lemma}[theorem]{Lemma}

\newtheorem{proposition}[theorem]{Proposition}
\newdef{definition}[theorem]{Definition}
\newdef{remark}[theorem]{Remark}
\newdef{example}[theorem]{Example}
\makeatletter
\def\@citex[#1]#2{\if@filesw\immediate\write\@auxout{\string\citation{#2}}\fi
  \def\@citea{}\@cite{\@for\@citeb:=#2\do
    {\@citea\def\@citea{,\linebreak[0]}\@ifundefined
       {b@\@citeb}{{\bf ?}\@warning
       {Citation `\@citeb' on page \thepage \space undefined}}%
\hbox{\csname b@\@citeb\endcsname}}}{#1}}
\makeatother

\newcommand{\naturalnumber}{\ensuremath{{  \mathbb{N} }}}



\newcommand{\up}{\mbox{\rm UP}}

\newcommand{\coup}{\mbox{\rm coUP}}

\newcommand{\p}{\mbox{\rm P}}

\newcommand{\np}{\mbox{\rm NP}}
\newcommand{\scriptnp}{\mbox{\protect\scriptsize\rm NP}}
\newcommand{\ip}{\mbox{\rm IP}}

\newcommand{\zpp}{\mbox{\rm ZPP}}
\newcommand{\rp}{\mbox{\rm R}}
\newcommand{\cor}{\mbox{\rm coR}}

\newcommand{\pair}[1]{\mathopen\langle{#1}\mathclose\rangle}

\newcommand{\sigmastar}{\ensuremath{\Sigma^\ast}}

\makeatletter%
\def\nottoobig#1{{\hbox{$\left#1\vcenter to1.111\ht\strutbox{}\right.\n@space$}}}
\makeatother%

\newcommand{\condition}{\,\nottoobig{|}\:}

\def\land{{\; \wedge \;}}

\newcommand\integers{{\mathbb{Z}}}

\newcommand{\seq}{\subseteq}

\def\nats{\naturalnumber}

\newcommand\Lora{\, \Longrightarrow \ }

\def\equalsdef{=}

\newcommand{\image}{\mbox{\rm{}image}}
\newcommand{\domain}{\mbox{\rm{}domain}}
\newcommand{\sigmabot}{\stackrel{\mbox{\tiny $\bot$}}{\sigma}}
\newcommand{\sigmabots}[1]{\stackrel{\mbox{\tiny $\bot$}}{\sigma}\!\!(#1)}

\newcommand{\certificate}[2]{{\tt Certificates}_{#1}({#2})}
\newcommand{\graphiso}{{\tt Graph\mbox{-}Isomorphism}}
\newcommand{\threecolor}{{\tt Graph\mbox{-}Three\mbox{-}Colorability}}

\newenvironment{construction}{\bigbreak\begin{block}}{\end{block}
    \bigbreak}

\newenvironment{block}{\begin{list}{\hbox{}}{\leftmargin 1em
    \itemindent -1em \topsep 0pt \itemsep 0pt \partopsep 0pt}}{\end{list}}

\newenvironment{algorithm}{\begin{list}
   {{\bf Step~\arabic{alg}}.}
   {\usecounter{alg}}}{\end{list}}
\newcounter{alg}

\begin{abstract}
In this tutorial, selected topics of cryptology and of 
computational complexity theory are presented. We give a brief overview
of the history and the foundations of classical cryptography, and then 
move on to modern public-key cryptography.  Particular attention is
paid to cryptographic protocols and the problem of constructing key
components of protocols such as one-way functions.  A function is
one-way if it is easy to compute, but hard to invert.  We discuss
the notion of one-way functions both in a cryptographic and in a
complexity-theoretic setting. We also consider interactive proof systems
and present some interesting zero-knowledge protocols. In a
zero-knowledge protocol one party can convince the other party of 
knowing some secret information without disclosing any bit of this
information.  Motivated by these protocols, we survey some
complexity-theoretic results on interactive proof systems and related
complexity classes. 
\end{abstract}






\begin{document}

\begin{bottomstuff} 
  Author's address: J. Rothe, Institut f\"ur Informatik,
  Heinrich-Heine-Universit\"at D\"usseldorf, 40225 D\"usseldorf, Germany. 
  Email address: ${\tt rothe@cs.uni\mbox{-}duesseldorf.de}$. \newline
  This version, which revises earlier versions of this tutorial, appears
  in {\em ACM Computing Surveys}, vol.~34, no.~4, December 2002. \newline
  This work was supported in part by grant
  NSF-INT-9815095/DAAD-315-PPP-g\"{u}-ab. 
\end{bottomstuff}

\maketitle

\sloppy

\clearpage

\tableofcontents

\section*{Outline of the Tutorial}
\addcontentsline{toc}{section}{Outline of the Tutorial}

This tutorial consists of five lectures on cryptography, based on the lecture
notes for a course on this subject given by the author in August, 2001, at the
11th Jyv\"askyl\"a Summer School in Jyv\"askyl\"a, Finland.  As the title
suggests, a particular focus of this tutorial is to emphasize the close
relationship between cryptography and complexity theory.  The material
presented here is not meant to be a comprehensive study or a complete survey
of (the intersection of) these fields.  Rather, five vivid topics from those
fields are chosen for exposition, and from each topic chosen, some gems---some
particularly important, central, beautiful results---are presented.  Needless
to say, the choice of topics and of results selected for exposition is
based on the author's personal tastes and biases.

The first lecture sketches the history and the classical foundations of
cryptography, introduces a number of classical, symmetric cryptosystems, and
briefly discusses by example the main objectives of the two opposing parts of
cryptology: cryptography, which aims at designing secure ways of encryption,
versus cryptanalysis, which aims at breaking existing cryptosystems.  Then, we
introduce the notion of perfect secrecy for cryptosystems, which dates back
to Claude Shannon's pioneering work~\cite{sha:j:secrecy} on coding and
information theory.

The second lecture presents the public-key cryptosystem RSA, 
which was invented by Rivest, Shamir, and Adleman~\cite{riv-sha-adl:j:rsa}.  
RSA is the first public-key cryptosystem developed in the
public sector.  To describe RSA, some background from number
theory is provided in as short a way as possible but to the extent necessary to
understand the underlying mathematics.  In contrast to the
information-theoretical approach of perfect secrecy, the security of RSA is
based on the assumption that certain problems from number theory are
computationally intractable.  Potential attacks on the RSA cryptosystem as
well as appropriate countermeasures against them are discussed.

The third lecture introduces a number of cryptographic protocols, including
the secret-key agreement protocols of Diffie and
Hellman~\cite{dif-hel:j:diffie-hellman} and of Rivest and Sherman
(see~\cite{rab-she:t-no-URL:aowf,rab-she:j:aowf}), 
ElGamal's public-key cryptosystem~\cite{gam:j:public-key}, Shamir's no-key
protocol, and the digital signature schemes of Rivest, Shamir, and
Adleman~\cite{riv-sha-adl:j:rsa}, ElGamal~\cite{gam:j:public-key}, and Rabi
and Sherman~\cite{rab-she:t-no-URL:aowf,rab-she:j:aowf}, respectively.  Again,
the underlying mathematics and, relatedly, security issues of these protocols
are briefly discussed.  

A remark is in order here.  The protocols presented here are among the most
central and important cryptographic protocols, with perhaps two exceptions:
the Rivest--Sherman and the Rabi--Sherman protocols.
While the secret-key agreement protocol of Diffie and
Hellman~\cite{dif-hel:j:diffie-hellman} is widely used in practice, that of
Rivest and Sherman (see~\cite{rab-she:t-no-URL:aowf,rab-she:j:aowf}) is not
(yet) used in applications and, thus, might appear somewhat exotic at first
glance.  An analogous comment applies to 
the Rabi--Sherman digital signature protocol.
However, from our point of view, there is some hope that this fact,
though currently true, might change in the near future.  In
Section~\ref{sec:discussion}, we will discuss the state of the art on 
the Diffie--Hellman protocol and the Rivest--Sherman protocol, 
and we will argue that recent progress of 
results in complexity theory may lead to a significant
increase in the cryptographic security and the applicability of the
Rivest--Sherman protocol.  One line of complexity-theoretic research 
that is relevant here is presented in Section~\ref{sec:aowf};
another line of research is 
Ajtai's breakthrough result~\cite{ajt:c:hard-instances-in-lattices} 
on the complexity of the shortest lattice vector problem (SVP, for short),
which is informally stated 
in Section~\ref{sec:discussion}.

The fourth lecture introduces interactive proof systems and zero-knowledge
protocols.  This area has rapidly developed and flourished in complexity
theory and has yielded a number of powerful results.
For example, Shamir's famous 
result~\cite{sha:j:ip} characterizes the power of interactive proof systems 
in terms of classical complexity classes: Interactive proof systems precisely
capture the class of problems solvable in polynomial space.
Also, the study of interactive proof systems is related to
probabilistically checkable proofs, which has yielded novel 
nonapproximability results for hard
optimization problems; see the survey~\cite{gol:b:taxonomy-of-proof-systems}.
Other results about interactive proof systems and the related zero-knowledge
protocols have direct applications in cryptography.  
In particular, zero-knowledge protocols enable one party to
convince another party of knowledge of 
some secret information without conveying
any bit of this information.  Thus, they are ideal technical tools for
authentication purposes.  We present two of the classic zero-knowledge
protocols: the Goldreich-Micali-Wigderson protocol for graph
isomorphism~\cite{gol-mic-wid:c:nothing,gol-mic-wid:j:nothing} and the
Fiat-Shamir protocol~\cite{fia-sha:c:fiat-shamir-zero-knowledge} that is based
on a number-theoretical problem.  For an in-depth treatment of zero-knowledge
protocols and many more technical details, the reader is referred to Chapter~4
of Goldreich's book~\cite{gol:b:foundations}.

The fifth lecture gives an overview on the progress of results that was
recently obtained by Hemaspaandra, Pasanen, and this
author~\cite{hem-rot:j:aowf,hem-pas-rot:c:strong-noninvertibility}.  Their
work, which is motivated by the Rivest--Sherman and the Rabi--Sherman protocols, 
studies properties of functions that are used in building these two
cryptographic protocols.  It is results about these functions that
may be useful in quantifying the security of these protocols.
In particular, the key building block of the Rivest--Sherman protocol is a
strongly noninvertible, 
associative one-way function.  
Section~\ref{sec:aowf} presents the
result~\cite{hem-rot:j:aowf} on how to construct such a function
from the assumption that $\p \neq \np$.
In addition, recent
results on strong
noninvertibility are surveyed, including the perhaps somewhat surprising
result that if $\p \neq \np$ then there exist strongly noninvertible functions
that in fact are invertible~\cite{hem-pas-rot:c:strong-noninvertibility}.  
These results are obtained in the {\em worst-case\/} complexity model, 
which is relevant and interesting in a
complexity-theoretic setting, but useless in applied cryptography.  For
cryptographic applications, one would need to construct such functions based
on the {\em average-case\/} complexity model, under plausible assumptions.
Hence, the most challenging open research question 
related to strongly noninvertible, associative
one-way functions is to find some evidence that they exist even in the
average-case model.  As noted above, our hope
of obtaining such a result is based on recent progress on the shortest
lattice vector problem accomplished
by~Ajtai~\cite{ajt:c:hard-instances-in-lattices}.  Roughly speaking, Ajtai 
proved that this problem is as hard in the average-case as it is in the
worst-case model.  
Based on this result, Ajtai and 
Dwork~\cite{ajt-dwo:c:public-key-system-worst-average-equivalence}
designed a public-key
cryptosystem whose security is based merely on worst-case assumptions.  
Ajtai's breakthrough results, his techniques, and their
cryptographic applications are not covered in this tutorial.
We refer to the nice surveys by Cai~\cite{cai:c:lattice-problems-survey} and,
more recently, by Kumar and Sivakumar~\cite{kum-siv:j:svp-survey}
and Nguyen and Stern~\cite{ngu-ste:c:two-faces-of-lattices} on the
complexity of~SVP and the use of lattices in crytography.

The tutorial is suitable for graduate students with some background in
computer science and mathematics and may also be accessible to interested
undergraduate students.  Since it is organized in five essentially
independent, self-contained lectures, it is also possible to present only a
proper subset of these lectures.  The only dependencies occurring between
lectures are that some of the number-theoretical background given in
Section~\ref{sec:rsa-system} is also used in Section~\ref{sec:protocols}, and
that the Rivest--Sherman secret-key agreement protocol and the 
Rabi--Sherman digital signature protocol presented in
Section~\ref{sec:protocols} motivate the investigations in
Section~\ref{sec:aowf}.  
This last section contains perhaps the technically most challenging material,
which in part is presented on an expert level with the intention of
guiding the reader towards an active field of current research.

There are a number of textbooks and monographs on cryptography that
cover various parts of the field in varying depth, such as the
books by 
Goldreich~\cite{gol:b:modern-cryptography,gol:b:foundations},
Salomaa~\cite{sal:b:public-key-cryptography},
Stinson~\cite{sti:b:cryptography}, and Welsh~\cite{wel:b:codes}.  Schneier's
book~\cite{sch:b:applied-cryptography} provides a very comprehensive
collection of literally all notions and concepts known in cryptography, which
naturally means that the single notions and concepts cannot be treated in
mathematical detail there, but the interested reader is referred to an
extraordinarily large bibliography for such an in-depth treatment.
Singh~\cite{sin:b:code-book} wrote a very charming, easy-to-read, interesting
book about the history of cryptography from its ancient roots to its modern
and even futuristic branches such as quantum cryptography.  An older but still
valuable source is Kahn's book~\cite{kah:b:codebreakers}.
We conclude this
list, without claiming it to be complete, with the books by
Bauer~\cite{bau:b:decrypted-secrets},
Beutelspacher et al.~\cite{beu-sch-wol:b:kryptographie,beu:b:cryptology},
and Buchmann~\cite{buc:b:kryptographie}.

\section{Cryptosystems and Perfect Secrecy}
\label{sec:classical-cryptography}

\subsection{Classical Cryptosystems}

The notion of a cryptosystem is formally defined as follows.

\begin{definition}[Cryptosystem]~
\label{def:cryptosystem}
\begin{itemize}
\item A {\em cryptosystem\/} is a quintuple $(\mathcal{P}, \mathcal{C},
  \mathcal{K}, \mathcal{E}, \mathcal{D})$ such that:
\begin{enumerate}
\item $\mathcal{P}$, $\mathcal{C}$, and $\mathcal{K}$ are finite sets, where

\begin{tabular}{rl}
\centering
$\mathcal{P}$ & is 
      the {\em plain text space\/} or {\em clear text space\/};\\
$\mathcal{C}$ & is the {\em cipher text space\/};\\
$\mathcal{K}$ & is the {\em key space}.
\end{tabular}

Elements of $\mathcal{P}$ are referred to as plain text (or clear text), and
elements of $\mathcal{C}$ are referred to as cipher text.  A {\em message\/}
is a string of plain text symbols.

\item $\mathcal{E} = \{E_k \condition k \in \mathcal{K}\}$ is a family of
  functions $E_k : \mathcal{P} \rightarrow \mathcal{C}$ that are used for
  encryption, and $\mathcal{D} = \{D_k \condition k \in \mathcal{K}\}$ is a
  family of functions $D_k : \mathcal{C} \rightarrow \mathcal{P}$ that are
  used for decryption.

\item For each key $e \in \mathcal{K}$, there exists a key $d \in \mathcal{K}$
  such that for each $p \in \mathcal{P}$:
\begin{equation}
\label{equ:correctness}
D_d (E_e (p)) = p .
\end{equation}
\end{enumerate}

\item A {\em cryptosystem\/} is called {\em symmetric\/} (or {\em
    ``private-key''\/}) if $d = e$, or if $d$ can at least be ``easily''
  computed from~$e$.
  
\item A {\em cryptosystem\/} is called {\em asymmetric\/} (or {\em
    ``public-key''\/}) if $d \neq e$, and it is ``computationally infeasible
  in practice'' to compute $d$ from~$e$.  Here, $d$ is the {\em private key},
  and $e$ is the {\em public key}.

\end{itemize}
\end{definition}

At times, different key spaces are used for encryption and for decryption,
which results in a slight modification of the above definition.  

We now present and discuss some examples of classical cryptosystems.  Consider
the English alphabet $\Sigma = \{\mbox{\rm{}A}, \mbox{\rm{}B}, \ldots ,
\mbox{\rm{}Z}\}$.  To carry out the arithmetic modulo $26$ with letters as if
they were numbers, we identify $\Sigma$ with $\integers_{26} = \{0, 1, \ldots
, 25\}$; thus, $0$ represents \mbox{\rm{}A} and $1$ represents~\mbox{\rm{}B},
and so on.  This encoding of the plain text alphabet by integers and the
decoding of $\integers_{26}$ back to $\Sigma$ is not part of the actual
encryption and decryption, respectively.  It will be used for the next three
examples.  Note that messages are elements of~$\sigmastar$, where $\sigmastar$
denotes the set of strings over~$\Sigma$.

\begin{example}[Caesar cipher, a monoalphabetic symmetric cryptosystem]~
  
  Let $\mathcal{K} = \integers_{26}$, and let $\mathcal{P} = \mathcal{C} =
  \Sigma$.  The {\em Caesar cipher\/} encrypts messages by shifting (modulo
  $26$) each character of the plain text by the same number $k$ of letters in
  the alphabet, where $k$ is the key.  Shifting each character of the cipher
  text back using the same key $k$ reveals the original message:
\begin{itemize}
\item For each $e \in \integers_{26}$, define the encryption function $E_e :
  \Sigma \rightarrow \Sigma$ by
\[
E_e (p) = (p + e) \mod 26,
\]
where addition with $e$ modulo $26$ is carried out character-wise, i.e., each
character $m_i \in \Sigma$ of a message $m \in \sigmastar$ is shifted by $e$
positions to $m_i + e \mod 26$.  For example, using the key $e = 11 =
\mbox{\rm{}L}$, the message {\rm{}``SUMMER''} will be encrypted as
{\rm{}``DFXXPC.''}

\item For each $d \in \integers_{26}$, define the decryption function $D_d :
\Sigma \rightarrow \Sigma$ by
\[
D_d (c) = (c - d) \mod 26,
\]
where subtraction by $e$ modulo $26$ again is carried out character-wise.
Hence, $d=e$.  For example, decrypting the cipher text {\rm{}``DNSZZW''} with
the key $d = 11$ reveals the plain text {\rm{}``SCHOOL.''}
\end{itemize}
\end{example}

Since the key space is very small, breaking the Caesar cipher is very easy.
It is vulnerable even to {\em ``cipher-text-only attacks,''} i.e., an attacker
given enough cipher text $c$ can easily check the $26$ possible keys to
see which one yields a meaningful plain text.  Note that the given cipher text
should contain enough letters to enable a unique decryption.

The Caesar cipher is a monoalphabetic cryptosystem, since 
it replaces each given
plain text letter, wherever in the message it occurs, by the same letter of
the cipher text alphabet.  In contrast, the French cryptographer and diplomat
Blaise de Vigen\`{e}re (1523--1596) proposed a polyalphabetic cryptosystem,
which is much harder to break.  Vigen\`{e}re's system builds on earlier work
by the Italian mathematician Leon Battista Alberti (born in 1404), the German
abbot Johannes Trithemius (born in 1492), and the Italian scientist Giovanni
Porta (born in 1535), see~\cite{sin:b:code-book}.  It works like the Caesar
cipher, except that the cipher text letter encrypting any given plain text
letter X varies with the position of X in the plain text.

More precisely, one uses for encryption and decryption a {\em Vigen\`{e}re
  square}, which consists of 26 rows with 26 columns each.  Every row contains
the 26 letters of the alphabet, shifted by one from row to row, i.e., the rows
and columns may be viewed as a Caesar encryption of the English alphabet with
keys $0$, $1$, $\ldots$,~$25$.  Given a message~$m \in \sigmastar$, one first
chooses a key $k \in \sigmastar$, which is written above the message~$m$,
symbol by symbol, possibly repeating $k$ if $k$ is shorter than~$m$ until 
every character of $m$ has a symbol above it.  
Denoting the $i$th letter of any string
$w$ by~$w_i$, each letter $m_i$ of $m$ is then encrypted as in the Caesar
cipher, using the row of the Vigen\`{e}re square that starts with~$k_i$, where
$k_i$ is the key letter right above~$m_i$.  Below, we describe the
Vigen\`{e}re system formally and give an example of a concrete encryption.
 
\begin{example}[Vigen\`{e}re cipher, a polyalphabetic symmetric cryptosystem]~
  
  For fixed~$n \in \nats$, let $\mathcal{K} = \mathcal{P} = \mathcal{C} =
  \integers_{26}^{n}$.  Messages $m \in \sigmastar$, where $\Sigma$ again is
  the English alphabet, are split into blocks of length $n$ and are encrypted
  block-wise.  The {\em Vigen\`{e}re cipher\/} is defined as follows.
\begin{itemize}
\item For each $e \in \integers_{26}^{n}$, define the encryption function $E_e
  : \integers_{26}^{n} \rightarrow \integers_{26}^{n}$ by
\[
E_e (p) = (p + e) \mod 26,
\]
where addition with $e$ modulo~$26$ is carried out character-wise, i.e., each
character $p_i \in \Sigma$ of a plain text $p \in \mathcal{P}$ is shifted by
$e_i$ positions to $p_i + e_i \mod 26$.

\item For each $d \in \integers_{26}^{n}$, define the decryption function $D_d
  : \integers_{26}^{n} \rightarrow \integers_{26}^{n}$ by
\[
D_d (c) = (c - d) \mod 26,
\]
where subtraction modulo~$26$ again is carried out character-wise.  As in the
Caesar cipher, $d = e$.
\end{itemize}
For example, choose the word $k =$~{\rm{}ENGLISH} to be the key.  Suppose we
want to encrypt the message
$m=$~{\rm{}FINNISHISALLGREEKTOGERMANS,}\footnote{From this example we not only
  learn how the Vigen\`{e}re cipher works, but also that using a language
  such as Finnish, which is not widely used, often makes illegal decryption
  harder, and thus results in a higher level of security.  This is not a
  purely theoretical observation.  During World War~II, the US Navy
  transmitted important messages using the language of the Navajos, a Native
  American tribe.  The ``Navajo Code'' 
  was never broken by the Japanese code-breakers, see~\cite{sin:b:code-book}.
} %
omitting the spaces between words.  Table~\ref{tab:vigenere} shows how each
plain text letter is encrypted, yielding the cipher text~$c$.  For instance,
the first letter of the message, {\rm{}``F,''} corresponds to the first letter
of the key, {\rm{}``E.''}~~Hence, the intersection of the {\rm{}``F''}-column
with the {\rm{}``E''}-row of the Vigen\`{e}re square gives the first letter,
{\rm{}``J,''} of the cipher text.
\begin{table}[ht]
\centering
\rm\small
\begin{tabular}{|l|cccccccccccccccccccccccccc|}
\hline
$k$ &               E &\hspace{-4mm} N &\hspace{-4mm} G &\hspace{-4mm} L 
    &\hspace{-4mm} I &\hspace{-4mm} S &\hspace{-4mm} H &\hspace{-4mm} E 
    &\hspace{-4mm} N &\hspace{-4mm} G &\hspace{-4mm} L &\hspace{-4mm} I 
    &\hspace{-4mm} S &\hspace{-4mm} H &\hspace{-4mm} E &\hspace{-4mm} N 
    &\hspace{-4mm} G &\hspace{-4mm} L &\hspace{-4mm} I &\hspace{-4mm} S 
    &\hspace{-4mm} H &\hspace{-4mm} E &\hspace{-4mm} N &\hspace{-4mm} G 
    &\hspace{-4mm} L &\hspace{-4mm} I \\\hline
$m$ &              F &\hspace{-4mm} I &\hspace{-4mm} N &\hspace{-4mm} N 
    &\hspace{-4mm} I &\hspace{-4mm} S &\hspace{-4mm} H &\hspace{-4mm} I 
    &\hspace{-4mm} S &\hspace{-4mm} A &\hspace{-4mm} L &\hspace{-4mm} L 
    &\hspace{-4mm} G &\hspace{-4mm} R &\hspace{-4mm} E &\hspace{-4mm} E 
    &\hspace{-4mm} K &\hspace{-4mm} T &\hspace{-4mm} O &\hspace{-4mm} G 
    &\hspace{-4mm} E &\hspace{-4mm} R &\hspace{-4mm} M &\hspace{-4mm} A 
    &\hspace{-4mm} N &\hspace{-4mm} S \\\hline
$c$ &              J &\hspace{-4mm} V &\hspace{-4mm} T &\hspace{-4mm} Y
    &\hspace{-4mm} Q &\hspace{-4mm} K &\hspace{-4mm} O &\hspace{-4mm} M
    &\hspace{-4mm} F &\hspace{-4mm} G &\hspace{-4mm} W &\hspace{-4mm} T
    &\hspace{-4mm} Y &\hspace{-4mm} Y &\hspace{-4mm} I &\hspace{-4mm} R
    &\hspace{-4mm} Q &\hspace{-4mm} E &\hspace{-4mm} W &\hspace{-4mm} Y
    &\hspace{-4mm} L &\hspace{-4mm} V &\hspace{-4mm} Z &\hspace{-4mm} G
    &\hspace{-4mm} Y &\hspace{-4mm} A\\
\hline
\end{tabular}
\caption{An example of encryption by the Vigen\`{e}re cipher.
\label{tab:vigenere}
}
\end{table}
\end{example}

Our last example of a classical, historically important cryptosystem is the
Hill cipher, which was invented by Lester Hill in 1929.  It is based on
linear algebra and, like the Vigen\`{e}re cipher, is an affine linear block
cipher.

\begin{example}[Hill cipher, a symmetric cryptosystem and a 
linear block cipher]~
  
  For fixed~$n \in \nats$, the key space $\mathcal{K}$ is the set of all
  invertible $n \times n$ matrices in $\integers_{26}^{n \times n}$.
  Again, $\mathcal{P} = \mathcal{C} =
  \integers_{26}^{n}$ and messages $m \in \sigmastar$ are split into blocks of
  length $n$ and are encrypted block-wise.  All arithmetic operations are
  carried out modulo~$26$.

The {\em Hill cipher\/} is defined as follows.
\begin{itemize}
\item For each $K \in \mathcal{K}$, define the encryption function $E_K
  : \integers_{26}^{n} \rightarrow \integers_{26}^{n}$ by
\[
E_K (p) = K \cdot p \mod 26,
\]
where $\cdot$ denotes matrix multiplication modulo~$26$.

\item Letting $K^{-1}$ denote the inverse matrix of~$K$, the decryption
  function $D_{K^{-1}} : \integers_{26}^{n} \rightarrow \integers_{26}^{n}$ is
  defined by
\[
D_{K^{-1}} (c) =  K^{-1} \cdot c \mod 26.
\]
Since $K^{-1}$ can easily be computed from~$K$, the Hill cipher is a
symmetric cryptosystem.  It is also the most general linear block cipher.

Concrete examples of messages encrypted by the Hill cipher can be found in,
e.g.,~{\rm{}\cite{sal:b:public-key-cryptography}}.
\end{itemize}
\end{example}

Affine linear block ciphers are easy to break by ``{\em known-plain-text
  attacks},'' i.e., for an attacker who knows some sample plain texts with the
corresponding encryptions, it is not too hard to find the key used to encrypt
these plain texts.  They are even more vulnerable to ``{\em chosen-plain-text
  attacks},'' where the attacker can choose some pairs of
corresponding plain texts and encryptions, which may be useful if there are
reasonable conjectures about the key used.  

The method of frequency counts is often useful for decrypting messages.  It
exploits the redundancy of the natural language used for plain text messages.
For example, in many languages the letter ``E'' occurs, statistically
significant, 
most frequently, with a percentage of $12.31\%$ in English, of $15.87\%$ in
French, and even of $18.46\%$ in German,
see~\cite{sal:b:public-key-cryptography}.  Some languages have other letters
that occur with the highest frequency; for example, ``A'' is the most frequent
letter in average Finnish texts, with a percentage of
$12.06\%$~\cite{sal:b:public-key-cryptography}.

In 1863, the German cryptanalyst
Friedrich Wilhelm Kasiski found a method to break the Vigen\`{e}re
cipher.
Singh~\cite{sin:b:code-book} attributes this achievement
  also to an unpublished work, done probably around~1854, by the British
  genius and eccentric Charles Babbage.
The books by Salomaa~\cite{sal:b:public-key-cryptography} and
Singh~\cite{sin:b:code-book} describe Kasiski's and Babbage's method.  
It marks a
breakthrough in the history of cryptanalysis, because previously the
Vigen\`{e}re cipher was considered unbreakable.  In particular, like similar
periodic cryptosystems with an unknown period, the Vigen\`{e}re cipher
appeared to resist cryptanalysis by counting and analysing the frequency
of letters in the cipher text.  Kasiski showed how to determine the period
from repetitions of the same substring in the cipher text.

In light of Kasiski's and Babbage's achievement, it is natural to ask
whether there exist any cryptosystems that guarantee {\em perfect secrecy}.
We turn to this question in the next section that describes some of the
pioneering work of Claude Shannon~\cite{sha:j:secrecy}, who laid the
foundations of modern coding and information theory.

\subsection{Conditional Probability and Bayes's Theorem}

To discuss perfect secrecy of cryptosystems in mathematical terms, we first
need some preliminaries from elementary probability theory.

\begin{definition}
Let $A$ and $B$ be events with $\mbox{\rm Pr}(B) > 0$.
\begin{itemize}
\item The {\em probability that $A$ occurs under the condition that $B$
    occurs\/} is defined by
\[
\mbox{\rm Pr}(A \, |\, B) = \frac{\mbox{\rm Pr}(A \cap B)}{\mbox{\rm Pr}(B)}.
\]
\item $A$ and $B$ are {\em independent\/} if $\mbox{\rm Pr}(A \cap B) =
  \mbox{\rm Pr}(A) \; \mbox{\rm Pr}(B)$ (equivalently, if $\mbox{\rm Pr}(A \,
  |\, B) = \mbox{\rm Pr}(A)$).
\end{itemize}
\end{definition}

\begin{lemma}[Bayes's Theorem]
  Let $A$ and $B$ be events with $\mbox{\rm Pr}(A) > 0$ and $\mbox{\rm Pr}(B)
  > 0$.  Then,
\[
\mbox{\rm Pr}(B) \; \mbox{\rm Pr}(A \, |\, B) = \mbox{\rm Pr}(A) \; \mbox{\rm
  Pr}(B \, |\, A).
\]
\end{lemma}

\begin{proof}
By definition,
\[
\mbox{\rm Pr}(B) \; \mbox{\rm Pr}(A \, |\, B) = \mbox{\rm Pr}(A \cap B) =
\mbox{\rm Pr}(B \cap A) = \mbox{\rm Pr}(A) \; \mbox{\rm Pr}(B \, |\, A).
\]
\end{proof}

\subsection{Perfect Secrecy: Shannon's Theorem}

Consider the following scenario:

\[
\begin{array}{ccc}
 & \psfig{file=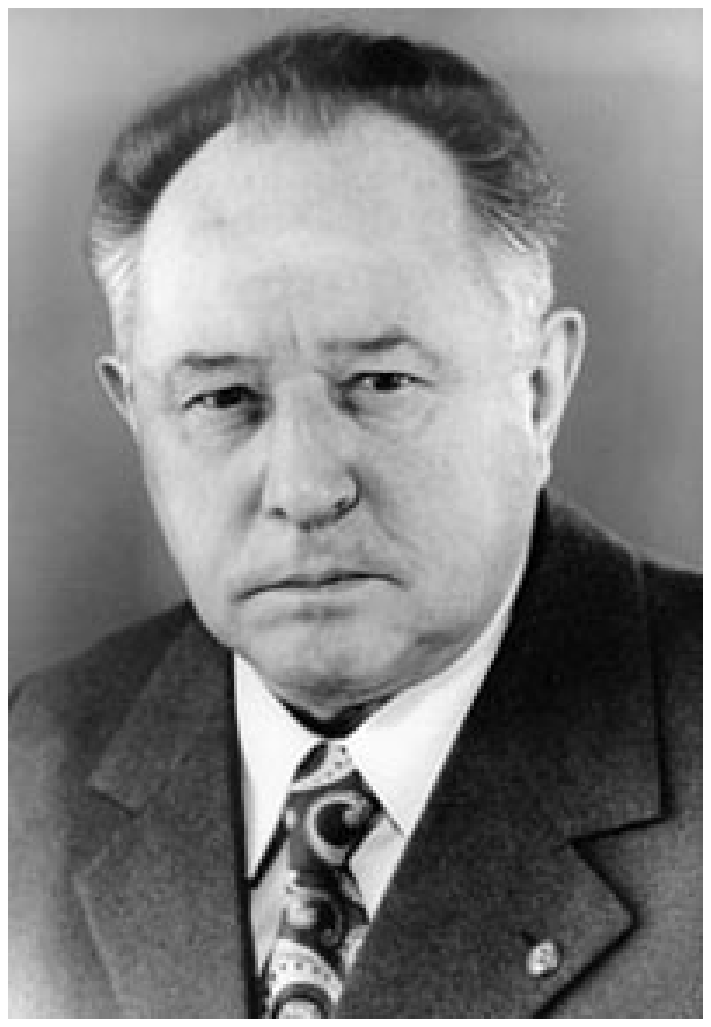,height=2cm} & \\
 & \mbox{{\bf\large Erich}} & \\[.2cm]
\psfig{file=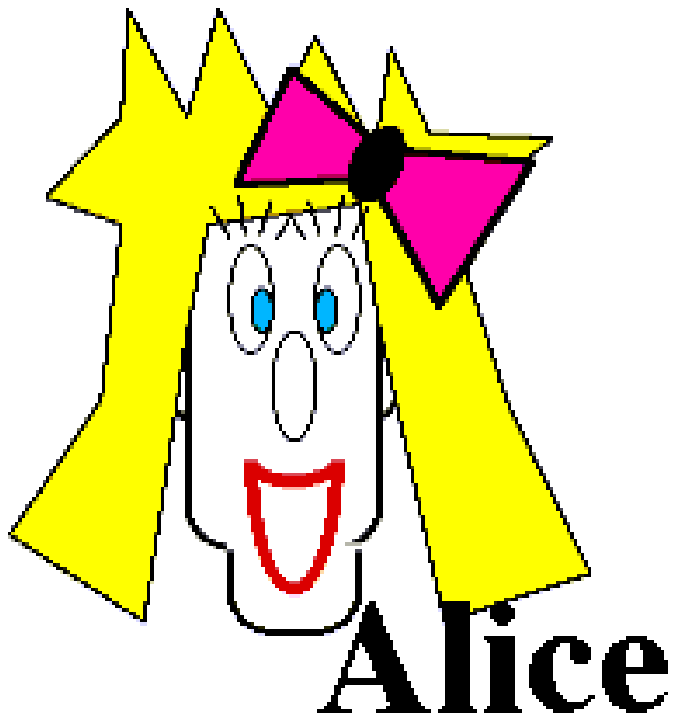,height=2cm} & 
\psfig{file=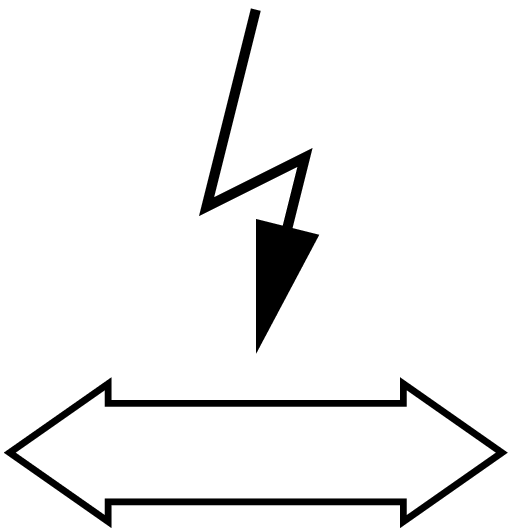,width=2cm} & 
\psfig{file=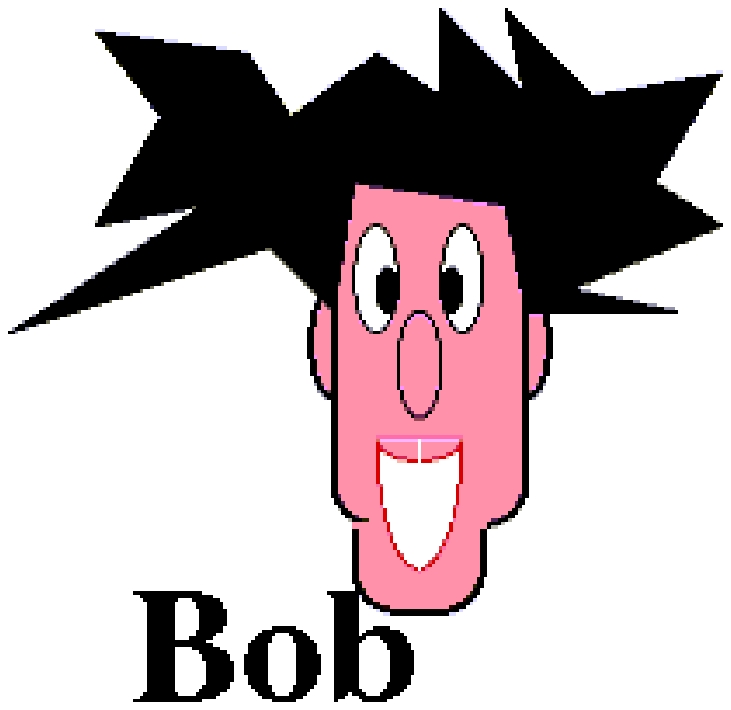,height=2cm} 
\end{array}
\]

Using a cryptosystem $(\mathcal{P}, \mathcal{C}, \mathcal{K}, \mathcal{E},
\mathcal{D})$, Alice and Bob are communicating over an insecure channel in the
presence of eavesdropper Erich.  Recall that $\mathcal{P}$, $\mathcal{C}$, and
$\mathcal{K}$ are finite sets.  Erich reads a cipher text,~$c \in
\mathcal{C}$, and tries to get some information about the corresponding plain
text,~$p \in \mathcal{P}$.  The plain texts are distributed on $\mathcal{P}$
according to a probability distribution $\mbox{\rm Pr}_{\mathcal{P}}$ that may
depend on the language used.  For each new plain text, Alice chooses a new key
from $\mathcal{K}$ that is independent of the plain text to be encrypted.  The
keys are distributed according to a probability distribution $\mbox{\rm
  Pr}_{\mathcal{K}}$ on~$\mathcal{K}$.  The distributions $\mbox{\rm
  Pr}_{\mathcal{P}}$ and $\mbox{\rm Pr}_{\mathcal{K}}$ induce a probability
distribution $\mbox{\rm Pr} = \mbox{\rm Pr}_{\mathcal{P} \times \mathcal{K}}$
on $\mathcal{P} \times \mathcal{K}$.  Thus, for each plain text $p$ and each
key~$k$,
\[
\mbox{\rm Pr}(p, k) = \mbox{\rm Pr}_{\mathcal{P}}(p) \; \mbox{\rm
  Pr}_{\mathcal{K}}(k)
\]
is the probability that the plain text $p$ is encrypted with the key~$k$,
where $p$ and $k$ are independent.

$\mbox{\rm Pr}(p) = \mbox{\rm Pr}_{\mathcal{P}}(p)$ is the probability that
the plain text $p$ will be encrypted.  
Similarly, $\mbox{\rm Pr}(k) = \mbox{\rm Pr}_{\mathcal{K}}(k)$ is the
probability that the key $k$ will be used. 
Let $c$ be another random variable whose distribution is determined by the
system used.  Then, $\mbox{\rm Pr}(p \, |\, c)$ is the probability that
  $p$ is encrypted under the condition that $c$ is received.
Erich knows the cipher text $c$, and he knows the probability
  distribution~$\mbox{\rm Pr}_{\mathcal{P}}$, since he knows the language used
  by Alice and Bob.

\begin{definition}
  A cryptosystem $(\mathcal{P}, \mathcal{C}, \mathcal{K}, \mathcal{E},
  \mathcal{D})$ provides {\em perfect secrecy\/} if and only if
\[
(\forall p \in \mathcal{P})\, (\forall c \in \mathcal{C})\, [\mbox{\rm Pr}(p
\, |\, c) = \mbox{\rm Pr}(p)].
\]
\end{definition}

That is, a cryptosystem achieves perfect secrecy if the event that some
plain text $p$ is encrypted and the event that some cipher text $c$ is received
are independent: Erich learns nothing about $p$ from knowing~$c$.  The
following example of a cryptosystem that does not provide perfect secrecy is due
to Buchmann~\cite{buc:b:kryptographie}.

\begin{example}[Perfect secrecy]

  Let $\mathcal{P}$, $\mathcal{C}$, and $\mathcal{K}$ be given such that:
\begin{itemize}
\item $\mathcal{P} = \{0,1\}$, where $\mbox{\rm Pr}(0) = \frac{1}{4}$ and
  $\mbox{\rm Pr}(1) = \frac{3}{4}$;
  
\item $\mathcal{K} = \{A,B\}$, where $\mbox{\rm Pr}(A) = \frac{1}{4}$ and
  $\mbox{\rm Pr}(B) = \frac{3}{4}$;
  
\item $\mathcal{C} = \{a,b\}$.
\end{itemize}

It follows that, for example, the probability that a ``$1$'' occurs and is
encrypted with the key $B$ is:
\[
\mbox{\rm Pr}(1,B) = \mbox{\rm Pr}(1) \cdot \mbox{\rm Pr}(B) = \frac{3}{4}
\cdot \frac{3}{4} = \frac{9}{16}.
\]
Let the encryption functions be given by:
\[
\begin{array}{cccc}
 E_A(0) = a; & E_A(1) = b; & E_B(0) = b; & E_B(1) = a.
\end{array}
\]
Hence, the probability that the cipher text $a$ occurs is:
\[
\mbox{\rm Pr}(a) = \mbox{\rm Pr}(0,A) + \mbox{\rm Pr}(1,B) = \frac{1}{16} +
\frac{9}{16} = \frac{5}{8}.
\]
Similarly, the probability that the cipher text $b$ occurs is:
\[
\mbox{\rm Pr}(b) = \mbox{\rm Pr}(1,A) + \mbox{\rm Pr}(0,B) = \frac{3}{16} +
\frac{3}{16} = \frac{3}{8}.
\]
Then, for each pair $(p, c) \in \mathcal{P} \times \mathcal{C}$, the
conditional probability $\mbox{\rm Pr}(p \, |\, c)$ is:
\begin{eqnarray*}
\mbox{\rm Pr}(0 \, |\,  a) = \frac{\mbox{\rm Pr}(0,A)}{\mbox{\rm Pr}(a)} = 
                        \frac{\frac{1}{16}}{\frac{5}{8}} = 
                        \frac{1}{10} ; & \hspace*{1cm} & 
\mbox{\rm Pr}(0 \, |\,  b) = \frac{\mbox{\rm Pr}(0,B)}{\mbox{\rm Pr}(b)} = 
                        \frac{\frac{3}{16}}{\frac{3}{8}} = 
                        \frac{1}{2} ;\\
\mbox{\rm Pr}(1 \, |\,  a) = \frac{\mbox{\rm Pr}(1,B)}{\mbox{\rm Pr}(a)} = 
                        \frac{\frac{9}{16}}{\frac{5}{8}} = 
                        \frac{9}{10} ; & \hspace*{1cm} & 
\mbox{\rm Pr}(1 \, |\,  b) = \frac{\mbox{\rm Pr}(1,A)}{\mbox{\rm Pr}(b)} = 
                        \frac{\frac{3}{16}}{\frac{3}{8}} = 
                        \frac{1}{2} .
\end{eqnarray*}
In particular, it follows that
\[
\mbox{\rm Pr}(0) = \frac{1}{4} \neq \frac{1}{10} = \mbox{\rm Pr}(0 \, |\,  a) ,
\]
and thus the given cryptosystem does not provide perfect secrecy: If Erich
sees the cipher text~$a$, he can be pretty sure that the encrypted plain text
was a~``$1$.''
\end{example}

\begin{theorem}[Shannon~\cite{sha:j:secrecy}]
  Let $S = (\mathcal{P}, \mathcal{C}, \mathcal{K}, \mathcal{E}, \mathcal{D})$
  be a cryptosystem with $||\mathcal{C}|| = ||\mathcal{K}||$ and $\mbox{\rm
    Pr}(p) > 0$ for each $p \in \mathcal{P}$.  Then, $S$ provides perfect
  secrecy if and only if
\begin{enumerate}
\item[{\rm{}(1)}] $\mbox{\rm Pr}_{\mathcal{K}}$ is the uniform distribution,
  and
\item[{\rm{}(2)}] for each $p \in \mathcal{P}$ and for each $c \in
  \mathcal{C}$, there exists a unique key $k \in \mathcal{K}$ with $E_k(p) =
  c$.
\end{enumerate}
\end{theorem}

\begin{proof}
  Assume that $S$ provides perfect secrecy.  We show that the conditions~(1)
  and~(2) hold.
  
  Condition~(2): Fix a plain text $p \in \mathcal{P}$.  Suppose that there is a
  cipher text $c \in \mathcal{C}$ such that for all $k \in \mathcal{K}$, it
  holds that $E_k(p) \neq c$.  Thus,
\[
\mbox{\rm Pr}(p) \neq 0 = \mbox{\rm Pr}(p \, |\, c),
\]
which implies that $S$ does not provide perfect secrecy, a contradiction.
Hence,
\[
(\forall c \in \mathcal{C})\, (\exists k \in \mathcal{K})\, [E_k(p) = c] .
\]
Now, $||\mathcal{C}|| = ||\mathcal{K}||$ implies that each cipher text $c \in
\mathcal{C}$ has a unique key $k$ with $E_k(p) = c$.

Condition~(1): Fix a cipher text $c \in \mathcal{C}$.  For $p \in
\mathcal{P}$, let $k(p)$ be the unique key $k$ with $E_k(p) = c$.  By
Bayes's theorem, for each $p \in \mathcal{P}$, we have:
\begin{equation}
\label{equ:bayes}
\mbox{\rm Pr}(p \, |\, c) = \frac{\mbox{\rm Pr}(c \, |\, p ) \; 
\mbox{\rm Pr}(p)}{\mbox{\rm Pr}(c)}
 = \frac{\mbox{\rm Pr}(k(p)) \; \mbox{\rm Pr}(p)}{\mbox{\rm Pr}(c)} .
\end{equation}
Since $S$ provides perfect secrecy, we have $\mbox{\rm Pr}(p \, |\, c) =
\mbox{\rm Pr}(p)$.  By Equation~(\ref{equ:bayes}), this implies $\mbox{\rm
  Pr}(k(p)) = \mbox{\rm Pr}(c)$, and this equality holds independently of~$p$.

Hence, the probabilities $\mbox{\rm Pr}(k)$ are equal for all $k \in
\mathcal{K}$, which implies $\mbox{\rm Pr}(k) = \frac{1}{||\mathcal{K}||}$.
Thus, $\mbox{\rm Pr}_{\mathcal{K}}$ is the uniform distribution.
  
Conversely, suppose that conditions~(1) and~(2) hold.  We show that $S$
provides perfect secrecy.  Let $k = k(p, c)$ be the unique key $k$ with $E_k(p)
= c$.  By Bayes's theorem, it follows that
\begin{eqnarray}
\label{equ:cond}
\mbox{\rm Pr}(p \, |\, c) & = &
\frac{\mbox{\rm Pr}(p) \; \mbox{\rm Pr}(c \, |\, p )}{\mbox{\rm Pr}(c)} 
\nonumber \\
                     & = & 
\frac{\mbox{\rm Pr}(p) \; \mbox{\rm Pr}(k(p, c))}
     {\sum_{q \in \mathcal{P}} \mbox{\rm Pr}(q) \; \mbox{\rm Pr}(k(q, c))} .
\end{eqnarray}
Since all keys are uniformly distributed, it follows that
\[
\mbox{\rm Pr}(k(p, c)) = \frac{1}{||\mathcal{K}||} .
\]
Moreover, we have that
\[
{\sum_{q \in \mathcal{P}} \mbox{\rm Pr}(q) \; \mbox{\rm Pr}(k(q, c))}
 = \frac{\sum_{q \in \mathcal{P}} \mbox{\rm Pr}(q)}{||\mathcal{K}||}
 = \frac{1}{||\mathcal{K}||} .
\]
Substituting this equality in Equation~(\ref{equ:cond}) gives:
\[
\mbox{\rm Pr}(p \, |\, c) = \mbox{\rm Pr}(p) .
\]
Hence, $S$ provides perfect secrecy.
\end{proof}

\subsection{Vernam's One-Time Pad}

The Vernam one-time pad is a symmetric cryptosystem that does provide perfect
secrecy.  It was invented by Gilbert Vernam in 1917,\footnote{Slightly
  differing from the system described here, Vernam's actual invention was a
  system with a finite period and hence did not provide perfect secrecy; see
  Kahn~\cite{kah:b:codebreakers} on this point.
}
and is defined as
follows.  Let $\mathcal{P} = \mathcal{C} = \mathcal{K} = \{0,1\}^n$ for some
$n \in \nats$.  For $k \in \{0,1\}^n$, define
\begin{itemize}
\item the encryption function $E_k : \{0,1\}^n \rightarrow \{0,1\}^n$ by
\[
E_k (p) = p \oplus k \mod 2\; \mbox{, and}
\]

\item the decryption function $D_k : \{0,1\}^n \rightarrow \{0,1\}^n$ by
\[
D_k (c) = c \oplus k  \mod 2,
\]
\end{itemize}
where $\oplus$ denotes bit-wise addition modulo~$2$.  The keys are uniformly
distributed on $\{0,1\}^n$.  Note that for each plain text $p$ a new key $k$
is chosen from~$\{0,1\}^n$.

By Shannon's Theorem, the one-time pad provides perfect secrecy, since for each
plain text $p \in \mathcal{P}$ and for each cipher text $c \in \mathcal{C}$,
there exists a unique key $k \in \mathcal{K}$ with $c = p \oplus k$, namely
the string $k = c \oplus p$.

However, the one-time pad has major disadvantages that make it impractical to
use in most concrete scenarios: To obtain perfect secrecy, every key can be
used only once, and it must be at least as long as the plain text to be
transmitted.  Surely, since for every communication a new secret key at least
as long as the plain text must be transmitted, this results in a vicious
circle.  Despite these drawbacks, for the perfect secrecy it provides, the
one-time pad has been used in real-world applications such as, allegedly, the
hotline between Moscow and Washington,
see~\cite[p.~316]{sim:j:symmetric-asymmetric-encryption}.

\section{RSA Cryptosystem}
\label{sec:rsa-system}

The RSA cryptosystem, named after its inventors Ron Rivest, Adi Shamir, and
Leonard Adleman, is the first public-key
cryptosystem~\cite{riv-sha-adl:j:rsa}.  It is still widely used in
cryptographic applications today.  Again, the scenario is that Alice and Bob
want to exchange messages over an insecure channel on which Erich is an
eavesdropper:

\[
\begin{array}{ccc}
 & \psfig{file=mielke.ps,height=2cm} & \\
 & \mbox{{\bf\large Erich}} & \\[.2cm]
\psfig{file=alice.ps,height=2cm} & 
\psfig{file=channel.eps,width=2cm} & 
\psfig{file=bob.ps,height=2cm} 
\end{array}
\]

In order to describe how the RSA cryptosystem works, we first need some
preliminaries from elementary number theory.

\subsection{Euler and Fermat's Theorems}

The {\em greatest common divisor\/} of two integers $a$ and $b$ is denoted by
$\mbox{gcd}(a,b)$.  For $n \in \nats$, define the set
\[
\integers_{n}^{\ast} = \{ i \condition 1 \leq i \leq n-1 \mbox{ and }
\mbox{gcd}(i,n) = 1\} .
\]

The {\em Euler function\/} $\phi$ is defined by $\phi(n) = ||
\integers_{n}^{\ast} ||$.  Note that $\integers_{n}^{\ast}$ is a group (with
respect to multiplication) of order~$\phi(n)$.  The following useful
properties of $\phi$ follow from the definition:
\begin{itemize}
\item $\phi(m \cdot n) = \phi(m) \cdot \phi(n)$ for all $m, n \in \nats$ with
  $\mbox{gcd}(m,n) = 1$, and

\item $\phi(p) = p - 1$ for all primes~$p$.
\end{itemize}
We will specifically use that $\phi(n) = (p-1)(q-1)$, where $p$ and $q$ are
primes and $n = pq$. 

Euler's Theorem below is a special case (for the group~$\integers_{n}^{\ast}$)
of Langrange's Theorem, which states that for each element $g$ of a finite
multiplicative group $G$ having order $|G|$ and the neutral element~$1$, it
holds that $g^{|G|} = 1$.

\begin{theorem}[Euler]
  For each $a \in \integers_{n}^{\ast}$, $a^{\phi(n)} \equiv 1 \mod n$.
\end{theorem}

The special case of Euler's Theorem with $n$ being a prime not dividing $a$ is
known as Fermat's Little Theorem.

\begin{theorem}[Fermat's Little Theorem]
\label{thm:fermat}
  If $p$ is a prime and $a \in \integers_{p}^{\ast}$, then $a^{p-1} \equiv 1
  \mod p$.
\end{theorem}

\subsection{RSA}
\label{sec:rsa}

\noindent
{\bf (1) Key generation:}
\begin{enumerate}
\item Bob chooses randomly two large primes $p$ and $q$ with $p \neq q$, and
  computes their product $n = pq$.

\item Bob chooses a number $e \in \nats$ with 
\begin{eqnarray}
\label{equ:e}
1 < e < \phi(n) = (p-1)(q-1) & \mbox{ and } & \mbox{gcd}(e, \phi(n)) = 1 .
\end{eqnarray}

\item Bob computes the unique number $d$ satisfying
\begin{eqnarray}
\label{equ:d}
1 < d < \phi(n) & \mbox{ and } & e \cdot d \equiv 1 \mod \phi(n) .
\end{eqnarray}
That is, $d$ is the inverse of $e$ modulo~$\phi(n)$.

\item The pair $(n,e)$ is Bob's {\em public key}, and $d$ is Bob's {\em
    private key}.
\end{enumerate}

\begin{figure}[!ht]
\centering
\fbox{
\begin{minipage}{4.5in}
\begin{construction}
\item {\bf Euclid's Algorithm (extended)} 
  \begin{block}
    \item {\bf Input:} Two integers, $b_0$ and~$b_1$.
    \item {\bf begin} $x_0 := 1$;  $y_0 := 0$;  $x_1 := 0$;  $y_1 := 1$;  
                      $i:= 1$;
    \begin{block}
      \item {\bf while} $b_i$ does not divide $b_{i-1}$ {\bf do}
      \begin{block}
        \item {\bf begin}
        \begin{block}
           \item $q_i := \left\lfloor \frac{b_{i-1}}{b_i}  \right\rfloor$;
           \item $b_{i+1} := b_{i-1} - q_i \cdot b_i$;
           \item $x_{i+1} := x_{i-1} - q_i \cdot x_i$;
           \item $y_{i+1} := y_{i-1} - q_i \cdot y_i$;
           \item $i := i+1$
        \end{block}
        \item {\bf end}
      \end{block}
      \item {\bf begin output} 
        \begin{block}
           \item $b := b_i$; 
                 \hfill $(\ast$ $b = \mbox{gcd}(b_0, b_1) = 1$ $\ast)$
           \item $x: = x_i$;
           \item $y := y_i$
                         \hfill $(\ast$ $y$ is the inverse of $b_1 \mod b_0$
                         $\ast)$
        \end{block}
      \item {\bf end output}
    \end{block}
    \item {\bf end}
  \end{block}
\end{construction}
\end{minipage}}
\caption{The extended algorithm of Euclid.
\label{fig:euklid}
}
\end{figure}

In order to generate two large primes (e.g., primes with 80 digits each)
efficiently, one can choose large numbers at random and test them for
primality. 
Since by the Prime Number Theorem, the number of primes not exceeding $N$ is
approximately $\frac{N}{\ln N}$, the odds of hitting a prime are good after a
reasonably small number of trials.  To verify the primality of 
the number picked, one usually
makes use of a randomized polynomial-time primality test such as the Monte
Carlo\footnote{A Monte Carlo algorithm is a randomized algorithm whose ``yes''
  answers are reliable, while its ``no'' answers may be erroneous with a
  certain error probability, or vice versa.  The corresponding complexity
  classes are called R and coR, respectively,
  see~\cite{gil:j:probabilistic-tms}.  In contrast, a Las Vegas algorithm may
  for certain sequences of coin flips halt without giving an answer at all,
  but whenever it gives an answer, this answer is correct.  The corresponding
  class, $\zpp = \rp \cap \cor$, was also defined by
  Gill~\cite{gil:j:probabilistic-tms}.}
algorithm of Rabin~\cite{rab:j:probabilistic-algorithms-for-primality} that is
related to a deterministic algorithm due to
Miller~\cite{mil:j:riemann-primes}; their primality test is known as the
Miller-Rabin test.  An alternative, though less popular Monte Carlo algorithm
was proposed by Solovay and
Strassen~\cite{sol-str:j:monte-carlo-for-primality}.  The reason why the 
Solovay-Strassen test is less popular than the Miller-Rabin test is that
it is less efficient and less accurate.
These two primality
tests, along with a careful complexity analysis and the required
number-theoretical background, can be found in, e.g., the books by
Stinson~\cite{sti:b:cryptography} and
Salomaa~\cite{sal:b:public-key-cryptography}.  Additional primality tests are
contained in~\cite{gol:b:foundations,buc:b:kryptographie}.

\begin{quote}
  {\em Note Added in Proof\/}: Quite recently, Agrawal et
  al.~\cite{agr-kay-sax:m:primes-in-p} designed a deterministic
  polynomial-time algorithm for primality.  Their breakthrough result is a
  milestone in complexity theory and solves a long-standing open problem.  It
  is unlikely, though, that this algorithm will have immediate consequences
  for cryptographic applications, since Agrawal et
  al.~\cite{agr-kay-sax:m:primes-in-p} note that their algorithm has a running
  time of roughly $n^{12}$ and thus is much less efficient than the
  probabilistic primality tests currently in use.
\end{quote}

We now argue that the keys can be computed efficiently.  In particular, the
inverse $d$ of $e$ modulo~$\phi(n)$ can be computed efficiently via the
extended algorithm of Euclid; see Figure~\ref{fig:euklid}.

\begin{lemma}
\label{lem:euklid}
On input $b_0 = \phi(n)$ and~$b_1 = e$, the extended algorithm of Euclid
computes in polynomial time integers $x$ and $y$ such that
\[
x \cdot \phi(n) + y \cdot e \equiv 1 \mod \phi(n) .
\]
Thus, $y$ is the inverse of $e$ modulo~$\phi(n)$, and Bob chooses
$d \equiv y \mod \phi(n)$ as his private key.
\end{lemma}

\begin{example}
  Bob chooses the primes $p = 11$ and $q = 23$, and computes their product $n
  = 253$ and $\phi(253) = 10 \cdot 22 = 220$.  The smallest possible $e$
  satisfying Equation~{\rm{}(\ref{equ:e})} is $e = 3$.  The extended algorithm
  of Euclid yields the following sequence of $b_i$, $x_i$, and~$y_i$:
\[
\begin{array}{|r||r|r|r|r|}
\hline\hline
i & b_i & x_i & y_i & q_i \\ 
\hline
0 & 220 & 1 & 0 & \mbox{--} \\
1 & 3 & 0 & 1 & 73 \\
2 & 1 & \mbox{\boldmath $1$} & \mbox{\boldmath $- 73$} & \mbox{--} \\ 
\hline\hline
\end{array}
\]
Since $1 \cdot 220 + (- 73) \cdot 3 = 220 - 219 \equiv 1 \mod 220$, the unique
value $d = -73 + 220 = 147$ computed by Bob satisfies
Equation~{\rm{}(\ref{equ:d})} and is the inverse of $e = 3$ modulo~$220$.
\end{example}

\medskip 
\noindent
{\bf (2) Encryption:} We assume that messages over some alphabet $\Sigma$ are
block-wise encoded as positive integers with a fixed block length.  Suppose
that~$m < n$ is the message Alice wants to send to Bob.
Alice knows Bob's public key $(n, e)$ and computes the encryption $c =
E_{(n,e)}(m)$ of~$m$, where the encryption function is defined by
\[
E_{(n,e)}(m) = m^e \mod n .
\]

Performed naively, this computation may require a large number of
multiplications, depending on the choice of~$e$.  To ensure efficient
encryption, we will employ a ``fast exponentiation'' algorithm called
``square-and-multiply,'' see Figure~\ref{fig:square-and-multiply} below.

\begin{figure}[!ht]
\centering
\fbox{
\begin{minipage}{4.5in}
\noindent
{\bf Square-and-Multiply Algorithm}
\begin{algorithm} 
\item[{\bf Input}:] $m, n, e \in \nats$, where $m < n$.
\item Let the binary expansion of the exponent $e$ be given by
\[
e = \sum_{i = 0}^{k} e_i 2^{i} , \quad \mbox{where $e_i \in \{0,1\}$.} 
\]
\item Successively compute $m^{2^i}$, where $0 \leq i \leq k$, using the
      equality
\[
m^{2^{i+1}} = \left( m^{2^i}\right)^{2}.
\]
It is not necessary to store the intermediate values of~$m^{2^i}$.

\item In the arithmetic modulo~$n$, compute
\begin{equation}
\label{equ:square-and-multiply}
m^e = \prod_{\stackrel{\mbox{\protect\scriptsize $i = 0$}}{e_i = 1}}^{k}
  m^{2^i}.
\end{equation}
\label{item:square-and-multiply:3}

\item[{\bf Output}:] $m^e$.
\end{algorithm}
\end{minipage}}
\caption{The square-and-multiply algorithm.
\label{fig:square-and-multiply}
}
\end{figure}

Equation~{\rm{}(\ref{equ:square-and-multiply})} in
Step~\ref{item:square-and-multiply:3} of Figure~\ref{fig:square-and-multiply}
is correct, since
\[
m^e = m^{\sum_{i = 0}^{k} e_i 2^{i}} = 
\prod_{i = 0}^{k} \left( m^{2^i}\right)^{e_i} = 
\prod_{\stackrel{\mbox{\protect\scriptsize $i = 0$}}{e_i = 1}}^{k} m^{2^i} .
\]

Hence, instead of $e$ multiplications, Alice need compute no more than $2
\log e$ multiplications.  Thus, the square-and-multiply method speeds up the
encryption exponentially.  

\begin{example}
Suppose Alice wants to compute $c = 6^{17} \mod 100$.  The binary expansion of
the exponent is $17 = 1 + 16 = 2^0 + 2^4$.
\begin{enumerate}
\item Alice successively computes:
\[
\begin{array}{lclcl}
6^{2^0} & =      & 6^1             & =      & 6 ;\\
6^{2^1} & =      & 6^2             & =      & 36 ;\\
6^{2^2} & =      & 36^2            & \equiv & -4 \mod 100 ; \\
6^{2^3} & \equiv & (-4)^2 \mod 100 & \equiv & 16 \mod 100 ; \\
6^{2^4} & \equiv & 16^2 \mod 100   & \equiv & 56 \mod 100 .
\end{array}
\]

\item Alice computes her cipher text
\[
\begin{array}{lclcl}
c & = & 6^{17} \mod 100 & \equiv & 6 \cdot 6^{2^4} \mod 100 \\
  &   &                 & \equiv & 6 \cdot 56 \mod 100 \\
  &   &                 & \equiv & 36 \mod 100 .
\end{array}
\]
Note that only four squarings and one multiplication are needed for her to
compute the cipher text.
\end{enumerate}
\end{example}

\medskip 
\noindent
{\bf (3) Decryption:} Let $c$, $0 \leq c < n$, be the cipher text sent to Bob;
$c$ is subject to eavesdropping by Erich.  Bob decrypts $c$ using his
private key $d$ and the following decryption function:
\[
D_d (c) = c^d \mod n .
\]
Again, the fast exponentiation algorithm described in
Figure~\ref{fig:square-and-multiply} ensures that the legal recipient Bob can
decrypt the cipher text efficiently.  Thus, the RSA protocol is feasible.  To
prove that it is correct, we show that Equation~{\rm{}(\ref{equ:correctness})}
is satisfied.

\begin{figure}[!htb]
\centering
\begin{tabular}{||c||c|c|c||}
\hline\hline
\parbox[t]{.5cm}{\bf Step} & 
\psfig{file=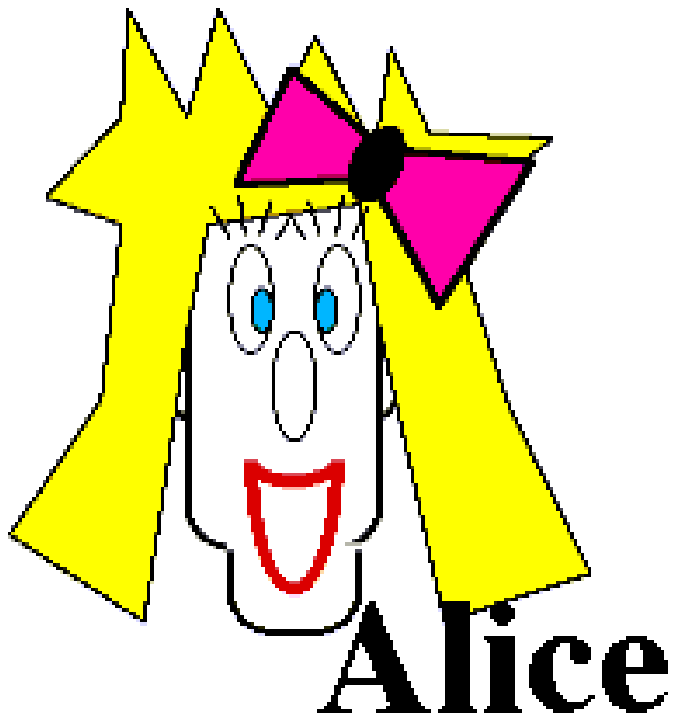,height=2cm} & 
\psfig{file=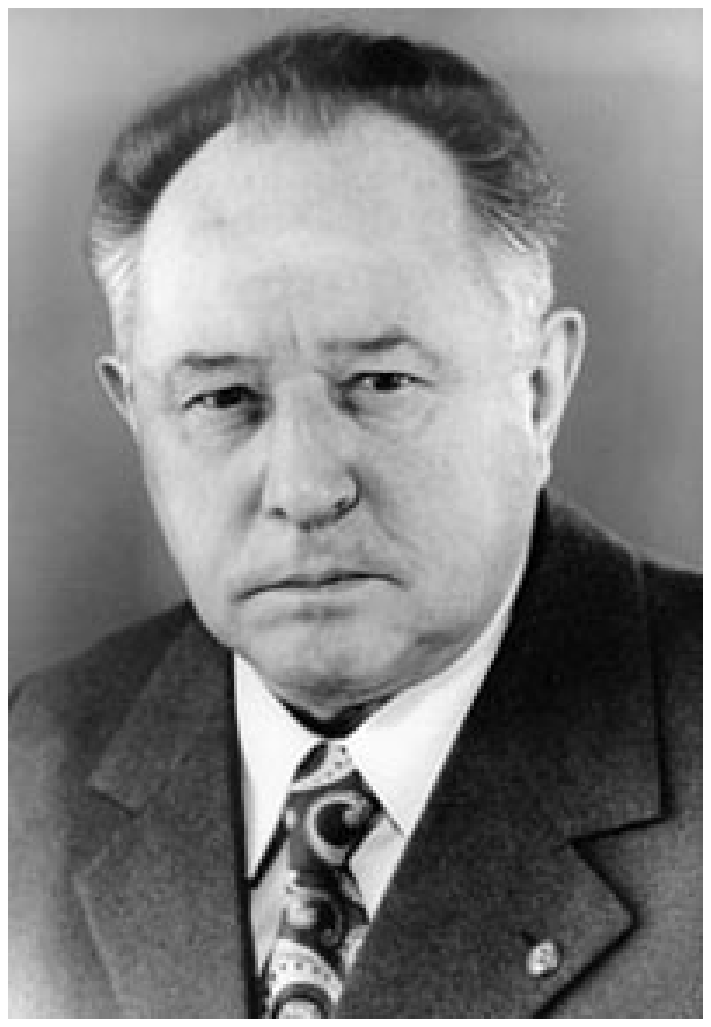,height=2cm} & 
\psfig{file=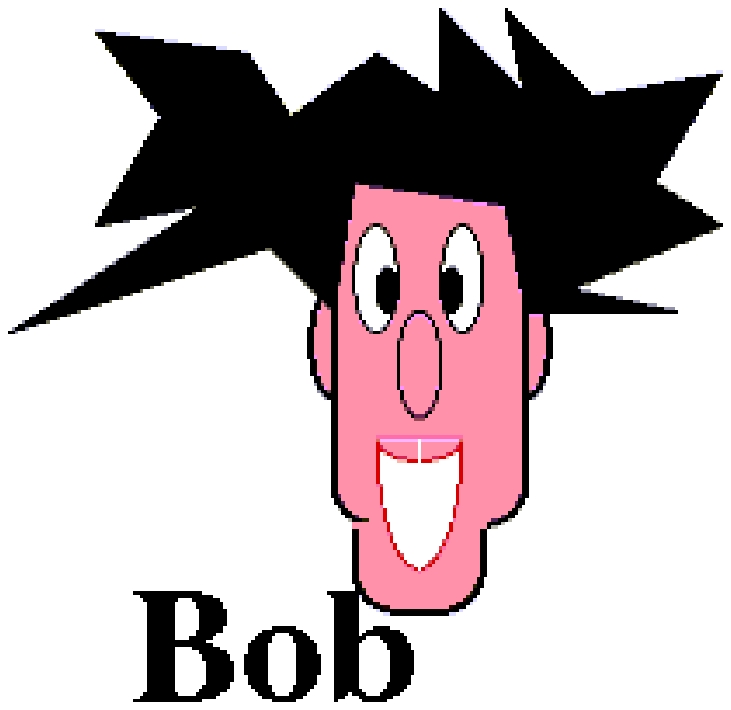,height=2cm} \\ \hline\hline
{\bf 1} & & & \parbox[t]{5.5cm}
  {chooses large primes $p$, $q$ at random, computes $n = pq$ 
   and $\phi(n) = (p-1)(q-1)$, his public key $(n, e)$ 
   with $e$ satisfying Eq.~{\rm{}(\ref{equ:e})}, and 
   his private key $d$ satisfying Eq.~{\rm{}(\ref{equ:d})}} \\ \hline
{\bf 2} & & 
\mbox{\huge $\stackrel{\mbox{\normalsize $(n, e)$}}{\Leftarrow}$} & \\ \hline
{\bf 3} & \parbox[t]{3cm}
  {encrypts message $m$ by computing 
\[
c = m^e \mod n
\]
}
 & & \\ \hline
{\bf 4} & & 
\mbox{\huge $\stackrel{\mbox{\normalsize $c$}}{\Rightarrow}$} & \\ \hline
{\bf 5} & 
 & & \parbox[t]{5.5cm}
  {decrypts cipher text $c$ by computing 
\[
m = c^d = \left(m^e\right)^d \mod n
\]
}
 \\ \hline\hline
\end{tabular}
\caption{The RSA protocol.
\label{fig:rsa}
}
\end{figure}

Figure~\ref{fig:rsa} summarizes the single steps of the RSA protocol and
displays the information communicated by Alice and Bob that is subject to
eavesdropping by Erich.

\begin{theorem}
  Let $(n, e)$ and $d$ be Bob's public and private key in the RSA
  protocol.  Then, for each message $m$ with $0 \leq m < n$,
\[
m = \left( m^e \right)^d \mod n .
\] 
That is, RSA is a public-key cryptosystem.
\end{theorem}

\begin{proof}
Since $e \cdot d \equiv 1 \mod \phi(n)$ by Equation~{\rm{}(\ref{equ:d})},
there exists an integer $t$ such that
\[
e \cdot d = 1 + t(p-1)(q-1) ,
\]
where $n = pq$.  It follows that
\[
\begin{array}{lclcl}
\left( m^e \right)^d & = & m^{e \cdot d}& = & m^{1 + t(p-1)(q-1)} \\
                     &   &              & = & m \left(m^{t(p-1)(q-1)}\right) \\
                     &   &              & = & m \left(m^{p-1}\right)^{t(q-1)}.
\end{array}
\]
Hence, we have
\begin{equation}
\label{equ:mmodp}
\left( m^e \right)^d \equiv m \mod p ,
\end{equation}
since if $p$ divides $m$ then both sides of Equation~{\rm{}(\ref{equ:mmodp})}
are $0 \mod p$, and if $p$ does not divide $m$ (i.e., $\mbox{gcd}(p,m) = 1$)
then by Fermat's Little Theorem, we have
\[
m^{p-1} \equiv 1 \mod p .
\]
By a symmetric argument, it holds that 
\[
\left( m^e \right)^d \equiv m \mod q .
\]
Since $p$ and $q$ are primes with $p \neq q$, it follows from the Chinese
Remainder Theorem (see, e.g., \cite{knu:b2:2} or~\cite{sti:b:cryptography})
that
\[
\left( m^e \right)^d \equiv m \mod n.
\]
Since $m < n$, the claim follows.
\end{proof}

\subsection{RSA Digital Signature Protocol}
\label{sec:digital-sign-rsa}

\begin{figure}[!htb]
\centering
\begin{tabular}{||c||c|c|c||}
\hline\hline
\parbox[t]{.5cm}{\bf Step} & 
\psfig{file=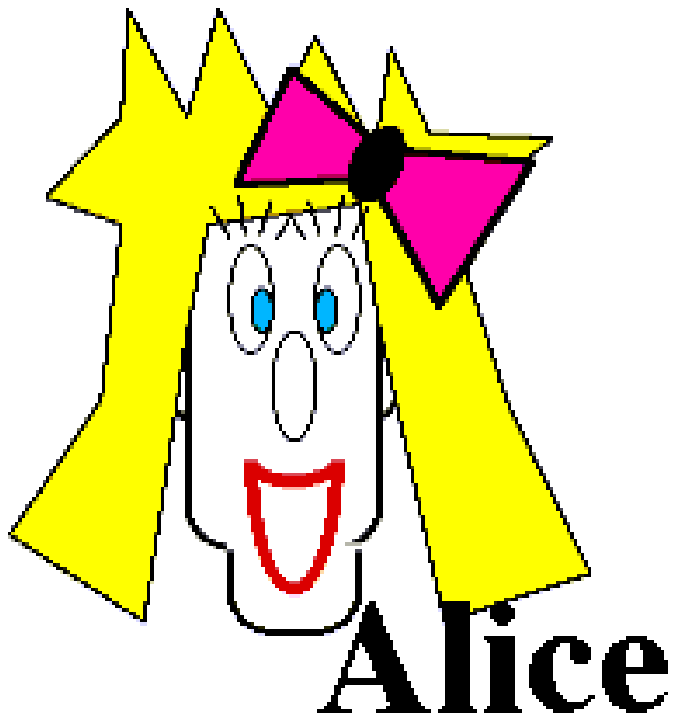,height=2cm} & 
\psfig{file=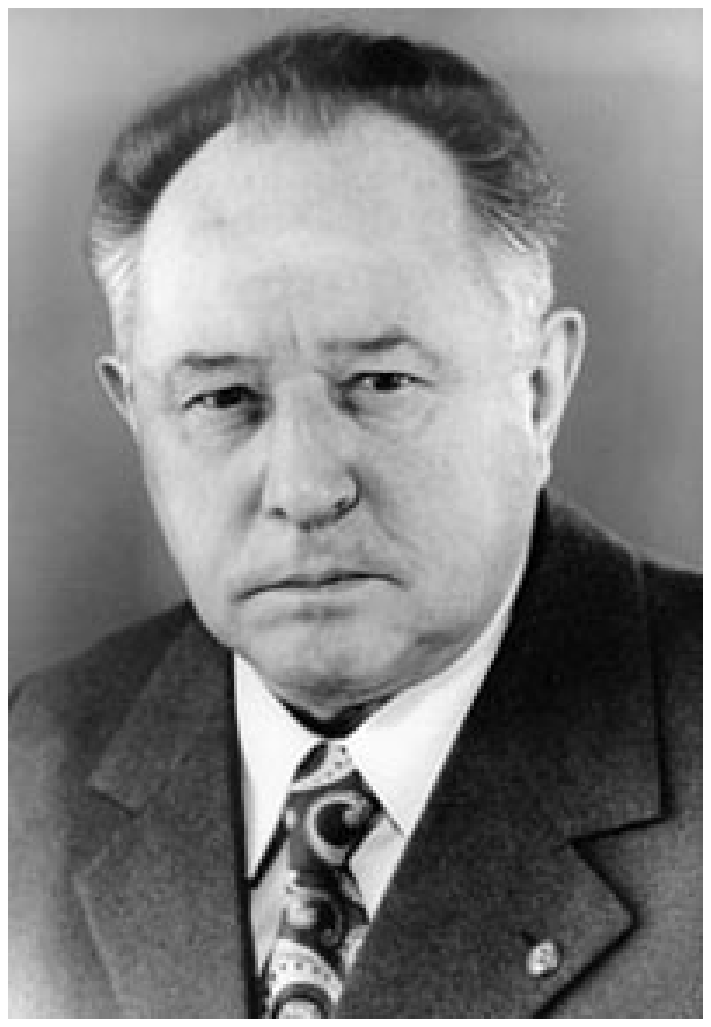,height=2cm} & 
\psfig{file=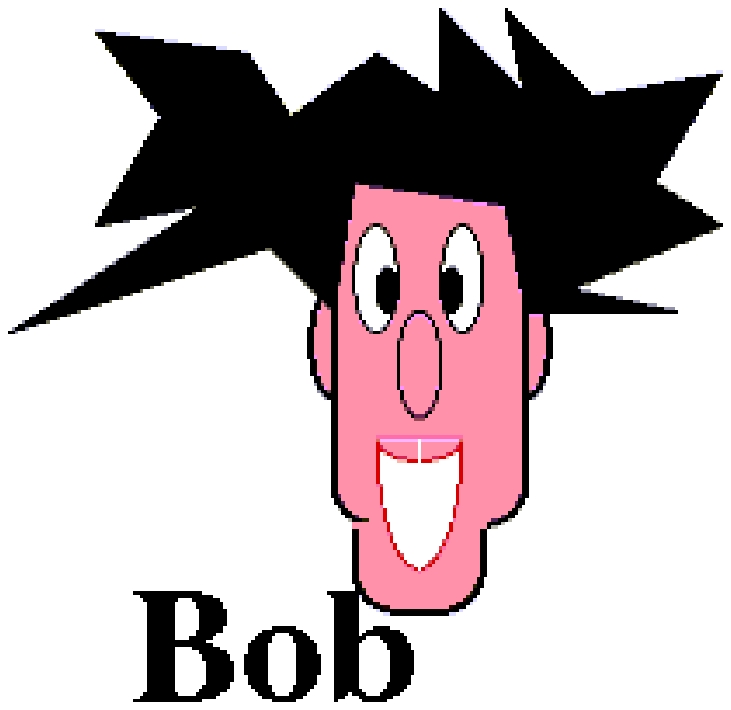,height=2cm} \\ \hline\hline
{\bf 1} & \parbox[t]{4cm}
  {chooses $n = pq$, her public key $(n, e)$, and her private key $d$ as in
    the RSA protocol, see Section~\ref{sec:rsa}} & & \\ \hline
{\bf 2} & \parbox[t]{4cm}
  {computes her signature
\[
\mbox{sig}_A (m) = m^d \mod n
\]
  for the message~$m$}
 & & \\ \hline
{\bf 3} & & 
\mbox{\huge $\stackrel{\mbox{\normalsize $m, \mbox{sig}_A (m)$}}{\Rightarrow}$} & \\ \hline
{\bf 4} & 
 & & \parbox[t]{4cm}
  {verifies Alice's signature by checking the congruence
\[
m \equiv \left(\mbox{sig}_A (m)\right)^e \mod n
\]}
 \\ \hline\hline
\end{tabular}
\caption{The RSA digital signature protocol.
\label{fig:rsa-digital-signature}
}
\end{figure}

The RSA public-key cryptosystem described in
Section~\ref{sec:rsa} can be modified so as to yield a digital signature
protocol.  Figure~\ref{fig:rsa-digital-signature} shows how the RSA digital
signature protocol works.  A chosen-plain-text attack on the RSA digital
signature scheme, and countermeasures to avoid it, are described in
Section~\ref{sec:security-rsa}.

\subsection{Security of RSA and Possible Attacks on RSA}
\label{sec:security-rsa}

The security of the RSA cryptosystem strongly depends on whether factoring
large integers is intractable.  It is widely believed that there is no
efficient factoring algorithm, since no such algorithm could be designed as
yet, despite considerable efforts in the past.  However, it is not known
whether the problem of factoring large integers is as hard as the problem of
cracking the RSA system.

Here is a list of potential attacks on the RSA system.  To preclude these
direct attacks, some care must be taken in choosing the primes $p$ and~$q$,
the modulus~$n$, the exponent~$e$, and the private key~$d$.  For further
background on the security of the RSA system and on proposed attacks to break
it, the reader is referred
to~\cite{bon:j:rsa-attacks,sha:j:rsa-for-paranoids,kal-rob:j:secure-use-rsa,moo:b:protocol-failures}.
For each attack on RSA that has been proposed in the literature to date, some
practical countermeasures are known, rules of thumb that prevent the success
of those attacks or, at least, that make their likelihood of success
negligibly small.

\begin{description}
\item[Factoring attacks:] The aim of the attacker Erich is to use the
  public key $(n, e)$ to recover the private key $d$ by factoring~$n$, i.e.,
  by computing the primes $p$ and $q$ with $n = pq$.  Knowing $p$ and~$q$, he
  can just like Bob compute $\phi(n) = (p-1)(q-1)$ and thus the inverse $d$ of
  $e$ modulo~$\phi(n)$, using the extended algorithm of Euclid; see
  Figure~\ref{fig:euklid} and Lemma~\ref{lem:euklid}.  There are various ways
  in which Erich might mount this type of attack on RSA\@.
\begin{itemize}
\item {\em Brute-force attack\/}: Erich might try to factor the modulus $n$
  simply by exhaustive search of the complete key space.  Choosing $n$
  sufficiently large will prevent this type of attack.  Currently, it is
  recommended to use moduli $n$ with at least 768 bits, i.e., the size of 512
  bits formerly in use no longer provides adequate protection today.  Of
  course, the time complexity of modular exponentiation grows rapidly with the
  modulus size, and thus there is a tradeoff between increasing the security
  of RSA and decreasing its efficiency.
  
  It is also generally accepted that those moduli $n$ consisting of prime
  factors $p$ and $q$ of roughly the same size are the hardest to factor.
  
\item {\em General-purpose factoring methods\/}: Examples of such general
  factoring algorithms are the {\em general number field sieve\/} (see,
  e.g.,~\cite{len-len:b:number-field-sieve}) or the older {\em quadratic
    sieve} (see, e.g.,~\cite{buc:b:kryptographie,sti:b:cryptography}). They
  are based on the following simple idea.  Suppose $n$ is the number to be
  factorized.  Using the respective ``sieve,'' one determines integers $a$ and
  $b$ such that
\begin{eqnarray}
\label{equ:sieve}
a^2 \equiv b^2 \mod n & \mbox{and} & a \not\equiv \pm b \mod n .
\end{eqnarray}
Thus, $n$ divides $a^2 - b^2 = (a-b)(a+b)$, but neither $a-b$ nor $a+b$.
Hence, $\mbox{gcd}(a-b,n)$ is a nontrivial factor of~$n$.  The general number
field sieve and the quadratic sieve differ in the specific way the integers
$a$ and $b$ satisfying Equation~(\ref{equ:sieve}) are found.

\item {\em Special-purpose factoring methods\/}: Depending on the form of
  the primes $p$ and~$q$, it might be argued that using special-purpose
  factoring methods such as Pollard's ``$p-1$
  method''~\cite{pol:j:factorization} may be more effective and more
  successful than using general-purpose factoring methods.  This
  potential threat led to the introduction of {\em strong primes\/} that
  resist such special-purpose factoring methods.  A strong prime $p$ is
  required to satisfy certain conditions such as that $p-1$ has a large
  factor~$r$ and $r-1$, in turn, has a large factor, etc.
  
\item {\em Elliptic curve method\/}: This factoring method was introduced
  by Lenstra~\cite{len:j:elliptic-curves}, and it has some success probability
  regardless of the form of the primes chosen.  Consequently, the most
  effective countermeasure against the elliptic curve method is to use primes
  of very large size.  This countermeasure simultaneously provides, with a
  very high probability, protection against all known types of special-purpose
  factoring methods.  In short, randomly chosen large primes are more
  important than strong primes.  Note that weak primes are believed to be
  rare; Pomerance and Sorenson~\cite{pom-sor:j:weak-primes} study the density
  of weak primes.
  
\item {\em Factoring on a quantum computer\/}: Last, we mention that Shor's
  algorithm for factoring large numbers on a quantum
  computer~\cite{shor:j:factoring-on-quantum-computer} poses a potential
  threat to the security of RSA and other cryptosystems whose security relies
  on the hardness of the factoring problem.  More precisely, Shor's efficient
  quantum algorithm determines the order of a given group element, a problem
  closely related to the factoring problem.  Using Miller's randomized
  reduction~\cite{mil:j:riemann-primes}, if one can efficiently compute the
  order of group elements, then one can efficiently solve the factoring
  problem.  However, the quantum computer is a theoretical construct
  currently.  Whether or not Shor's quantum factoring algorithm will be a
  practical threat remains to be seen in the future.
\end{itemize}

\item[Superencryption:] Early on Simmons and
  Norris~\cite{sim-nor:j:comments-on-rsa} proposed an attack on RSA called
  superencryption.  This attack is based on the observation that a sufficient
  number of encryptions will eventually recover the original message, since
  the RSA encryption function is an injective mapping onto a finite set, which
  makes the graph of the function a union of disjoint cycles.  This attack is
  a threat to the security of RSA, provided that the number of encryptions
  required is small.  Luckily, superencryption is not a practical attack if
  the primes are large and are chosen at random.
  
\item[Wiener's attack:] Wiener~\cite{wie:j:cryptanalysis-of-rsa} proposed an
  attack on the RSA system by a continued fraction approximation, using the
  public key $(n, e)$ to provide sufficient information to recover the private
  key~$d$.  More precisely, Wiener proved that if the keys in the RSA system
  are chosen such that $n = pq$, where $q < p < 2q$, and $d < \frac{1}{3}
  \sqrt[4]{n}$, then given the public key $(n, e)$ with $ed \equiv 1 \mod
  \phi(n)$ the private key $d$ can be computed in linear time.
  
  Here is a proof sketch of Wiener's result (see~\cite{bon:j:rsa-attacks}).
  Since $ed \equiv 1 \mod \phi(n)$, there exists a $k$ such that $ed - k
  \phi(n) = 1$, which implies that $\frac{k}{d}$ is an approximation of
  $\frac{e}{\phi(n)}$:
\begin{eqnarray}
\label{equ:wiener}
\left| \frac{e}{\phi(n)} - \frac{k}{d} \right| & = & 
\left| \frac{1}{d \phi(n)} \right| . 
\end{eqnarray}
Erich does not know~$\phi(n)$, but he can use $n$ in place of~$\phi(n)$.
Using $ed - k \phi(n) = 1$ and the easily verified fact that $|n - \phi(n)| <
3 \sqrt{n}$, in place of Equation~(\ref{equ:wiener}) we now have
\[
\left| \frac{e}{n} - \frac{k}{d} \right|
 = \left| \frac{1 - k(n - \phi(n))}{d n} \right| 
 \leq \left| \frac{3k\sqrt{n}}{d n} \right| 
 = \frac{3k}{d \sqrt{n}}.
\]
Since $k \phi(n) = ed -1 < ed$ and $e < \phi(n)$, we have $k < d < \frac{1}{3}
  \sqrt[4]{n}$.  Hence,
\[
\left| \frac{e}{n} - \frac{k}{d} \right|
 < \frac{1}{d\sqrt[4]{n}}
 < \frac{1}{2d^2}.
\]
There are at most $\log n$ fractions $\frac{k}{d}$ with $d < n$ approximating
$\frac{e}{n}$ so tightly, and they can be obtained by computing the $\log n$
convergents of the continued fraction expansion of $\frac{e}{n}$
(see~\cite[Thm.~177]{har-wri:b:number}).  Since $ed - k \phi(n) = 1$, we have
$\mbox{gcd}(k, d) = 1$, so $\frac{k}{d}$ is a reduced fraction.

Note that this attack is efficient and practical, and thus is a concern, only
if the private key $d$ is chosen to be small relative to~$n$.  For example, if
$n$ is a 1024 bits number, then $d$ must be at least 256 bits long in order to
prevent Wiener's attack.  A small value of~$d$, however, enables fast
decryption and in particular is desirable for low-power devices such as
``smartcards.''~~Therefore, Wiener proposed certain techniques that avoid his
attack.  

The first technique is to use a large encryption exponent,
say~$\tilde{e} = e + \ell \phi(n)$ for some large~$\ell$.  For a large
enough~$\tilde{e}$, the factor $k$ in the above proof is so large that
Wiener's attack cannot be mounted, regardless of how small $d$ is.

The second technique uses the Chinese Remainder Theorem to speed up
decryption, even if $d$ is not small.  Let $d$ be a large decryption exponent
such that both $d_p \equiv d \mod p-1$ and $d_q \equiv d \mod q-1$ are small.
Then, one can decrypt a given cipher text $c$ as follows.  Compute $m_p =
c^{d_p} \mod p$ and $m_q = c^{d_q} \mod q$, and use the Chinese Remainder
Theorem to obtain the unique solution $m$ modulo $n = pq$ of the two equations
$m = m_p \mod p$ and $m = m_q \mod q$.  The point is that although $d_p$ and
$d_q$ are small, $d$ can be chosen large enough to resist Wiener's attack.

Boneh and Durfee~\cite{bon-dur:j:improving-wiener-attack} recently improved
Wiener's result: Erich can efficiently compute $d$ from $(n, e)$ provided that
$d < n^{0.292}$.

\item[Small-message attack:] RSA encryption is not effective if both the
  message $m$ to be encrypted and the exponent $e$ to be used for encryption
  are small relative to the modulus~$n$.  In particular, if $c = m^e < n$ is
  the cipher text, then $m$ can be recovered from $c$ by ordinary root
  extraction.  Thus, either the public exponent should be large or the
  messages should always be large.  It is this latter suggestion that is more
  useful, for a small public exponent is often preferred in order to speed up
  the encryption and to preclude Wiener's attack.
 
\item[Low-exponent attack:] One should take precautions, though, not to choose
  the public exponent too small.  A preferred value of $e$ that has been used
  often in the past is $e = 3$.  However, if three parties participating in
  the same system encrypt the same message $m$ using the same public
  exponent~$3$, although perhaps different moduli $n_1$, $n_2$, and~$n_3$,
  then one can easily compute $m$ from the three cipher texts:
\begin{eqnarray*}
c_1 & = & m^3 \mod n_1 \\
c_2 & = & m^3 \mod n_2 \\
c_3 & = & m^3 \mod n_3 .
\end{eqnarray*}
In particular, the message $m$ must be smaller than the moduli, and so $m^3$
will be smaller than $n_1 n_2 n_3$.  Using the Chinese Remainder Theorem (see,
e.g., \cite{knu:b2:2,sti:b:cryptography}), one can compute the unique solution
\[
c = m^3 \mod n_1  n_2 n_3 = m^3 .
\]
Hence, one can compute $m$ from $c$ by ordinary root extraction.

More generally, suppose that $k$ related plain texts are encrypted with the
same exponent~$e$:
\begin{eqnarray*}
c_1 & = & (a_1 m + b_1)^e \mod n_1 \\
c_2 & = & (a_2 m + b_2)^e \mod n_2 \\
 & \vdots & \\ 
c_k & = & (a_k m + b_k)^e \mod n_k ,
\end{eqnarray*}
where $a_i$ and $b_i$, $1 \leq i \leq k$, are known and $k > \frac{e(e+1)}{2}$
and $\min(n_i) > 2^{e^2}$.  Then, an attacker can solve for $m$ in polynomial
time using lattice reduction techniques.  This observation is due to Johan
H{\aa}stad~\cite{has:j:solving-low-degree-equations}, and his ``broadcast
attack'' has been strengthened by Don
Coppersmith~\cite{cop:j:low-exponent-rsa-attacks}.  This attack is a
concern if the messages are related in a known way.  Padding the messages with
pseudorandom strings prior to encryption prevents mounting this attack in
practice, see, e.g.,~\cite{kal-rob:j:secure-use-rsa}.  If the messages are
related in a known way, they should not be encrypted with many RSA keys.

A recommended value of $e$ that is commonly used today is $e = 2^{16} +1$.
One advantage of this value for $e$ is that its binary expansion has only two
ones, which implies that the square-and-multiply algorithm of
Figure~\ref{fig:square-and-multiply} requires very few
operations,\footnote{How many exactly?}  
and so is very efficient.

\item[Forging RSA signatures:] This attack is based on the fact that the RSA
  encryption function is a homomorphism: if $(n, e)$ is the public key and
  $m_1$ and $m_2$ are two messages then
\begin{equation}
\label{equ:homomorph-rsa}
m_{1}^{e} \cdot m_{2}^{e} \equiv \left(m_{1} \cdot m_{2}\right)^{e} \mod n .
\end{equation}
Another identity that can easily be verified is:
\begin{equation}
\label{equ:identity-rsa}
\left( m \cdot r^e \right)^d \equiv m^d \cdot r \mod n .
\end{equation}
In particular, these identities can be used to mount an attack on the digital
signature scheme based on the RSA algorithm, see
Figure~\ref{fig:rsa-digital-signature} and Section~\ref{sec:digital-sign-rsa}.
Given previous message-signature pairs $(m_1 , \mbox{sig}_A (m_1)), \ldots ,
(m_k , \mbox{sig}_A (m_k))$, Erich can use the
congruences~(\ref{equ:homomorph-rsa}) and~(\ref{equ:identity-rsa}) to compute
a new message-signature pair $(m, \mbox{sig}_A (m))$ by
\begin{eqnarray*}
m & = & r^e \prod_{i = 1}^{k} m_i^{e_i} \mod n ; \\
\mbox{sig}_A (m) & = & 
r \prod_{i = 1}^{k} \left(\mbox{sig}_A (m_i) \right)^{e_i} \mod n ,
\end{eqnarray*}
where $r$ and the $e_i$ are arbitrary.  Hence, Erich can forge Alice's
signature without knowing her private key, and Bob will not detect the
forgery, since $m \equiv \left(\mbox{sig}_A (m)\right)^e \mod n$.  Note that,
in Equation~(\ref{equ:homomorph-rsa}), even if $m_{1}$ and~$m_{2}$ are
meaningful plain texts, $m_{1} \cdot m_{2}$ usually is not.  Thus, Erich can
forge Alice's signature only for messages that may or may not be useful.
However, he might choose the messages $m_i$ so as to generate a meaningful
message $m$ with a forged digital signature.  This {\em chosen-plain-text
  attack\/} can again be avoided by pseudorandom padding techniques that
destroy the algebraic relations between messages.  Pseudorandom padding is
also a useful countermeasure against the following {\em chosen-cipher-text
  attack\/}: Erich intercepts some cipher text~$c$, chooses $r \in \nats$ at
random, and computes $c \cdot r^e \mod n$, which he sends to the legitimate
receiver Bob.  By Equation~(\ref{equ:identity-rsa}), Bob will decrypt the
string $\hat{c} = c^d \cdot r \mod n$, which is likely to look like a random
string.  Erich, however, if he were to get his hands on~$\hat{c}$, could
obtain the original message $m$ by multiplying by~$r^{-1}$, the inverse of $r$
modulo~$n$, i.e., by computing $m = r^{-1} \cdot c^d \cdot r \mod n$.
\end{description}

\section{Protocols for Secret-Key Agreement, Public-Key Encryption, and
  Digital Signatures}
\label{sec:protocols}

Consider again a scenario where Alice and Bob want to exchange messages over
an insecure channel such as a public telephone line, and where Erich is an
eavesdropper: 

\[
\begin{array}{ccc}
 & \psfig{file=mielke.ps,height=2cm} & \\
 & \mbox{{\bf\large Erich}} & \\[.2cm]
\psfig{file=alice.ps,height=2cm} & 
\psfig{file=channel.eps,width=2cm} & 
\psfig{file=bob.ps,height=2cm} 
\end{array}
\]
This is why Alice and Bob want to encrypt their messages.  For efficiency
purposes, they decide to use a symmetric cryptosystem in which they both
possess the same key for encryption and for decryption; recall
Definition~\ref{def:cryptosystem}.  But then, how can they agree on a joint
secret key when they can communicate only over an insecure channel?  If they
were to send an encrypted message containing the key to be used in subsequent
communications, which key should they use to encrypt {\em this\/} message?

This paradoxical situation is known as the {\em secret-key agreement\/}
problem, and it was considered to be unsolvable since the beginning of
cryptography.  It was quite a surprise when in 1976 Whitfield Diffie and
Martin Hellman~\cite{dif-hel:j:diffie-hellman} did solve this long-standing,
seemingly paradoxical problem by proposing the first secret-key agreement
protocol.  We describe their protocol in Section~\ref{sec:diffie-hellman}.
Interestingly, it was the Diffie--Hellman protocol that inspired Rivest,
Shamir, and Adleman to invent the RSA system.  That is, Diffie and Hellman's
key idea to solve the secret-key agreement problem opened the door to modern
public-key cryptography, which no longer requires sending secret keys over
insecure channels.  

Strangely enough, the reverse happened in the nonpublic sector.  The
Communications Electronics Security Group (CESG) of the
British Government Communications Head Quarters (GCHQ) claims to have invented
the RSA public-key cryptosystem prior to Rivest, Shamir, and Adleman and the
Diffie--Hellman secret-key agreement scheme independently of Diffie and
Hellman.  And they did so in reverse order. James Ellis first discovered the
principle possibility of public-key cryptography in the late sixties.  In
1973, Clifford Cocks developed the mathematics necessary to realize Ellis's
ideas and formulated what four years later became known as the RSA system.
Soon thereafter, inspired by Ellis's and Cocks's work, Malcolm Williamson
invented what became known as the Diffie--Hellman secret-key agreement scheme,
around the same time Diffie and Hellman succeeded.  None of the results of
Ellis, Cocks, and Williamson became known to the public then.  The full
story---or what of it is publicly known by now---is told in Singh's
book~\cite{sin:b:code-book}.

Section~\ref{sec:elgamal} shows how to modify the Diffie--Hellman protocol in
order to obtain a public-key cryptosystem.  This protocol is due to Taher
ElGamal~\cite{gam:j:public-key}.  Just like the Diffie--Hellman protocol,
ElGamal's cryptosystem is based on the difficulty of computing discrete
logarithms.

Section~\ref{sec:no-key} gives an interesting protocol due to an unpublished
work of Adi Shamir.  In this protocol, keys do not need to be agreed upon
prior to exchanging encrypted messages.

Another cryptographic task is the generation of {\em digital signatures\/}:
Alice wants to sign her encrypted messages to Bob in a way that allows Bob to
verify that Alice was indeed the sender of the message.  Digital signature
protocols are used for the authentication of documents such as email messages.
The goal is to preclude Erich from forging Alice's messages and her signature.
Digital signature protocols are described in
Section~\ref{sec:digital-sign-rsa} (RSA digital signatures), in
Section~\ref{sec:elgamal} (ElGamal digital signatures) and in
Section~\ref{sec:riv-rab-she} (Rabi and Sherman digital signatures).

\subsection{Diffie and Hellman's Secret-Key Agreement Protocol}
\label{sec:diffie-hellman}

\begin{figure}[!htb]
  \centering
\begin{tabular}{||c||c|c|c||}
\hline\hline
\parbox[t]{.5cm}{\bf Step} & 
\psfig{file=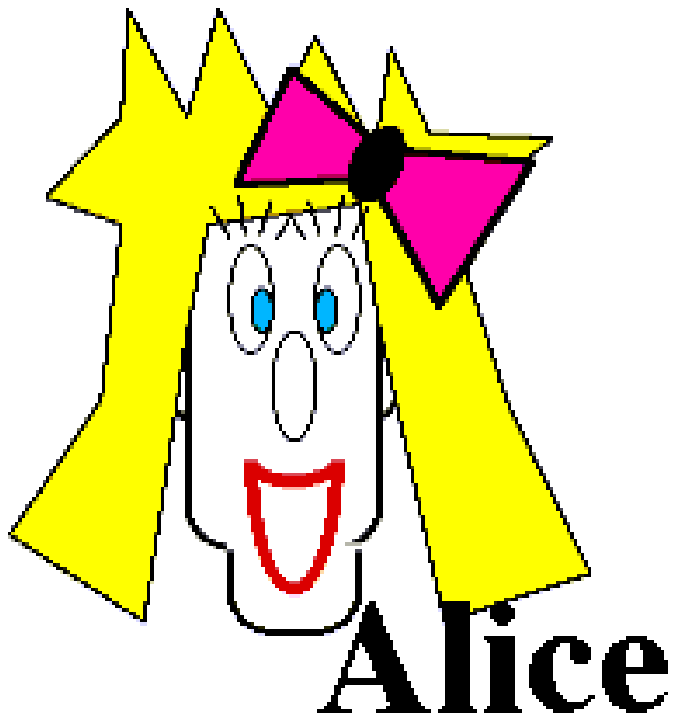,height=2cm} & 
\psfig{file=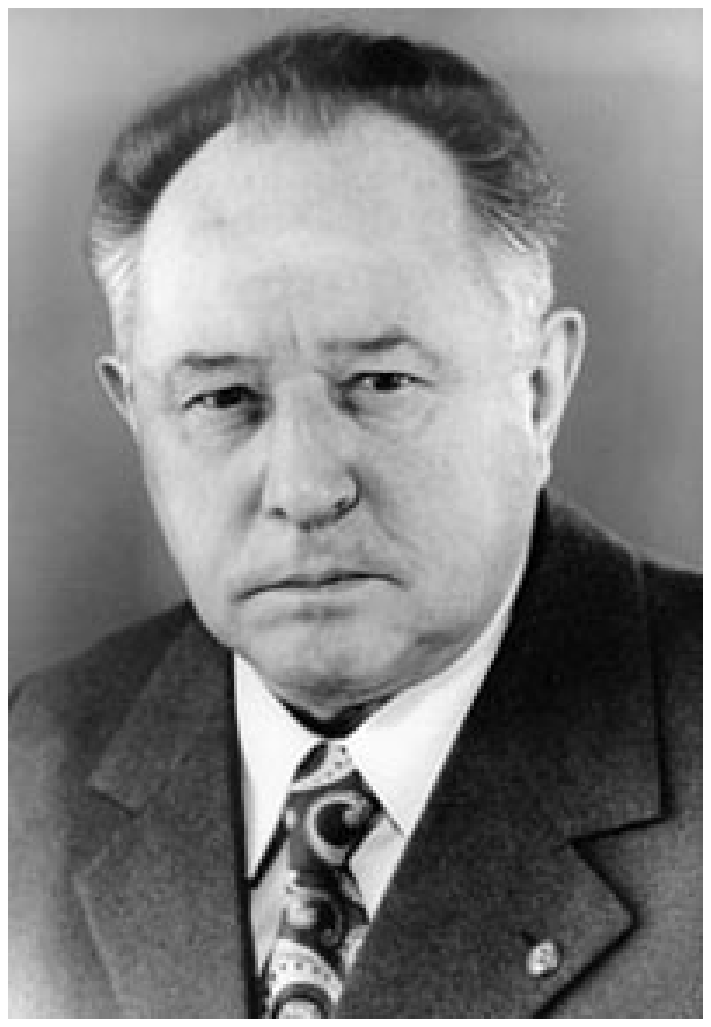,height=2cm} & 
\psfig{file=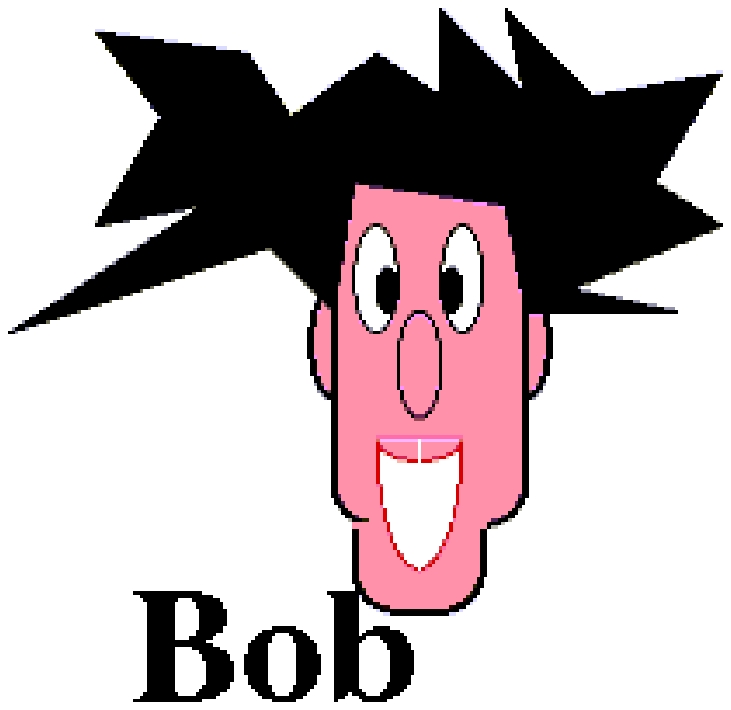,height=2cm} \\ \hline\hline
{\bf 1} & \multicolumn{3}{c||}
  {Alice and Bob agree upon a large prime~$p$ and a primitive root 
   $g$ of~$p$;} \\
 & \multicolumn{3}{c||}
  {$p$ and $g$ are public} \\ \hline
{\bf 2} & \parbox[t]{4cm}
  {chooses a large number $a$ at random, computes $\alpha = g^a \mod p$}
  & & \parbox[t]{4cm}
  {chooses a large number $b$ at random, computes $\beta = g^b \mod p$}
  \\ \hline
{\bf 3} & & 
\mbox{\huge $\stackrel{\mbox{\normalsize $\alpha$}}{\Rightarrow}$} & \\ 
 & & 
\mbox{\huge $\stackrel{\mbox{\normalsize $\beta$}}{\Leftarrow}$} & \\ \hline
{\bf 4} & \parbox[t]{4cm}
  {computes her key 
\[
k_A = \beta^a \mod p
\]
}
 & & \parbox[t]{4cm}
  {computes his key 
\[
k_B = \alpha^b \mod p
\]
} \\ \hline\hline
\end{tabular}
\caption{The Diffie--Hellman secret-key agreement protocol.
\label{fig:diffie-hellman}
}
\end{figure}

Figure~\ref{fig:diffie-hellman} shows how the Diffie--Hellman secret-key
agreement protocol works.  It is based on the modular exponential function
with base $g$ and modulus~$p$, where $p$ is a prime and $g$ is a primitive
root of $p$ in~$\integers_{p}^{\ast}$, the cyclic group of prime residues
modulo~$p$; recall that $\integers_{p}^{\ast}$ has order $\phi(p) = p-1$.
The formal definition is as follows.

\begin{definition}
\begin{itemize}
\item For $n \in \nats$, a {\em primitive root of~$n$\/} is any element $a \in
  \integers_{n}^{\ast}$ satisfying that, for each $d$ with $1 \leq d <
  \phi(n)$, it holds that
\[
a^d \not\equiv 1 \mod n .
\]
Equivalently, a primitive root of $n$ is a generator of~$\integers_{n}^{\ast}$.

\item Let $p$ be a prime, and let $g$ be a primitive root of~$p$.  
  The function
  $\alpha_{(g, p)} : \integers_{p-1} \rightarrow \integers_{p}^{\ast}$
  that is defined by
\[
\alpha_{(g, p)}(a) = g^a \mod p .
\]
is called the {\em modular exponential function with base $g$ and
  modulus~$p$}.  Its inverse function, which for fixed $p$ and $g$ maps
$\alpha_{(g, p)}(a)$ to~$a = \log_g \alpha \mod p$, is called the {\em
  discrete logarithm}.
\end{itemize}
\end{definition}

As noted above, every primitive root of $p$ generates the
entire group~$\integers_{p}^{\ast}$.  Moreover, 
$\integers_{p}^{\ast}$ has precisely
$\phi(p-1)$ primitive roots.  For example, $\integers_{5}^{\ast} = \{1, 2, 3,
4\}$ and $\integers_{4}^{\ast} = \{1, 3\}$, so $\phi(4) = 2$, and the two
primitive roots of~$5$ in $\integers_{5}^{\ast}$ are $2$ and~$3$, since
\[
\begin{array}{llll}
 2^1 = 2;\ & 2^2 = 4;\ & 2^3 \equiv 3 \mod 5;\ & 2^4 \equiv 1 \mod 5;\  \\
 3^1 = 3;\ & 3^2 \equiv 4 \mod 5;\ & 3^3 \equiv 2 \mod 5;\
 & 3^4 \equiv 1 \mod 5  .\
\end{array}
\]
Not every integer has a primitive root: $8$ is the smallest such example.  It
is known from elementary number theory that an integer $n$ has a primitive
root if and only if $n$ is $1$ or $2$ or~$4$, or is of the form $q^k$ or
$2q^k$ for some odd prime~$q$.

The protocol from Figure~\ref{fig:diffie-hellman} works, since
\[
k_A = \beta^a = g^{ba} = g^{ab} = \alpha^b = k_B .
\]
Thus, the keys computed by Alice and Bob indeed are the same.

Computing discrete logarithms is considered to be a very hard problem: no
efficient algorithms are known for solving it.  In contrast, the modular
exponential function can be computed efficiently, using the fast
exponentiation algorithm ``square-and-multiply'' described as
Figure~\ref{fig:square-and-multiply}.  That is why modular exponentiation is
considered to be a candidate for a 
``one-way function,'' i.e., a function that is easy to
compute but hard to invert.  Things are bad.  It is currently not known
whether or not one-way functions exist.  Things are worse.  Although they are
not known to exist, one-way functions play a key role in cryptography, and the
security of many cryptosystems is based on the assumption that one-way
functions do exist.  We will discuss the notion of one-way functions in more
detail in Section~\ref{sec:aowf}.

If Erich is listening carefully to Alice and Bob's communication in the
Diffie--Hellman protocol (see Figure~\ref{fig:diffie-hellman}), he knows $p$,
$g$, $\alpha$, and~$\beta$.  He wants to compute their joint secret key, $k_A
= k_B$.  This problem is known as the {\em Diffie--Hellman problem}.  If Erich
could solve the discrete logarithm problem efficiently, he could easily
compute $a = \log_g \alpha \mod p$ and $b = \log_g \beta \mod p$ and, thus,
$k_A = \beta^a \mod p$ and $k_B = \alpha^b \mod p$.  That is, the
Diffie--Hellman problem is no more difficult than the discrete logarithm
problem. The converse question---of whether the Diffie--Hellman problem is as
hard as the discrete logarithm problem---is still an unproven conjecture.
Fortunately, as noted above, the discrete logarithm problem is viewed as being
intractable, so this attack is very unlikely to be a practical threat.  On the
other hand, it is the only known attack for computing the keys directly from
$\alpha$ and $\beta$ in the Diffie--Hellman protocol.  Note, however, that no
proof of security for this protocol has been established up to date.

Note also that computing the keys $k_A = k_B$ directly from $\alpha$ and
$\beta$ is not the only possible attack on the Diffie--Hellman protocol.  For
example, it is vulnerable to the {\em Man-in-the-middle attack}.  Unlike 
passive attacks against the underlying mathematics of a
cryptosystem, in which an eavesdropper tries to gain information without 
affecting the protocol, the Man-in-the-middle attack is an active attack, in
which an eavesdropper attempts to alter the protocol to his own advantage.
That is, Erich, as the ``man in the middle,'' might
pretend to be Alice when communicating with Bob, and he might pretend to be
Bob when communicating with Alice.  He could intercept $\alpha = g^a \mod p$
that Alice sends to Bob and he could also intercept $\beta = g^b \mod p$ that
Bob sends to Alice, passing on his own values $\alpha_E$ in place of $\alpha$
to Bob and $\beta_E$ in place of $\beta$ to Alice.  That way Erich could
compute two (possibly distinct) keys, one for communicating with Alice, the
other one for communicating with Bob, without them having any clue that they
in fact are communicating with him.  
Thus, Alice and Bob cannot be certain of the authenticity of their respective
partners in the communication.  In Section~\ref{sec:zero-knowledge}, we
will introduce {\em zero-knowledge protocols}, which can be used to ensure
proper authentication.

\begin{figure}[!htp]
  \centering
\begin{tabular}{||c||c|c|c||}
\hline\hline
\parbox[t]{.5cm}{\bf Step} & 
\psfig{file=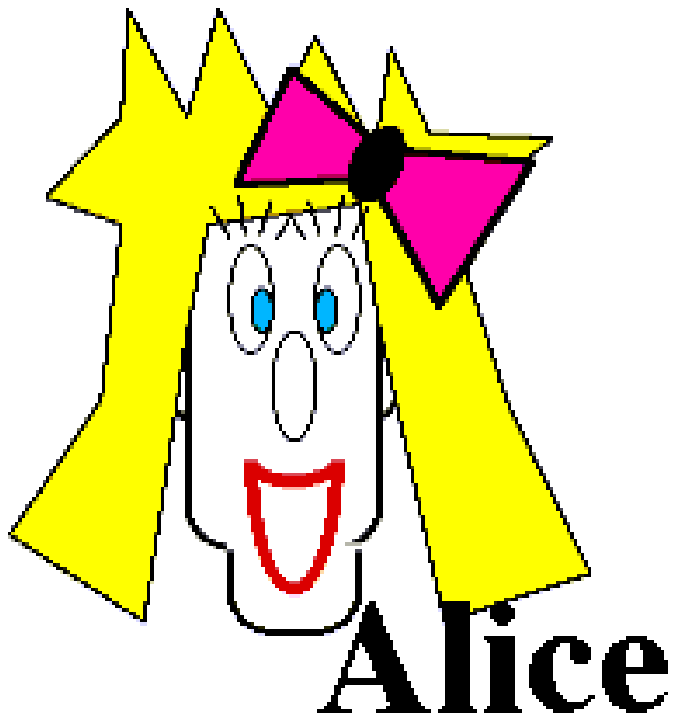,height=2cm} & 
\psfig{file=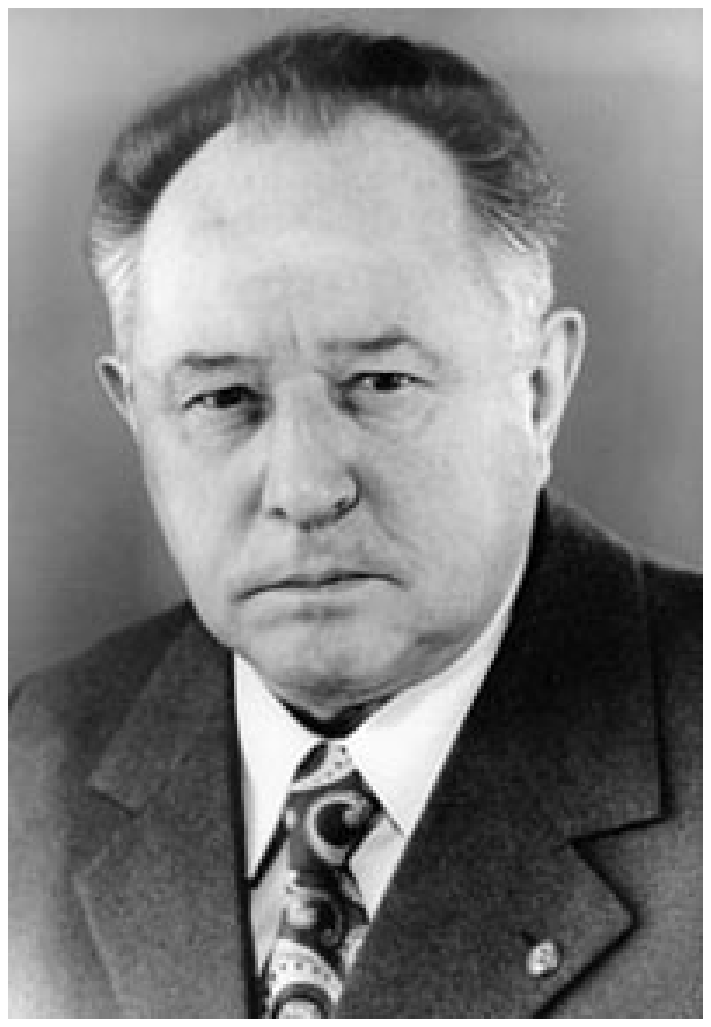,height=2cm} & 
\psfig{file=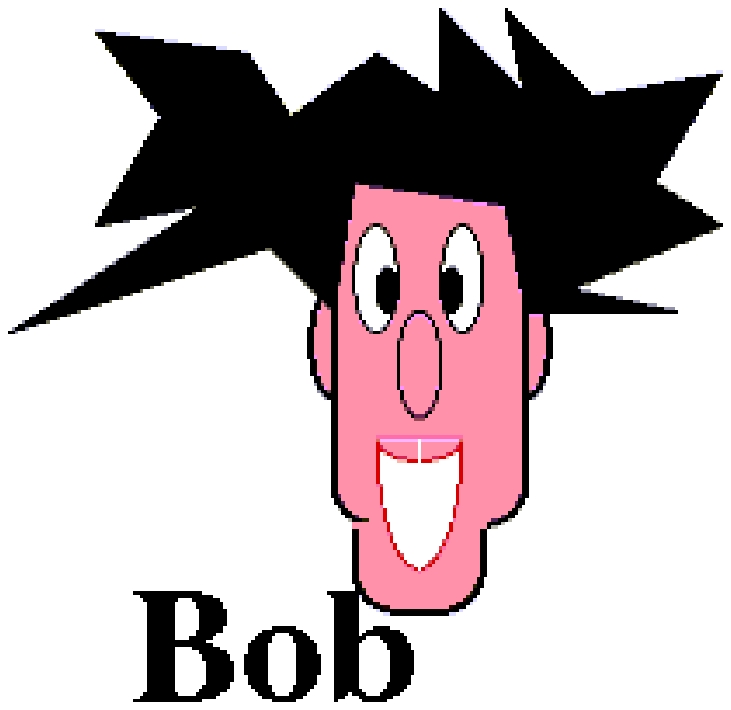,height=2cm} \\ \hline\hline
{\bf 1} & \multicolumn{3}{c||}
  {Alice and Bob agree upon a large prime~$p$ and a primitive root 
   $g$ of~$p$;} \\
 & \multicolumn{3}{c||}
  {$p$ and $g$ are public} \\ \hline
{\bf 2} & & & \parbox[t]{4cm}
  {chooses a large number $b$ at random as his private key
    and computes $\beta = g^b \mod p$}
  \\ \hline
{\bf 3} & & 
\mbox{\huge $\stackrel{\mbox{\normalsize $\beta$}}{\Leftarrow}$} & \\ \hline
{\bf 4} & \parbox[t]{4cm}
  {chooses a large number $a$ at random, computes $\alpha = g^a \mod p$, the
    key $k = \beta^a \mod p$, and the cipher text $c = E_k(m)$, where $m$ is
    the message to be sent} & & \\ \hline
{\bf 5} & & 
\mbox{\huge $\stackrel{\mbox{\normalsize $\alpha$, $c$}}{\Rightarrow}$} & \\ \hline
{\bf 6} & & & \parbox[t]{4cm}
  {computes $k = \alpha^b \mod p$ and 
   $m = D_k(c)$} \\ \hline\hline
\end{tabular}
\caption{A public-key cryptosystem based on the Diffie--Hellman protocol, which
  uses the encryption and decryption algorithms $E_k$ and $D_k$ of a given
  symmetric cryptosystem.
\label{fig:diffie-hellman-public-key}
}
\end{figure}

By slightly modifying the Diffie--Hellman protocol, it is possible to obtain a
public-key cryptosystem.  The variant of the Diffie--Hellman protocol
presented here 
in fact is a ``hybrid cryptosystem,'' a public-key cryptosystem making use of
a given symmetric cryptosystem.  Such hybrid systems are often useful in
practice, for they combine the advantages of asymmetric and symmetric
cryptosystems.  Symmetric systems are usually more efficient than public-key
systems.

The protocol works as follows.  Alice and Bob agree on a large prime~$p$ and a
primitive root $g$ of~$p$, which are public.  They also agree on some
symmetric cryptosystem $S = (\mathcal{P}, \mathcal{C}, \mathcal{K},
\mathcal{E}, \mathcal{D})$ with encryption functions $\mathcal{E} = \{E_k
\condition k \in \mathcal{K}\}$ and decryption functions $\mathcal{D} = \{D_k
\condition k \in \mathcal{K}\}$.  The subsequent steps of the protocol are
shown in Figure~\ref{fig:diffie-hellman-public-key}.  The message to be sent
is encrypted using the symmetric system~$S$, and the symmetric key $k$ used in
this encryption is transmitted in a Diffie--Hellman-like fashion.  This
modification of the original Diffie--Hellman protocol is the standard usage of
Diffie--Hellman.

The system in Figure~\ref{fig:diffie-hellman-public-key}
modifies the original Diffie--Hellman protocol in the following way.
While in the Diffie--Hellman scheme Alice and Bob {\em simultaneously\/}
compute and send their ``partial keys'' $\alpha$ and~$\beta$, respectively,
they do so {\em sequentially\/} in the protocol in
Figure~\ref{fig:diffie-hellman-public-key}.   That is, Alice must
wait for Bob's value~$\beta$, his public key, 
to be able to compute the key $k$ with which she
then encrypts her message~$m$ via the symmetric cryptosystem~$S$.  
Moreover,
Bob generates, once and for all, his public $\beta$ 
for possibly several communications with Alice, and also for possibly several
users other than Alice who might want to communicate with him.  In contrast,
Alice has to generate her $\alpha$ anew again and again every time she
communicates with Bob, just like in the original Diffie--Hellman protocol.
This modification of Diffie--Hellman is usually
referred to as Predistributed Diffie--Hellman.  In a {\em key
  distribution scheme}, one party chooses a key and then transmits it to
another party or parties over an insecure channel. In contrast, in a {\em
  secret-key agreement scheme\/} such as the original Diffie--Hellman protocol
from Figure~\ref{fig:diffie-hellman}, two or more parties jointly compute, by
communicating over an insecure channel, a shared secret key, which depends on
inputs from both or all parties.

\subsection{ElGamal's Public-Key Cryptosystem and Digital Signature Protocol}
\label{sec:elgamal}

Taher ElGamal~\cite{gam:j:public-key} developed a public-key cryptosystem and
a digital signature protocol that are based on the Diffie--Hellman protocol.
In fact, the variant of Diffie--Hellman presented in
Figure~\ref{fig:diffie-hellman-public-key} is somewhat reminiscent of the
original ElGamal public-key cryptosystem, which we will now describe.

\begin{figure}[!htp]
  \centering
\begin{tabular}{||c||c|c|c||}
\hline\hline
\parbox[t]{.7cm}{\bf Step} & 
\psfig{file=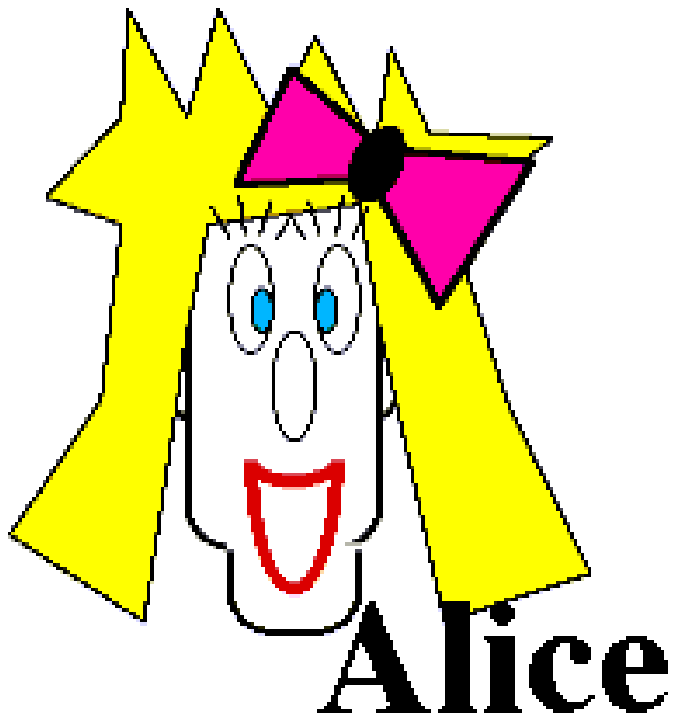,height=2cm} & 
\psfig{file=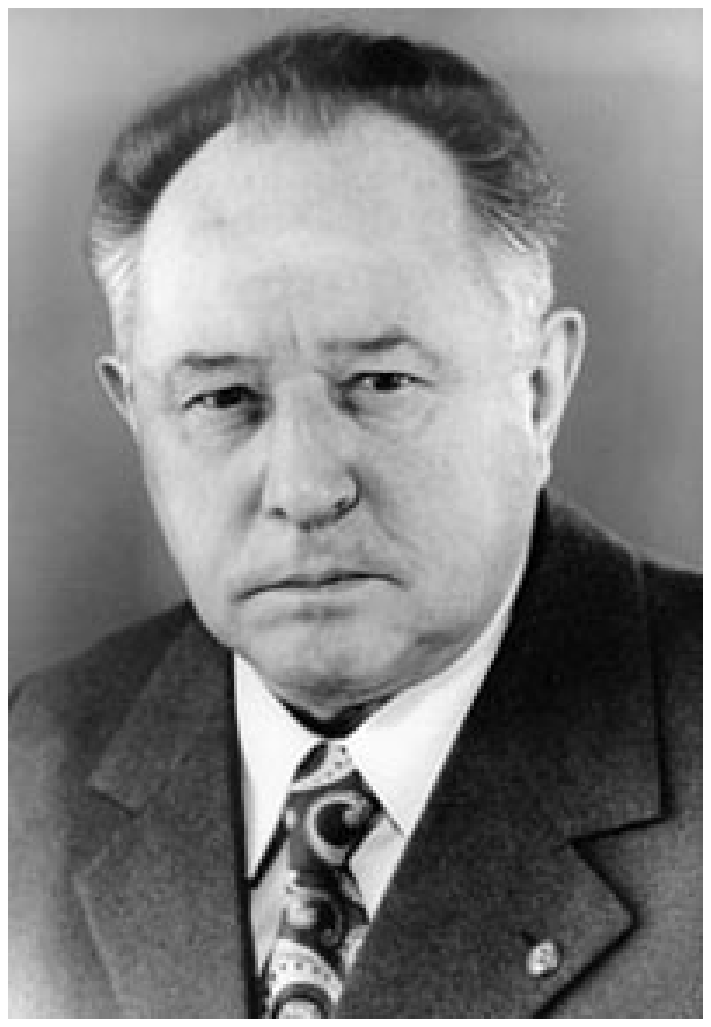,height=2cm} & 
\psfig{file=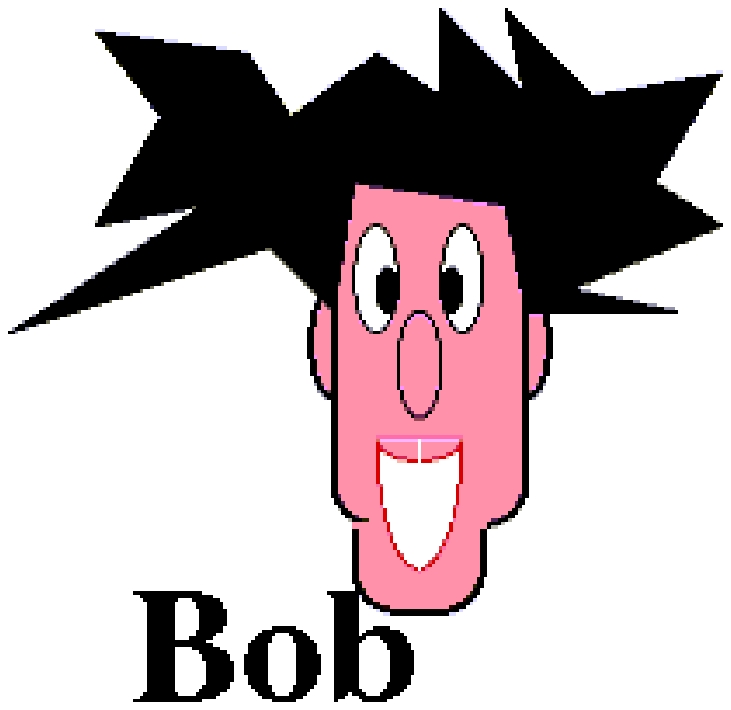,height=2cm} \\ \hline\hline
{\bf 1} & \multicolumn{3}{c||}
  {Alice and Bob agree upon a large prime~$p$ and a primitive root 
   $g$ of~$p$;} \\
 & \multicolumn{3}{c||}
  {$p$ and $g$ are public} \\ \hline
{\bf 2} & & & \parbox[t]{4cm}
  {chooses $b \in \integers^{\ast}_{p-1}$ at random and computes $\beta = g^b
    \mod p$; \\ $b$ is private and $\beta$ is public}
  \\ \hline
{\bf 3} & & 
\mbox{\huge $\stackrel{\mbox{\normalsize $\beta$}}{\Leftarrow}$} & \\ \hline
{\bf 4} & \parbox[t]{4cm}
  {picks a secret $a \in \integers^{\ast}_{p-1}$ at random,
  computes $\alpha = g^a \mod p$ and $c = m \beta^a \mod p$, 
  where $m$ is the message to be sent
} & &  \\ \hline
{\bf 5} & & 
\mbox{\huge $\stackrel{\mbox{\normalsize $\alpha$, $c$}}{\Rightarrow}$} & \\
\hline 
{\bf 6} & & & \parbox[t]{4cm}
  {computes $x = p - 1 - b$ and decrypts by computing
\[
m = c \alpha^x \mod p
\]
} \\ \hline\hline
\end{tabular}
\caption{The ElGamal public-key cryptosystem.
\label{fig:ElGamal}
}
\end{figure}

Figure~\ref{fig:ElGamal} shows ElGamal's public-key cryptosystem.  After Alice
and Bob have agreed on a prime $p$ and a primitive root $g$ of~$p$, Bob picks
a random value $b \in \integers^{\ast}_{p-1}$ and computes his public key
$\beta = g^b \mod p$.  If Alice wants to send him a message $m \in
\integers^{\ast}_{p}$, she looks up $\beta$ and ``disguises'' $m$ by
multiplying it with $\beta^a$ modulo~$p$, where $a \in \integers^{\ast}_{p-1}$
is a random number she has picked.  This yields the first part $c$ of the
cipher text, the second part is $\alpha = g^a \mod p$.  She sends both $c$ and
$\alpha$ to Bob.  To decrypt, Bob first computes $x = p - 1 - b$.  Since $1
\leq b \leq p-2$, it follows that $1 \leq x \leq p-2$.  Bob then can recover
the original plain text $m$ by computing:
\[
c \alpha^x \equiv 
m \beta^a g^{a(p - 1 - b)} \equiv 
m g^{ba + a(p-1) - ab} \equiv 
m \left(g^{p-1}\right)^a \equiv
m \mod p .
\]

Just as in the Diffie--Hellman protocol, the security of the ElGamal protocol
is based on the difficulty of computing discrete logarithms.  Although it is
not known whether breaking the ElGamal protocol is as hard as solving the
discrete logarithm problem, it can be shown that breaking the ElGamal protocol
is precisely as hard as solving the Diffie--Hellman problem.  To prevent known
attacks on the ElGamal cryptosystem, the prime $p$ should be chosen large
enough (at least 150 digits long) and such that $p-1$ has at least one large
prime factor.

\begin{figure}[!htp]
  \centering
\begin{tabular}{||c||c|c|c||}
\hline\hline
\parbox[t]{.5cm}{\bf Step} & 
\psfig{file=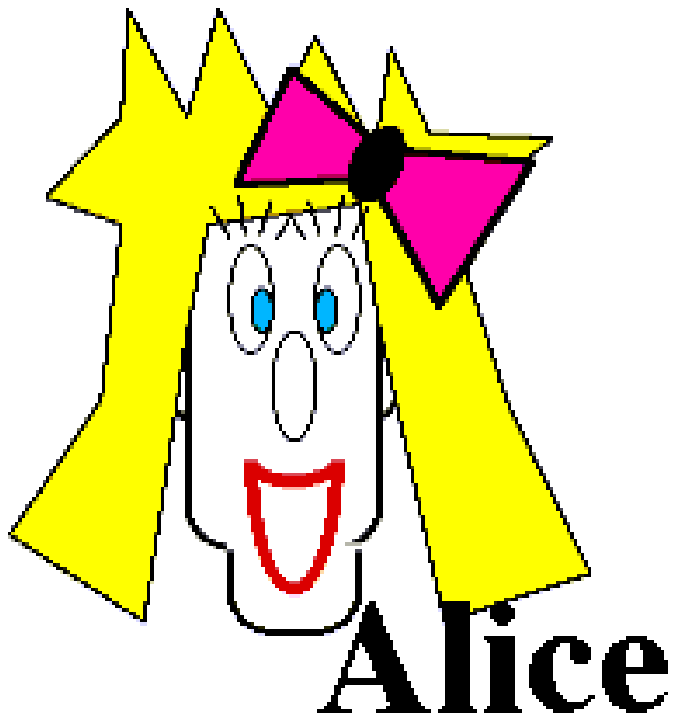,height=2cm} & 
\psfig{file=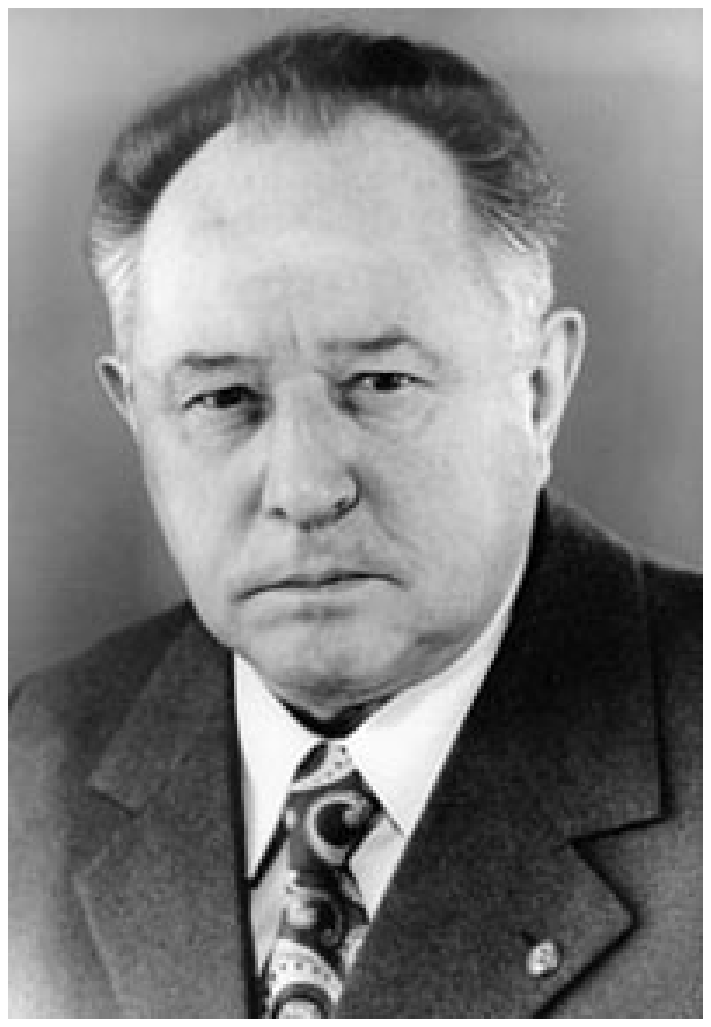,height=2cm} & 
\psfig{file=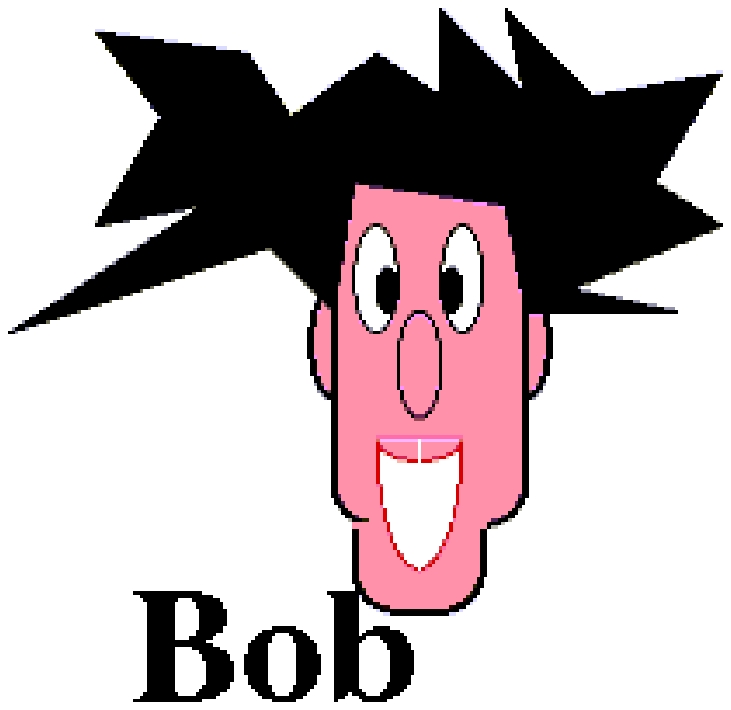,height=2cm} \\ \hline\hline
{\bf 1} & \multicolumn{3}{c||}
  {Alice and Bob agree upon a large prime~$p$ and a primitive root 
   $g$ of~$p$;} \\
 & \multicolumn{3}{c||}
  {$p$ and $g$ are public} \\ \hline
{\bf 2} & & & \parbox[t]{4cm}
  {chooses $b$ and $\beta = g^b \mod p$ as in Fig.~\ref{fig:ElGamal};
   chooses a number $r$ with $\mbox{gcd}(r, p-1) = 1$, 
   computes $\rho = g^r \mod p$ and $s$ according to 
   Eq.~{\rm{}(\ref{equ:elgamal-digital-signature})} and his signature
\[
\mbox{sig}_B (m) = (\rho, s)
\]
}
  \\ \hline
{\bf 3} & & 
\mbox{\huge $\stackrel{\mbox{\normalsize $\beta$, $m$, 
             $\mbox{sig}_B (m)
$}}{\Leftarrow}$} & \\ \hline
{\bf 4} & \parbox[t]{4cm}
  {verifies Bob's signature by checking that 
   Eq.~(\ref{equ:elgamal-dig-dig-check}) holds:
\[
g^m \equiv \beta^{\rho} \cdot \rho^s \mod p .
\]
} & & \\ \hline\hline
\end{tabular}
\caption{The ElGamal digital signature protocol.
\label{fig:ElGamal-digital-signature}
}
\end{figure}

ElGamal's system can be modified so as to yield a digital signature protocol.
A particularly efficient variant of this protocol that is due to an idea of
Schnorr~\cite{sch:c:signature} is now the United States ``Digital Signature
Standard''~\cite{NIST:1991:DSS,NIST:1992:DSS}.

The ElGamal digital signature protocol is presented in
Figure~\ref{fig:ElGamal-digital-signature}.  Suppose that Bob wants to send a
message $m$ to Alice.  To prove that he indeed is the sender, he wants to sign
the message in a way that Alice can verify.
Let a large prime $p$ and a primitive root $g$ of $p$
be given as in the ElGamal public-key cryptosystem, see
Figure~\ref{fig:ElGamal}.  As in that protocol, Bob chooses his private $b$
and computes $\beta = g^b \mod p$.  In addition, he now chooses a number $r$
coprime with $p-1$, and he computes $\rho = g^r \mod p$ and a solution $s$ to
the congruence
\begin{equation}
\label{equ:elgamal-digital-signature}
b \cdot \rho + r \cdot s \equiv m \mod p-1
\end{equation}
using the extended algorithm of Euclid, see Figure~\ref{fig:euklid} and
Lemma~\ref{lem:euklid}.

Bob keeps $b$ and $r$ secret, and he sends along with his message $m$ his
digital signature $\mbox{sig}_B (m) = (\rho, s)$ and the value $\beta$ to
Alice.

Alice checks the validity of the signature by verifying the congruence
\begin{equation}
\label{equ:elgamal-dig-dig-check}
g^m \equiv \beta^{\rho} \cdot \rho^s \mod p .
\end{equation}
The protocol is correct, since by Fermat's Little Theorem (see
Theorem~\ref{thm:fermat}) and by
Equation~(\ref{equ:elgamal-digital-signature}), it holds that
\[
g^m \equiv g^{b \cdot \rho + r \cdot s} \equiv 
\beta^{\rho} \cdot \rho^s \mod p .
\]
Note that the public verification key, which consists of the values $p$, $g$,
and~$\beta$, is computed just once and can be used to verify any message that
is signed with $p$, $g$, $b$, and~$\beta$.  However, a new value of $r$ is
chosen every time a message is signed.

\subsection{Shamir's No-Key Protocol}
\label{sec:no-key}

\begin{figure}[!htp]
\centering
\begin{tabular}{||c||c|c|c||}
\hline\hline
\parbox[t]{.5cm}{\bf Step} & 
\psfig{file=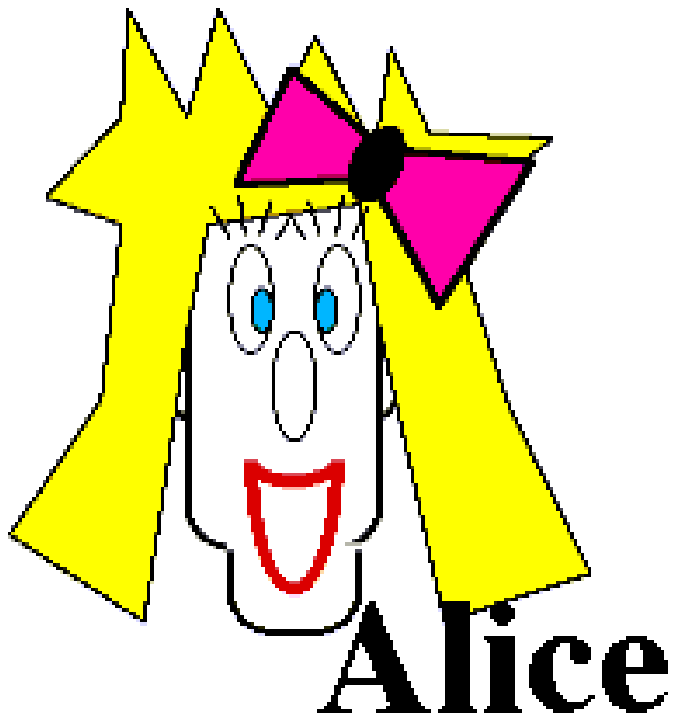,height=2cm} & 
\psfig{file=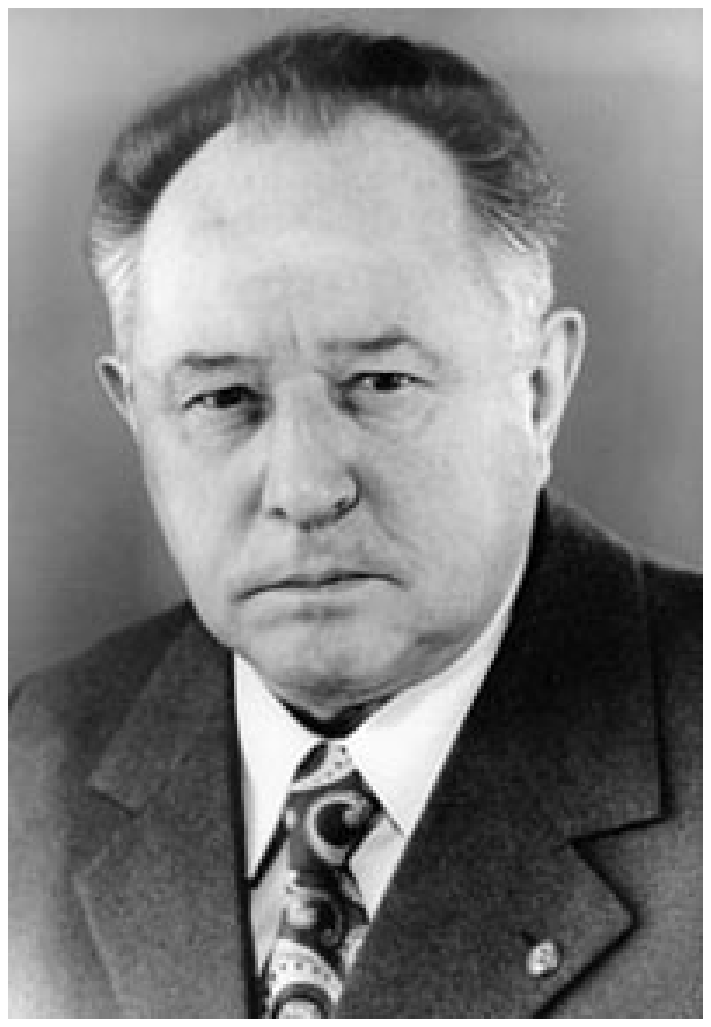,height=2cm} & 
\psfig{file=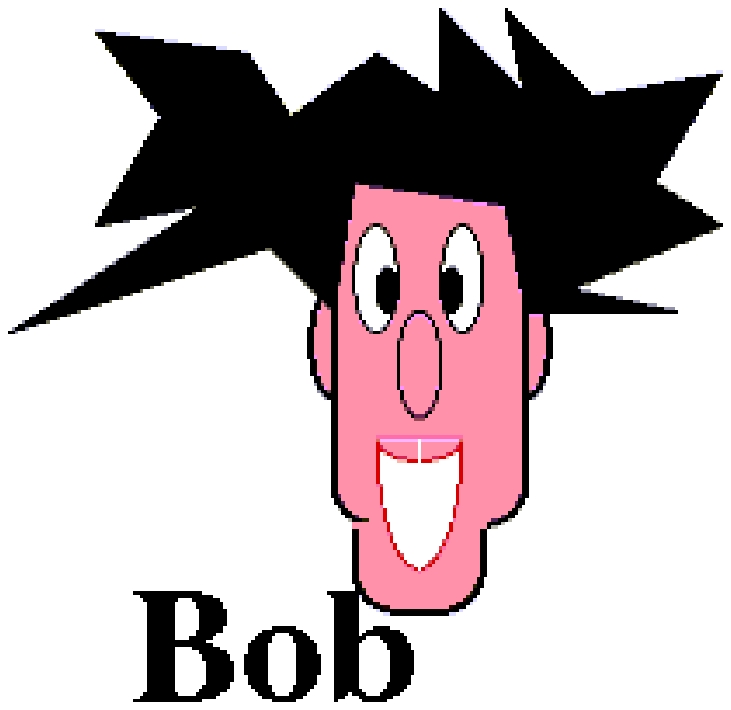,height=2cm} \\ \hline\hline
{\bf 1} & \multicolumn{3}{c||}
  {Alice and Bob agree upon a large prime~$p$, which is public} \\ \hline
{\bf 2} & \parbox[t]{4cm}
  {computes $x = m^a \mod p$, \\
   where $m$ is the message} & & \\ \hline
{\bf 3} & & 
\mbox{\huge $\stackrel{\mbox{\normalsize $x$}}{\Rightarrow}$} & \\ \hline
{\bf 4} & 
 & & \parbox[t]{4cm}
  {computes $y = x^b \mod p$} \\ \hline
{\bf 5} & & 
\mbox{\huge $\stackrel{\mbox{\normalsize $y$}}{\Leftarrow}$} & \\ \hline
{\bf 6} & \parbox[t]{4cm}
  {computes $z = y^{a^{-1}} \mod p$} & &  \\ \hline
{\bf 7} & & 
\mbox{\huge $\stackrel{\mbox{\normalsize $z$}}{\Rightarrow}$} & \\ \hline
{\bf 8} & 
 & & \parbox[t]{4cm}
  {computes $m = z^{b^{-1}} \mod p$} \\ \hline\hline
\end{tabular}
\caption{Shamir's no-key protocol.
\label{fig:shamir-no-keys}
}
\end{figure}

Adi Shamir proposed a 
cryptosystem by which Alice and Bob can
exchange messages that are encrypted by Alice's and Bob's individual secret
keys, yet in which there is no need for Alice and Bob to previously agree on a
{\em joint\/} secret key.  This clever idea is described in an unpublished
paper of Shamir, and it is again based on the modular exponentiation function
and the difficulty of efficiently computing discrete logarithms that was
useful for the Diffie--Hellman secret-key agreement protocol described in
Section~\ref{sec:diffie-hellman}.
The Shamir protocol is often called Massey-Omura in the literature.
Both inventors were preceded by Malcolm Williamson from GCHQ who developed the
same protocol in the nonpublic sector around 1974.

Figure~\ref{fig:shamir-no-keys} shows how Shamir's no-key protocol works.  In
this protocol, let
$m$ be the message that Alice wants to send to Bob.  First, Alice and Bob
agree on a large prime~$p$.  Alice generates a pair $(a,a^{-1})$ satisfying
\[
a a^{-1} \equiv 1 \mod p-1,
\]
where $a^{-1}$ is the inverse of $a$ modulo $p-1$.  Recall from
Section~\ref{sec:rsa} that, given a prime $p$ and an integer $a \in
\integers_{p}^{\ast}$, the inverse $a^{-1}$ of $a$ modulo $p-1$ can easily be
computed.  Similarly, Bob generates a pair $(b,b^{-1})$ satisfying
\[
b b^{-1} \equiv 1 \mod p-1,
\]
where $b^{-1}$ is the inverse of $b$ modulo $p-1$. See
Figure~\ref{fig:shamir-no-keys} for the rest of the steps.

The protocol is correct, since for all messages $m$, $1 \leq m \leq p$, it
holds that:
\begin{eqnarray*}
m \equiv m^{a a^{-1}} \mod p & \mbox{ and } & m \equiv  m^{b b^{-1}} \mod p .
\end{eqnarray*}
Hence, looking at Figure~\ref{fig:shamir-no-keys}, we obtain
\[
z^{b^{-1}} \equiv y^{a^{-1} b^{-1}} \equiv x^{b a^{-1} b^{-1}} \equiv 
m^{a b a^{-1} b^{-1}} \equiv m \mod p ,
\]
so Step~8 of Figure~\ref{fig:shamir-no-keys} is correct.

Note that modular exponentiation is used here 
both  for encryption and decryption.  
The key property for this protocol to work is that modular
exponentiation is symmetric in the exponents, i.e., for all $a$ and $b$, it
holds that
\[
\alpha_{(g, p)}(a \cdot b) \equiv g^{a \cdot b} \equiv g^{b \cdot a}
\mod p .
\]

\subsection{Rivest, Rabi, and Sherman's Secret-Key Agreement and 
  Digital Signature Protocols}
\label{sec:riv-rab-she}

\begin{figure}[!htp]
\centering
\begin{tabular}{||c||c|c|c||}
\hline\hline
\parbox[t]{.5cm}{\bf Step} & 
\psfig{file=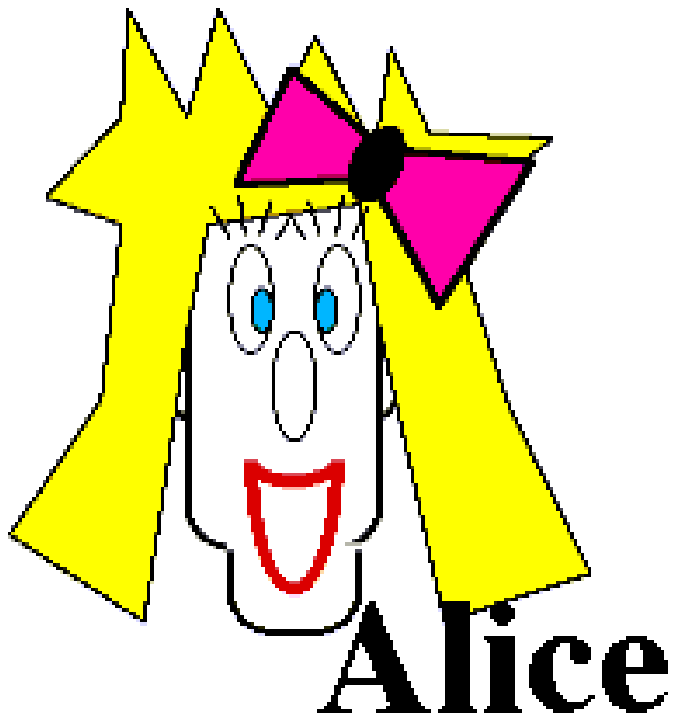,height=2cm} & 
\psfig{file=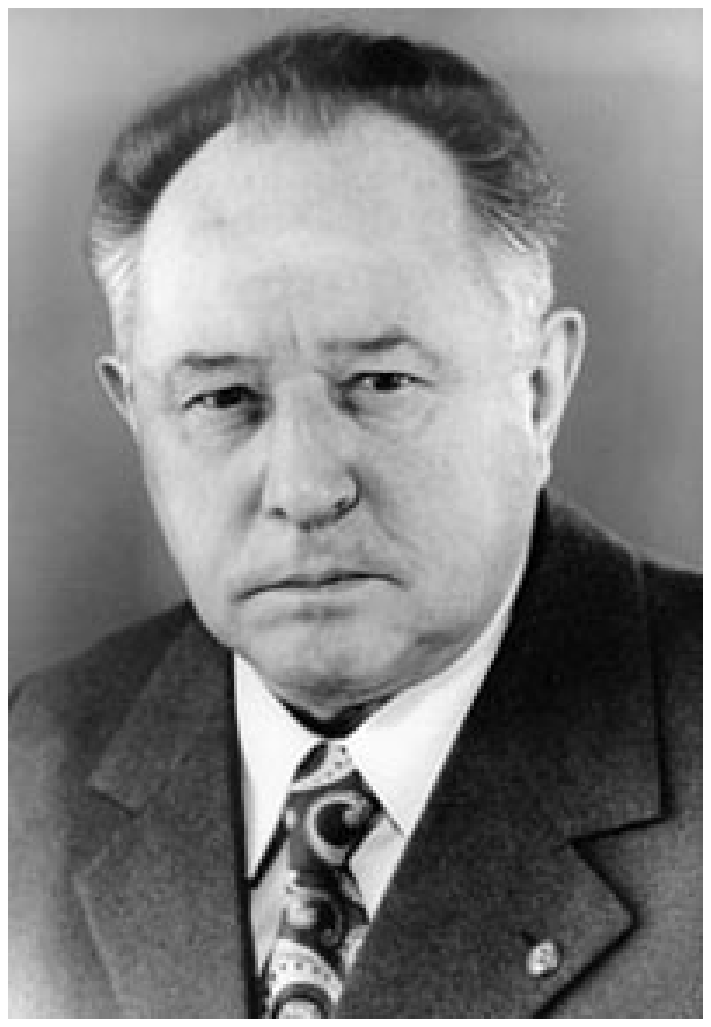,height=2cm} & 
\psfig{file=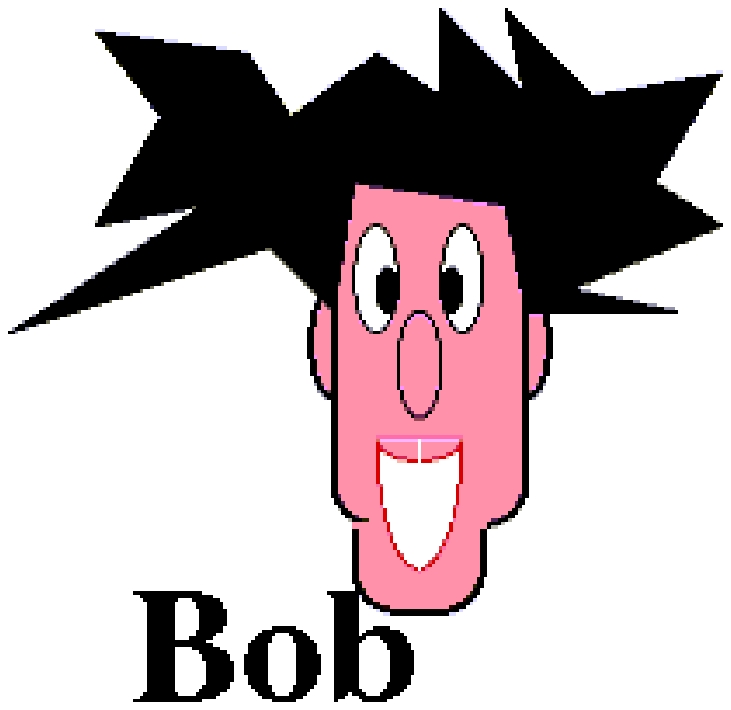,height=2cm} \\ \hline\hline
{\bf 1} & \parbox[t]{4cm}
  {chooses two large numbers $x$ and $y$ at random, keeps $x$ secret, and
    computes $\sigma(x,y)$} & & \\ \hline
{\bf 2} & & 
\mbox{\huge 
$\stackrel{\mbox{\normalsize $y$, $\sigma(x,y)$}}{\Rightarrow}$} & \\ \hline
{\bf 3} & 
 & & \parbox[t]{4cm}
  {chooses a large number $z$ at random, keeps $z$ secret and computes 
   $\sigma(y,z)$}
 \\ \hline
{\bf 4} & & 
\mbox{\huge 
$\stackrel{\mbox{\normalsize $\sigma(y,z)$}}{\Leftarrow}$}
 & \\ \hline
{\bf 5} & \parbox[t]{4cm}
  {computes her key 
\[
k_A = \sigma(x,\sigma(y,z))
\]
} & & \parbox[t]{4cm}
  {computes his key 
\[
k_B =  \sigma(\sigma(x,y),z)
\]
} \\ \hline\hline
\end{tabular}
\caption{The Rivest--Sherman secret-key agreement protocol, which uses 
a strongly noninvertible, associative one-way function~$\sigma$.
\label{fig:rivest-sherman-secret-key}
}
\end{figure}

\begin{figure}[!htp]
\centering
\begin{tabular}{||c||c|c|c||}
\hline\hline
\parbox[t]{.5cm}{\bf Step} & 
\psfig{file=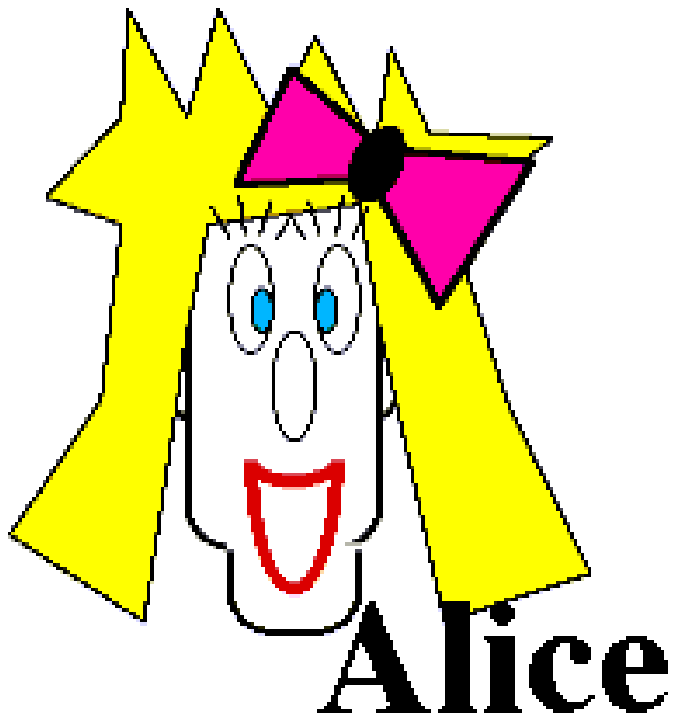,height=2cm} & 
\psfig{file=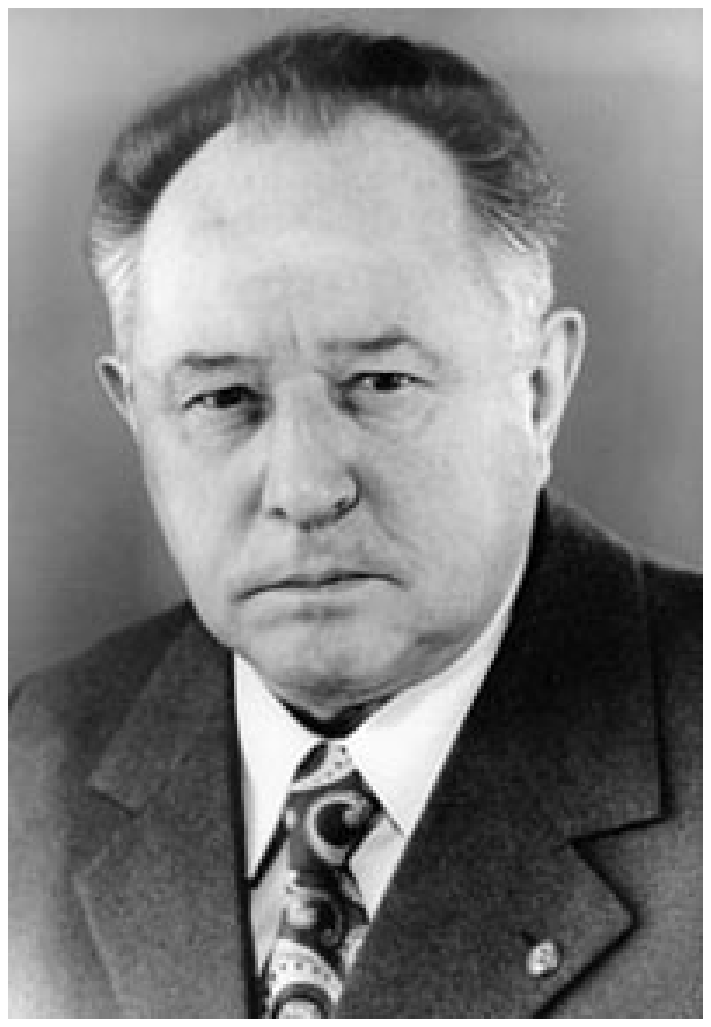,height=2cm} & 
\psfig{file=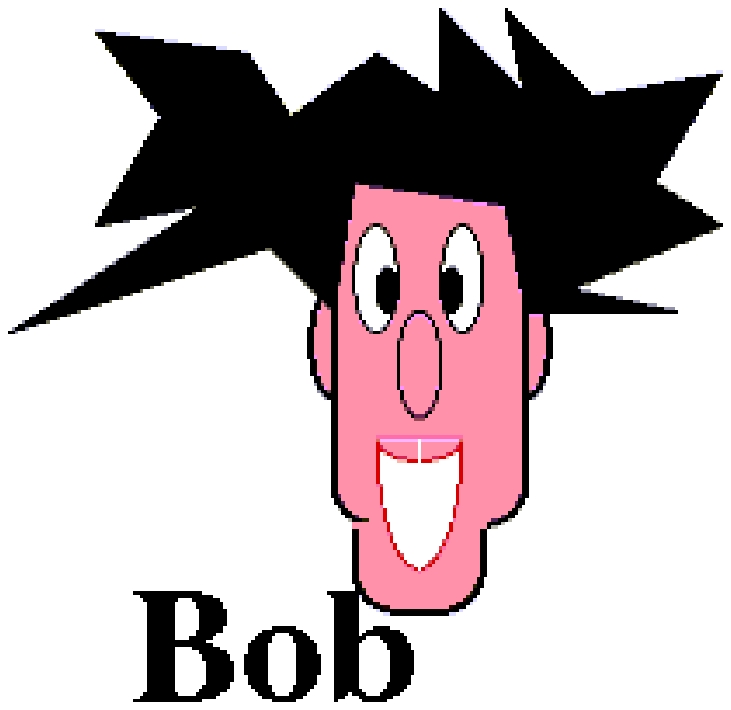,height=2cm} \\ \hline\hline
{\bf 1} & \parbox[t]{4cm}
  {chooses two large numbers $x_A$ and $y_A$ at random, keeps $x_A$ secret, and
    computes $\sigma(x_A, y_A)$} & & \\ \hline
{\bf 2} & & 
\mbox{\huge 
$\stackrel{\mbox{\normalsize 
$y_A$, $\sigma(x_A, y_A)$}}{\Rightarrow}$} & \\ \hline
{\bf 3} & 
 \parbox[t]{4cm}
  {computes her signature 
\[
\mbox{sig}_A (m) = \sigma(m, x_A) 
\]
for the message $m$}
 & & \\ \hline
{\bf 4} & & 
\mbox{\huge 
$\stackrel{\mbox{\normalsize 
$m$, $\mbox{sig}_A (m)$}}{\Rightarrow}$} & \\ \hline
{\bf 5} & & & \parbox[t]{4cm}
  {verifies Alice's signature by checking whether
$\sigma(m, \sigma(x_A, y_A))$ equals $\sigma(\sigma(m, x_A), y_A)$
} \\ \hline\hline
\end{tabular}
\caption{The Rabi--Sherman digital signature protocol, which uses a strongly
  noninvertible, associative one-way function~$\sigma$.
\label{fig:rabi-sherman-digital-signature}
}
\end{figure}

Ron Rivest, Muhammad Rabi, and Alan Sherman developed secret-key agreement and
digital signature protocols.  The secret-key agreement protocol from
Figure~\ref{fig:rivest-sherman-secret-key} is attributed to Rivest and Sherman
in~\cite{rab-she:t-no-URL:aowf,rab-she:j:aowf}.  The digital signature
protocol from Figure~\ref{fig:rabi-sherman-digital-signature} is due to Rabi
and Sherman~\cite{rab-she:t-no-URL:aowf,rab-she:j:aowf}.

Here is a brief, intuitive explanation of how these protocols work.  The key
building block of both protocols is a {\em total, strongly noninvertible,
  associative one-way function}.  As mentioned earlier, one-way functions are
theoretical constructs not known to exist.  However, there are plausible
assumptions under which one-way functions of various types can be constructed.
In Section~\ref{sec:aowf}, under a quite plausible complexity-theoretic
assumption, we will see how to construct a concrete candidate for a total,
strongly noninvertible, associative one-way function.  For now, assume that
$\sigma$ {\em is\/} such a function.  That is, $\sigma$ is a total two-ary
(i.e., two-argument) function mapping pairs of positive integers to positive
integers such that:
\begin{itemize}
\item $\sigma$ is {\em associative}, i.e., the equation $\sigma(x,\sigma(y,z))
  = \sigma(\sigma(x,y),z)$ holds for all $x, y, z \in \nats$.
  
\item $\sigma$ is {\em strongly noninvertible}, i.e., $\sigma$ is hard to
  invert even if in addition to the function value one of the arguments is
  given.  
\end{itemize}

Look at Rivest and Sherman's secret-key agreement protocol in
Figure~\ref{fig:rivest-sherman-secret-key}.  Since $\sigma$ is associative,
we have:
\[
k_A = \sigma(x,\sigma(y,z)) = \sigma(\sigma(x,y),z) = k_B ,
\]
and thus the keys computed by Alice and Bob indeed are the same.  On the other
hand, if Erich was listening carefully, he knows not only two function values,
$\sigma(x,y)$ and $\sigma(y,z)$, but he also knows~$y$, the first argument of
$\sigma(y,z)$ and the second argument of $\sigma(x,y)$.  That is why $\sigma$
must be strongly noninvertible, in order to prevent the direct attack that
Erich computes Alice's secret number $x$ from $\sigma(x,y)$ and $y$ or Bob's
secret number $z$ from $\sigma(y,z)$ and~$y$, in which case he could easily
obtain their joint secret key, $k_A = k_B$.  Analogous comments apply to Rabi
and Sherman's digital signature protocol presented in
Figure~\ref{fig:rabi-sherman-digital-signature}.

\subsection{Discussion of Diffie--Hellman versus Rivest--Sherman}
\label{sec:discussion}

While the secret-key agreement protocol of Diffie and
Hellman~\cite{dif-hel:j:diffie-hellman} is widely used in practice, that of
Rivest and Sherman (see~\cite{rab-she:t-no-URL:aowf,rab-she:j:aowf}) is not
(yet) used in applications and, thus, might appear somewhat exotic at first
glance.  Note, however, that neither the Diffie--Hellman nor the Rivest--Sherman
protocol has a proof of security up to date.  So, let us digress for a moment
to compare the state of the art on these two protocols.

\begin{itemize}
\item While the Diffie--Hellman protocol uses a concrete function, the
  Rivest--Sherman protocol is based on an unspecified, ``abstract'' function
  that is described only by listing the properties it should satisfy.  That is
  not to say that Rivest--Sherman is an abstract version of Diffie--Hellman.
  Rather, the Rivest--Sherman protocol may be seen as an alternative to the
  Diffie--Hellman protocol.  The advantage of Rivest and Sherman's approach is
  that it is more flexible, as it does not depend on a single function.

\item The security of the Diffie--Hellman scheme is based on the (unproven, yet
  plausible) assumption that computing discrete logarithms is a
  computationally intractable task.  
  
  In contrast, the Rivest--Sherman scheme uses a candidate for a strongly
  noninvertible, associative one-way function (see
  Section~\ref{sec:definitions} for the formal definition) as its key building
  block.  Although it is not known whether such functions exist, it has been
  shown recently by Hemaspaandra and this author~\cite{hem-rot:j:aowf} that
  they do exist in the worst-case model under the (unproven, yet plausible)
  assumption that $\p \neq \np$, where P denotes the class of polynomial-time
  solvable problems, and NP denotes the class of problems that can be solved
  nondeterministically in polynomial time. Section~\ref{sec:aowf} presents
  this result and a sketch of its proof.
  
\item Breaking Diffie--Hellman is not even known to be as hard as computing
  discrete logarithms, even though some nice progress in this direction has
  been made recently by Maurer and
  Wolf~\cite{mau-wol:j:breaking-diffie-hellman-and-discrete-log}, who
  established conditions for relating the hardness of breaking Diffie--Hellman
  to that of computing discrete logarithms.  Again, their results rest on
  unproven, yet plausible assumptions.  In particular, let $\nu(p)$ denote the
  minimum, taken over all numbers $d$ in the interval $[p - 2 \sqrt{p} + 1 ,\,
  p + 2 \sqrt{p} + 1]$, of the largest prime factors of~$d$.  The
  ``smootheness assumption'' says that $\nu(p)$ is polynomial in $\log p$.
  Why is this assumption plausible?  The idea is that numbers in the
  Hasse-Weil interval (which are sizes of elliptic curves) are smooth with the
  same probability as random numbers of the same length, and these
  probabilities are independent.  Under this smoothness assumption, Maurer and
  Wolf~\cite{mau-wol:j:breaking-diffie-hellman-and-discrete-log} proved that
  breaking Diffie--Hellman and computing the discrete logarithm are
  polynomial-time equivalent tasks in the underlying cyclic group, where
  the equivalence is nonuniform.

  Similarly, even if strongly noninvertible, associative one-way functions
  were known to exist, one could not conclude that the Rivest--Sherman protocol
  is secure; rather, strong noninvertibility merely precludes certain types of
  direct attacks~\cite{rab-she:j:aowf,hem-rot:j:aowf}.  Moreover, strongly
  noninvertible, associative one-way functions could be constructed so far
  only in the {\em worst-case\/} complexity model, assuming $\p \neq \np$.
  Although this result is relevant and interesting in a complexity-theoretic
  setting, it has no direct implications in applied cryptography.  
  For cryptographic
  applications, one would need to construct such functions based on the {\em
    average-case\/} complexity model, under plausible assumptions.  
\end{itemize}

As noted in the outline of the tutorial, there is some hope for obtaining such
a strong result by combining Hemaspaandra and Rothe's~\cite{hem-rot:j:aowf}
technique on constructing strongly noninvertible, associative one-way functions
in the worst case with Ajtai's~\cite{ajt:c:hard-instances-in-lattices} 
techniques on constructing hard instances of lattice problems.  
The shortest lattice vector problem, denoted by~SVP, is the problem of 
finding a shortest lattice vector in the lattice generated by a given
lattice basis.  Roughly speaking,
Ajtai~\cite{ajt:c:hard-instances-in-lattices} proved that the problem SVP is
as hard in the average-case as it is in the worst-case complexity model.

More precisely, Ajtai constructed an infinite family $\{\Lambda_n\}_{n \geq
  1}$ of lattices, where each $\Lambda_n$ is represented by a basis as an
instance of SVP, and he showed the following result: Suppose one can compute
in polynomial time, for each~$n$, an approximately shortest vector in a
lattice $\Lambda_i$ {\em randomly\/} chosen from $\{\Lambda_n\}_{n \geq 1}$,
with non-negligible probability.  Then, the length of a shortest vector in
{\em every\/} lattice from $\{\Lambda_n\}_{n \geq 1}$ can be estimated to
within a fixed polynomial factor in polynomial time with probability close to
one.  However, since the best approximation factor known to be achieved by
polynomial-time algorithms is essentially exponential, and since the best
algorithms known to achieve polynomial-factor approximations run in
exponential time, it follows that, as mentioned above, ``SVP is as hard in the
average-case as it is in the worst-case model.''~~In this regard, the SVP is a
unique problem; for no other problem in NP that is believed to be outside P
such a strong connection is known to hold.

Based on the worst-case/average-case equivalence of SVP, Ajtai and
Dwork~\cite{ajt-dwo:c:public-key-system-worst-average-equivalence} designed a
public-key cryptosystem whose cryptographic security depends only on
worst-case complexity assumptions.  However, the worst-case hardness of SVP
(in the Euclidean norm) had remained an open problem for a long time.  Solving
this problem, Ajtai~\cite{ajt:c:worst-case-hardness-of-svp} established the
NP-hardness of SVP under randomized reductions.  His result was strengthened
by Micciancio~\cite{mic:j:svp-is-np-hard-to-approximate}, who also simplified
Ajtai's proof.  Since the construction of strongly noninvertible, associative
one-way functions in~\cite{hem-rot:j:aowf} is based on the assumption $\p \neq
\np$, it seems reasonable to consider the NP-hard problem SVP to be a good
candidate for achieving strongly noninvertible, associative one-way functions
even in the technically more demanding average-case model.

The complexity of~SVP and the use of lattices in crytography are covered in
the surveys by Cai~\cite{cai:c:lattice-problems-survey}, Kumar and
Sivakumar~\cite{kum-siv:j:svp-survey}, and Nguyen and
Stern~\cite{ngu-ste:c:two-faces-of-lattices}.  Interestingly, lattices are
useful both in breaking existing cryptosystems like RSA (e.g., the
low-exponent attacks of H{\aa}stad~\cite{has:j:solving-low-degree-equations}
and Coppersmith~\cite{cop:j:low-exponent-rsa-attacks}), see
Section~\ref{sec:security-rsa}) and in designing secure cryptosystems (e.g.,
the Ajtai-Dwork public-ley cryptosystem).

\section{Interactive Proof Systems and Zero-Knowledge Protocols}
\label{sec:zero-knowledge}

In Section~\ref{sec:diffie-hellman}, we mentioned the Man-in-the-middle
attack on the Diffie--Hellman secret-key agreement protocol.  Imagine that Bob
has just agreed with his partner on a joint secret key via a public telephone
line.  Of course, he assumes it was Alice he was talking to.  Bob was so
clever to use the Diffie--Hellman protocol, and so he thinks that Erich does
not have a clue about what secret key they have chosen:
\[
\begin{array}{ccc}
 & \mbox{\bf ???} & \\
 & \psfig{file=mielke.ps,height=2cm} & \\
 & \mbox{{\bf\large Erich}}  & \\[.2cm]
\psfig{file=alice.ps,height=2cm} & 
\psfig{file=channel.eps,width=2cm} & 
\psfig{file=bob.ps,height=2cm} 
\end{array}
\]

But Erich was even smarter.  Here is what really happened:
\[
\begin{array}{ccc}
\psfig{file=mielke.ps,height=2cm} \ \ 
\mbox{{\bf\large Erich}} & 
\psfig{file=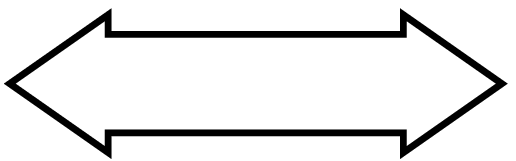,width=2cm} & 
\psfig{file=bob.ps,height=2cm} 
\end{array}
\]

This situation raises the issue of {\em authentication\/}: How can Bob be
certain that it in fact was Alice he was communicating with, and not Erich
pretending to be Alice?  In other words, how can Alice prove her identity to
Bob beyond any doubt?

In Section~\ref{sec:protocols}, we have seen how to use digital signatures for
the authentication of documents such as email messages.  In this section, our
goal is to achieve authentication of an {\em individual\/} rather than a
document.  One way to achieve this goal is to assign to Alice's identity some
secret information such as her PIN (``{\em P\/}ersonal {\em I\/}dentifaction
{\em N\/}umber'') or any other private information that nobody else knows.  We
refer to the information proving Alice's identity as Alice's {\em secret}.

But here's another catch.  Alice would like to convince Bob of her identity by
proving that she knows her secret.  Ideally, however, she should not disclose
her secret because then it wouldn't be a secret anymore: If Bob, for example,
knew Alice's secret, he could pretend to be Alice when communicating with
somebody else.  So the question is:
\begin{quote}
  {\em How can one prove the knowledge of a secret without telling the
    secret?\/}
\end{quote}
That is precisely what zero-knowledge protocols are all about.

\subsection{Interactive Proof Systems}

Zero-knowledge protocols are a special form of interactive proof systems,
which we will describe first.  Interactive proof systems were introduced by
Shafi Goldwasser, Silvio Micali, and Charles
Rackoff~\cite{gol-mic-rac:c:interactive-proof-systems,gol-mic-rac:j:interactive-proof-systems}.
Independently, Babai and
Moran~\cite{bab-mor:j:arthur-merlin,bab:c:trading} developed the essentially
equivalent notion of Arthur-Merlin games.

As in the previous protocols, we consider the communication between two
parties, the ``prover'' Alice and the ``verifier'' Bob:
\[
\begin{array}{ccc}
\mbox{\bf Prover} & & \mbox{\bf Verifier} \\
\psfig{file=alice.ps,height=2cm} & 
\psfig{file=channel-no-blitz.eps,width=2cm} & 
\psfig{file=bob.ps,height=2cm}
\end{array}
\]
For now, we are not interested in the security aspects that may arise when the
communication is eavesdropped; rather, we are concerned with the following
communication problem: Alice and Bob want to jointly solve a given
problem~$L$, i.e., they want to decide whether or not any given instance
belongs to~$L$.  For concreteness, consider the graph isomorphism problem.

\begin{definition}
  The vertex set of any graph $G$ is denoted by~$V(G)$, and the edge set of
  $G$ is denoted by~$E(G)$.  Let $G$ and $H$ be undirected, simple graphs,
  i.e., graphs with no reflexive or multiple edges.
  
  An {\em isomorphism\/} between $G$ and $H$ is a bijective mapping $\pi$
  from $V(G)$ onto $V(H)$ such that, for all $i,j \in V(G)$,
\[
\{i,j\} \in E(G) \,\Longleftrightarrow\, \{\pi(i),\pi(j)\} \in E(H).
\]  
$\graphiso$ denotes the set of all pairs of isomorphic graphs.
\end{definition}

The graph isomorphism problem is to determine whether or not any two given
graphs are isomorphic.  This problem belongs to NP, and since there is no
efficient algorithm known for solving it, it is widely considered to be a
hard, intractable problem.  However, it is not known to be complete
for~NP, i.e., it is not known whether this problem belongs to the hardest NP
problems.  In fact, due to its ``lowness'' properties, it is doubted that the
graph isomorphism problem is NP-complete.  A set $A$ is low for a complexity
class~$\mathcal{C}$ if it does not yield any additional computational power
when used as an oracle by the machines representing the class~$\mathcal{C}$,
i.e., if $\mathcal{C}^{A} = \mathcal{C}$.  Sch\"{o}ning~\cite{sch:j:gi} showed
that $\graphiso$ is in the second level of the low hierarchy within~NP, i.e.,
it is low for $\np^{\scriptnp}$, the second level of the polynomial hierarchy.
It follows that if $\graphiso$ were NP-complete then the polynomial hierarchy
would collapse, which is considered unlikely.  Moreover, K\"{o}bler et
al.~\cite{koe-sch-tor:j:pplow} proved $\graphiso$ low for~PP, probabilistic
polynomial time.

Therefore, it is conjectured that the graph isomorphism problem might be
neither in P nor NP-complete, and this is what makes this problem so
interesting for complexity theoreticians.  Of course, proving this conjecture
would immediately prove P different from NP; so, such a proof seems beyond
current techniques.  For more complexity-theoretic background on the graph
isomorphism problem, we refer to the book by K\"{o}bler, Sch\"{o}ning, and
Tor\'{a}n~\cite{koe-sch-tor:b:graph-iso}.

We mention in passing that (language versions of) the factoring problem and
the discrete logarithm problem are not known to be NP-complete either.  Unlike
the graph isomorphism problem, however, no lowness properties are known for
these two problems.  Grollmann and Selman~\cite{gro-sel:j:complexity-measures}
have shown that a language version of the discrete logarithm problem is
contained in~UP, which denotes Valiant's class ``unambiguous polynomial
time''~\cite{val:j:checking}.  NP-complete problems are very unlikely to
belong to UP, so this result gives some evidence against the NP-completeness
of the discrete logarithm problem.

Returning to Alice and Bob's communication problem, their task is to decide
whether or not any given pair $(G, H)$ of graphs is isomorphic.  Alice, the
prover, tries to {\em prove\/} them isomorphic by providing Bob with an
isomorphism $\pi$ between $G$ and~$H$.  She intends to convince Bob {\em no
  matter whether or not $G$ and $H$ in fact are isomorphic}.  But Bob is
impatient.  To accept the input, he wants to be convinced with overwhelming
probability that the proof provided by Alice indeed is correct.  Even worse,
he is convinced only if {\em every potential prover strategy\/} Alice might
come up with yields an overwhelming success probability.  If Alice can
accomplish this then Bob accepts the input, otherwise he rejects it.

To formalize this intuition, imagine Alice and Bob to be Turing machines.
Alice, the prover, is an all-powerful Turing machine with no computational
limitation whatsoever.  Bob, the verifier, is a randomized Turing machine
working in polynomial time, but capable of making random moves by flipping an
unbiased coin.  In Definition~\ref{def:interactive-proof-system} below, in
case of acceptance, it is enough that Alice finds one sufficient strategy to
convince Bob.  In case of rejection, however, rather than considering every
potential prover strategy of Alice, it is useful to quantify over all
possible provers that may replace Alice.

For the definition of randomized Turing machines, we refer to any textbook on
complexity theory such
as~\cite{bal-dia-gab:b:sctI:95,bov-cre:b:complexity,hem-ogi:b:companion,pap:b:complexity,pap:b:complexity}.
Essentially, every nondeterministic Turing machine can be viewed as a
randomized Turing machine by defining a suitable probability measure on the
computation trees of the machine.

\begin{definition}[Interactive Proof System]
{\rm{}\cite{gol-mic-rac:c:interactive-proof-systems,gol-mic-rac:j:interactive-proof-systems}}
\label{def:interactive-proof-system}
\begin{enumerate}
\item An {\em interactive proof system\/} (or ``{\em IP protocol\/}'') $(A,B)$
  is a protocol between Alice, the prover, and Bob, the verifier.  Alice runs
  a Turing machine $A$ with no limit on its resources, while Bob runs a
  polynomial-time randomized Turing machine~$B$.  Both access the same input
  on a joint input tape, and they are equipped with private work tapes for
  internal computations.  They also share a read-write communication tape to
  exchange messages.  Alice does not see Bob's random choices.  Let $\mbox{\rm
    Pr}((A, B)(x) = 1)$ denote the probability (according to the random
  choices made in the communication) that Bob accepts the input~$x$; i.e., for
  a particular sequence of random bits, ``$(A, B)(x) = 1$'' denotes the event
  that Bob is convinced by Alice's proof for $x$ and accepts.

\item An {\em interactive proof system $(A, B)$ accepts a set $L$\/} if and
  only if for each~$x$:
\begin{eqnarray}
x \in L & \Lora & (\exists A)\, 
                    [\mbox{\rm Pr}((A, B)(x) = 1) \geq \frac{3}{4}]; 
\label{eq:ip-acceptance}
\\
x \not\in L & \Lora & (\forall \widehat{A})\, 
                    [\mbox{\rm Pr}((\widehat{A}, B)(x) = 1) \leq \frac{1}{4}],
\label{eq:ip-rejection}
\end{eqnarray}
where in {\rm{}(\ref{eq:ip-acceptance})} we quantify over the prover
strategies (or ``proofs'') for $x$ of the prescribed Turing machine~$A$,
whereas in {\rm{}(\ref{eq:ip-rejection})} we quantify over the proofs
$\widehat{A}$ for $x$ of any prover (i.e., any Turing machine of unlimited
computational power) that may replace the fixed Turing machine~$A$.

\item $\ip$ denotes the class of all sets that can be accepted by an
  interactive proof system.
\end{enumerate}
\end{definition}

Note that the acceptance probabilities of at least $\frac{3}{4}$ if $x \in L$
(respectively, of at most $\frac{1}{4}$ if $x \not\in L$) are chosen at will.
By probability amplification 
techniques~\cite{pap:b:complexity,bal-dia-gab:b:sctI:95,bov-cre:b:complexity},
one can use any constants
$\frac{1}{2} + \epsilon$ and $\frac{1}{2} - \epsilon$, respectively, where
$\epsilon > 0$.  It is even possible to make the error probability as small as
$2^{-p(|x|)}$, for any fixed polynomial~$p$.  Better yet, Goldreich, Mansour,
and Sipser~\cite{gol-man-sip:c:provers-that-never-fail} have shown that one
can even require the acceptance probability of exactly~$1$ if $x \in L$,
without changing the class~$\ip$.

In the literature, verifier and prover are sometimes referred to as {\em
  Arthur\/} and {\em Merlin}.  In fact, the Arthur-Merlin games introduced by
Babai and Moran~\cite{bab-mor:j:arthur-merlin,bab:c:trading} are
nothing else than the interactive proof systems of Goldwasser et
al.~\cite{gol-mic-rac:c:interactive-proof-systems,gol-mic-rac:j:interactive-proof-systems}.
One difference between Definition~\ref{def:interactive-proof-system} and the
definition of Arthur-Merlin games is that the random bits chosen by Arthur are
public (i.e., they are known to Merlin), while they are private to Bob in
Definition~\ref{def:interactive-proof-system}.  However, Goldwasser and
Sipser~\cite{gol-sip:j:private} have shown that the privacy of the verifier's
random bits does not matter: Arthur-Merlin games are equivalent to interactive
proof systems.

What if Bob has run out of coins?  That is, what if he behaves
deterministically when verifying Alice's proof for ``$x \in L$''? Due to her
unlimited computational power, Alice can provide proofs of unlimited length,
i.e., of length not bounded by any function in the length of~$x$.  However,
since Bob is a polynomial-time Turing machine, it is clear that he can check
only proofs of length polynomially in~$|x|$.  It follows that IP, when
restricted to deterministic polynomial-time verifiers, is just a cumbersome
way of defining the class~$\np$.  Hence, since $\graphiso$ belongs to~NP, it
must also belong to the (unrestricted) class~$\ip$.  We omit presenting an
explicit IP protocol for $\graphiso$ here, but we refer to
Section~\ref{sec:graphiso}, where in
Figure~\ref{fig:goldreich-micali-wigderson} an IP protocol for $\graphiso$
with an additional property is given: it is a zero-knowledge protocol.

But what about the complement of $\graphiso$?  Does there exist an interactive
proof system that decides whether or not two given graphs are {\em
  non\/}-isomorphic?  Note that even though Alice is all-powerful
computationally, she may run into difficulties when she is trying to prove
that the graphs are non-isomorphic.  Consider, for example, two non-isomorphic
graphs with 1000 vertices each.  A proof of that fact seems to require Alice
to show that none of the $1000!$ possible permutations is an isomorphism
between the graphs.  Not only would it be impossible for Bob to check such a
long proof in polynomial time, also for Alice it would be literally impossible
to write this proof down.  After all, $1000!$ is approximately $4 \cdot
10^{2567}$.  This number exceeds the number of atoms in the entire visible
universe,\footnote{Dark matter excluded.} which is
currently estimated to be around $10^{77}$, by a truly astronomical factor.

That is why the following result of Goldreich, Micali, and
Wigderson~\cite{gol-mic-wid:c:nothing,gol-mic-wid:j:nothing} was a bit of a
surprise.

\begin{theorem}
{\rm{}\cite{gol-mic-wid:c:nothing,gol-mic-wid:j:nothing}} \quad
\label{thm:non-gi-proof-system}
$\overline{\graphiso}$ is in~$\ip$.
\end{theorem}

\begin{proof} 
  Figure~\ref{fig:goldreich-micali-wigderson-graphnoniso} shows the
  interactive proof system for the graph non-isomorphism problem.

\begin{figure}[!htb]
\centering
\begin{tabular}{||c||c|c|c||}
\hline\hline
\parbox[t]{.5cm}{\bf Step} & 
\psfig{file=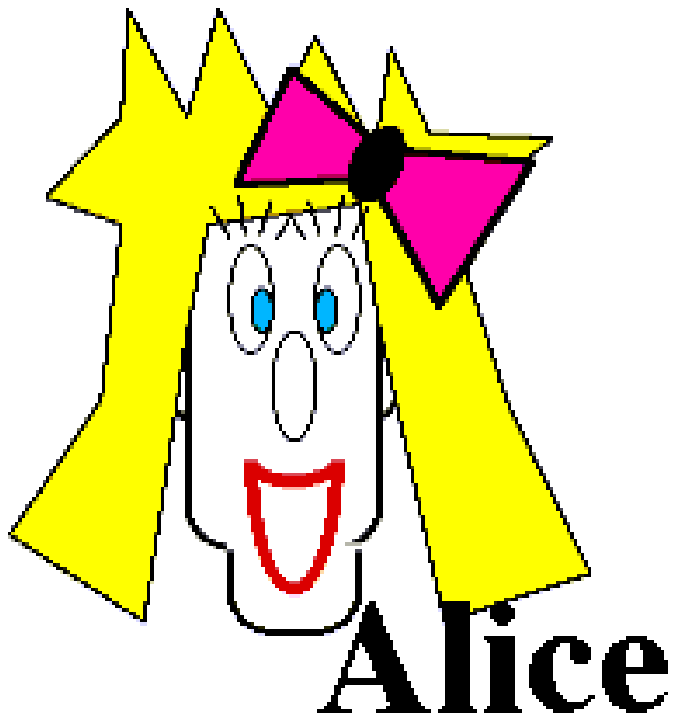,height=2cm} & 
\psfig{file=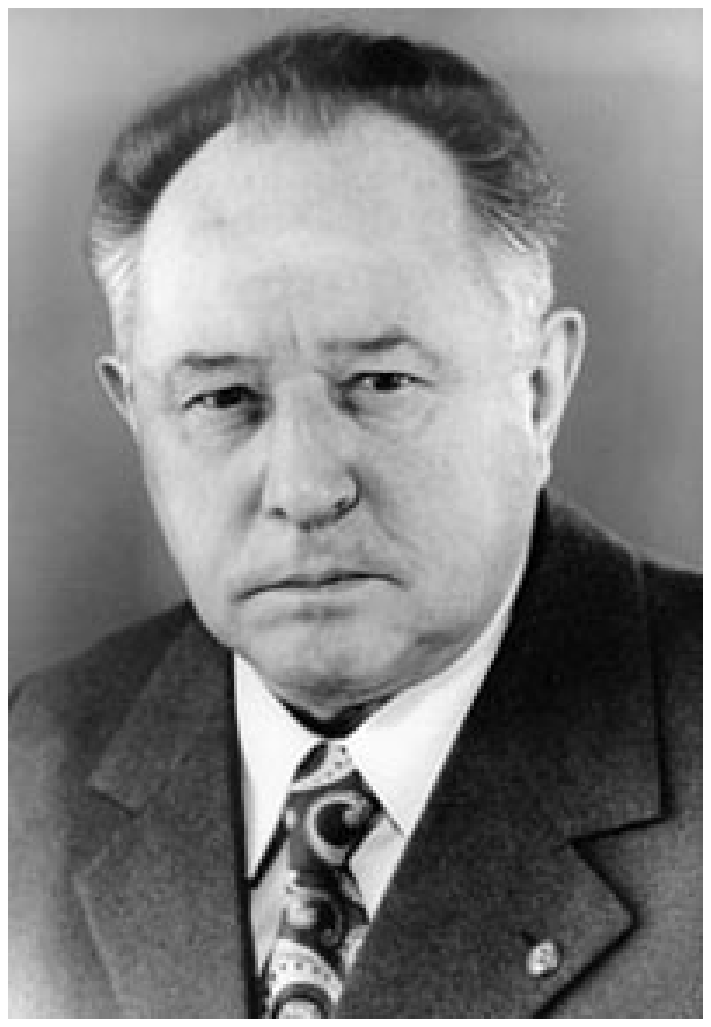,height=2cm} & 
\psfig{file=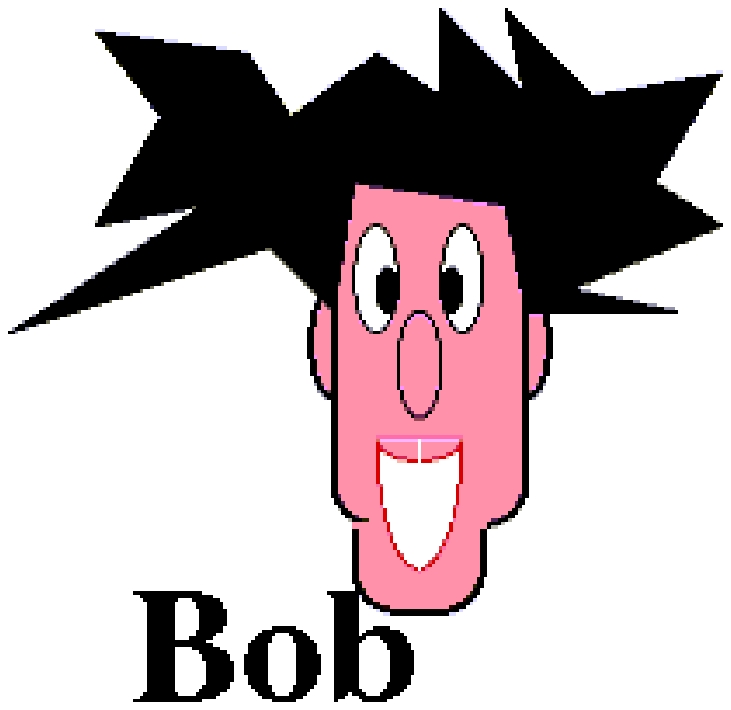,height=2cm} \\ \hline\hline
 & \multicolumn{3}{c||}
  {{\bf Input}: Two graphs $G_1$ and $G_2$} \\ \hline
{\bf 1} & & & \parbox[t]{4cm}
  {randomly chooses a permutation $\pi$ on $V(G_1)$ and a bit 
   $b \in \{1,2\}$, and computes $H = \pi(G_b)$} \\ \hline
{\bf 2} & & 
\mbox{\huge 
$\stackrel{\mbox{\normalsize $H$}}{\Leftarrow}$} & \\ \hline
{\bf 3} & 
 \parbox[t]{4cm}
  {determines $a \in \{1,2\}$ such that $G_a$ and $H$ are isomorphic} & &
\\ \hline
{\bf 4} & & 
\mbox{\huge 
$\stackrel{\mbox{\normalsize $a$}}{\Rightarrow}$} & \\ \hline
{\bf 5} & & & \parbox[t]{4cm}
  {accepts if and only if $a = b$}  \\ \hline\hline
\end{tabular}
\caption{The Goldreich-Micali-Wigderson IP protocol 
for $\overline{\graphiso}$.
\label{fig:goldreich-micali-wigderson-graphnoniso}
}
\end{figure}

Let us check that the implications (\ref{eq:ip-acceptance}) and
(\ref{eq:ip-rejection}) from Definition~\ref{def:interactive-proof-system} do
hold.  Suppose that $G_1$ and $G_2$ are non-isomorphic.  Then, it is easy for
Alice to determine that graph $G_b$, $b \in \{1,2\}$, to which $H$ is
isomorphic.  So she sends $a = b$, and Bob accepts with probability~$1$.  That
is,
\begin{eqnarray*}
(G_1, G_2) \in \overline{\graphiso} & \Lora & 
(\exists A)\, [\mbox{\rm Pr}((A, B)(G_1, G_2) = 1) = 1].
\end{eqnarray*}

Now suppose that $G_1$ and $G_2$ are isomorphic.  Then, no matter what clever
strategy Alice applies, her chance of answering correctly (i.e., with $a=b$)
is no better than~$\frac{1}{2}$ because she does not see Bob's random bit $b$
and so can do no better than guessing.  That is,
\begin{eqnarray*}
(G_1, G_2) \not\in \overline{\graphiso} & \Lora & 
(\forall \widehat{A})\, [\mbox{\rm Pr}((\widehat{A}, B)(G_1, G_2) = 1) 
\leq \frac{1}{2}].
\end{eqnarray*}
Note that the acceptance probability of $\leq \frac{1}{2}$ above is not yet
the acceptance probability of $\leq \frac{1}{4}$ required in
(\ref{eq:ip-rejection}) of Definition~\ref{def:interactive-proof-system}.
However, as mentioned above, standard probability amplification
techniques yield an error probability as close to zero as one desires.  We
leave the details to the reader.
\end{proof}

By definition, IP contains all of~$\np$. The above result shows that
IP also contains a problem from coNP, the class of complements of NP problems,
which is unlikely to be contained in~$\np$.  
So, the question arises of how big the
class IP actually is.  A famous result of Adi Shamir~\cite{sha:j:ip} settled
this question: IP equals PSPACE, the class of problems that can be decided in
polynomial space.

\subsection{Zero-Knowledge Protocols}
\label{sec:zero}

Recalling the issue of authentication mentioned at the beginning of this
section, we are now ready to define zero-knowledge protocols.  

As mentioned above, $\graphiso$ is in~$\ip$.  To prove that the two given
graphs are isomorphic, Alice simply sends an isomorphism $\pi$ to Bob, which
he then checks deterministically in polynomial time.  Suppose, however, that
Alice wants to keep the isomorphism $\pi$ secret.  On the one hand, she does
not want to disclose her secret; on the other hand, she wants to prove to Bob
that she knows it.  What she needs is a very special IP protocol that conveys
nothing about her secret isomorphism, and yet proves that the graphs are
isomorphic.  The next section will present such a zero-knowledge protocol for
$\graphiso$.

But what is a zero-knowledge protocol and how can one formalize it?  The
intuition is this.  Imagine that Alice has a twin sister named Malice who
looks just like her.  However, Malice does not know Alice's secret.  Moreover,
Malice does not have Alice's unlimited computational power; rather, just as
the verifier Bob, she only operates like a randomized polynomial-time Turing
machine.  Still, she tries to simulate Alice's communication with Bob.  An IP
protocol has the {\em zero-knowledge property\/} if the information
communicated in Malice's simulated protocol cannot be distinguished from the
information communicated in Alice's original protocol.  Malice, not knowing
the secret, cannot put any information about the secret into her simulated
protocol, and yet she is able to generate that clone of the original protocol
that looks just like the original to an independent observer.  Consequently,
the verifier Bob (or any other party such as Erich) cannot extract any
information from the original protocol.  In short, if there's nothing in
there, you can't get anything out of it.

\begin{definition}[Zero-Knowledge Protocols]
{\rm{}\cite{gol-mic-rac:c:interactive-proof-systems,gol-mic-rac:j:interactive-proof-systems}} \quad
\label{def:zero-knowledge}
Let $(A,B)$ be an interactive proof system accepting a problem~$L$.  We say
$(A,B)$ is a {\em zero-knowledge protocol for $L$\/} if and only if there
exists a simulator Malice such that the following holds:
\begin{itemize} 
\item Malice runs a randomized polynomial-time Turing machine $M$ to simulate
  the prover Alice in her communication with Bob, thus yielding a simulated
  protocol $(M,B)$;
\item for each~$x \in L$, the tuples $(a_1, a_2, \ldots , a_k)$ and $(m_1,
  m_2, \ldots , m_k)$ representing the communication in $(A,B)$ and
  in~$(M,B)$, respectively, are identically distributed over the coin tosses
  of $A$ and $B$ in $(A,B)$ and of $M$ and $B$ in~$(M,B)$, respectively.
\end{itemize}
\end{definition}

The above definition is called ``honest-verifier perfect zero-knowledge''
in the literature.  That is, (a)~one assumes that the verifier is {\em honest},
and (b)~one requires that the information communicated in the simulated
protocol {\em perfectly\/}
coincides with the information communicated in the original protocol.

Assumption~(a) is not quite realistic for most cryptographic applications.
A dishonest verifier might alter the protocol to his own advantage.
Therefore, one should modify the definition above to require that for {\em
  each\/} verifier $B^{\ast}$ there exists a simulator $M^{\ast}$ generating
a simulated protocol not distinguishable from the original one.
However, honest-verifier zero-knowledge protocols with public random bits can
always be transformed to protocols that have the zero-knowledge property
also in the presence of dishonest verifiers.

Regarding assumption~(b), there are several other notions of zero-knowledge
that are weaker than perfect zero-knowledge, such as ``statistical
zero-knowledge'' and ``computational zero-knowledge.''~~In a {\em statistical
  zero-knowledge protocol\/} (also known as {\em almost-perfect zero-knowledge
  protocol\/}), one requires that the information communicated in the original
and in the simulated protocol be indistinguishable by certain statistical
tests.  In a {\em computational zero-knowledge protocol}, one merely requires
that the information communicated in the original and in the simulated
protocol be computationally indistinguishable, i.e., for each randomized
polynomial-time Turing machine, the probability of detecting differences in
the corresponding distributions is negligibly small.

In the latter model, Goldreich, Micali, and
Wigderson~\cite{gol-mic-wid:c:nothing,gol-mic-wid:j:nothing} showed what is
considered by far the most important result on zero-knowledge: Every problem
in NP has a computational zero-knowledge protocol under the plausible
assumption that there exist cryptographically secure bit-commitment schemes.
The key idea is a computational zero-knowledge protocol for $\threecolor$, a
well-known NP-complete problem.  In contrast, it seems
unlikely~\cite{bra-cre:c:sorting-out-zero-knowledge} that such a strong claim
can be proven for the perfect zero-knowledge model presented in
Definition~\ref{def:zero-knowledge}.

For more information about interactive proof systems and zero-knowledge, we
refer to the books by 
Goldreich~\cite[Chapter~4]{gol:b:foundations}, Köbler
et al.~\cite[Chapter~2]{koe-sch-tor:b:graph-iso},
Papadimitriou~\cite[Chapter~12.2]{pap:b:complexity}, Balc\'{a}zar et
al.~\cite[Chapter~11]{bal-dia-gab:b:sctII}, and Bovet et
al.~\cite[Chapter~10]{bov-cre:b:complexity} and to the surveys by Oded
Goldreich~\cite{gol:j:zero-knowledge-survey}, Shafi
Goldwasser~\cite{gol:c:interactive-proof-systems}, and Joan
Feigenbaum~\cite{fei:j:interactive-proof-systems}.

\subsection{Zero-Knowledge Protocol for the Graph Isomorphism Problem}
\label{sec:graphiso}

Oded Goldreich, Silvio Micali, and Avi
Wigderson~\cite{gol-mic-wid:c:nothing,gol-mic-wid:j:nothing} proposed a
zero-knowledge protocol for the graph isomorphism problem.  This result was
quite a surprise, since previously zero-knowledge protocols were known only
for problems contained both in NP and coNP\@. It is considered to be
unlikely that NP equals coNP; in particular, it is considered to be unlikely
that $\graphiso$ is in coNP\@.

\begin{theorem}
{\rm{}\cite{gol-mic-wid:c:nothing,gol-mic-wid:j:nothing}} \quad
\label{thm:gi-is-zero-knowledge}
$\graphiso$ has a zero-knowledge protocol.
\end{theorem}

\begin{proof} 
  Figure~\ref{fig:goldreich-micali-wigderson} shows the
  Goldreich-Micali-Wigderson protocol.  One difference to the protocol for the
  graph non-isomorphism problem in
  Figure~\ref{fig:goldreich-micali-wigderson-graphnoniso} is that now Alice
  too makes random choices.  
  
  Alice's secret is the isomorphism $\pi$ she has chosen.  The protocol is
  correct, since Alice knows her secret $\pi$ and also her random
  permutation~$\rho$.  Hence, she can easily compute the
  isomorphism $\sigma$ with $\sigma(G_b) = H$ to prove her identity to Bob.
  When doing so, she does not have to disclose her secret $\pi$ to Bob in
  order to convince him of her identity.  In particular,
\begin{eqnarray*}
(G_1, G_2) \in \graphiso & \Lora & 
(\exists A)\, [\mbox{\rm Pr}((A, B)(G_1, G_2) = 1) = 1],
\end{eqnarray*}
so the implication~(\ref{eq:ip-acceptance}) from
Definition~\ref{def:interactive-proof-system} holds.  Since Alice herself has
chosen two isomorphic graphs, the case $(G_1, G_2) \not\in \graphiso$ does not
occur, so the implication~(\ref{eq:ip-rejection}) from
Definition~\ref{def:interactive-proof-system} trivially holds if the protocol
is implemented properly.  Thus, the protocol is an interactive proof system
for $\graphiso$.

Recall that Alice wants to prove her identity via this protocol.  Suppose that
Erich or Malice want to cheat by pretending to be Alice.  They do not know her
secret isomorphism~$\pi$, but they do know the public isomorphic graphs $G_1$
and~$G_2$.  They want to convince Bob that they know Alice's secret, which
corresponds to~$(G_1,G_2)$.  If, by coincidence, Bob's bit $b$ equals their
previously chosen bit~$a$, they win.  However, if $b \neq a$, computing
$\sigma = \rho \circ \pi$ or $\sigma = \rho \circ \pi^{-1}$ requires knowledge
of~$\pi$.  Without knowing~$\pi$, computing $\pi$ from the public graphs $G_1$
and $G_2$ seems to be impossible for them, since $\graphiso$ is a hard
problem, too hard even for randomized polynomial-time Turing machines.  Thus,
they will fail provided that the graphs are chosen large enough.  

\begin{figure}[!htb]
\centering
\begin{tabular}{||c||c|c|c||}
\hline\hline
\parbox[t]{.5cm}{\bf Step} & 
\psfig{file=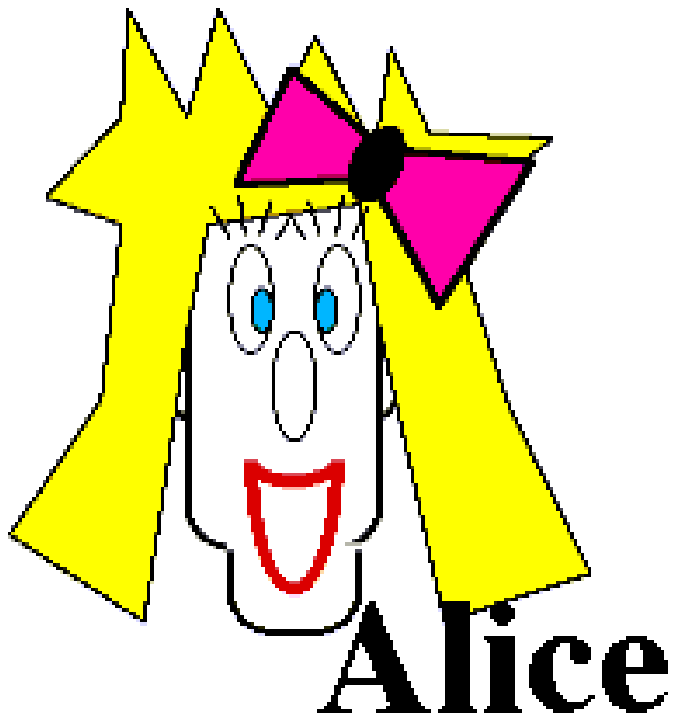,height=2cm} & 
\psfig{file=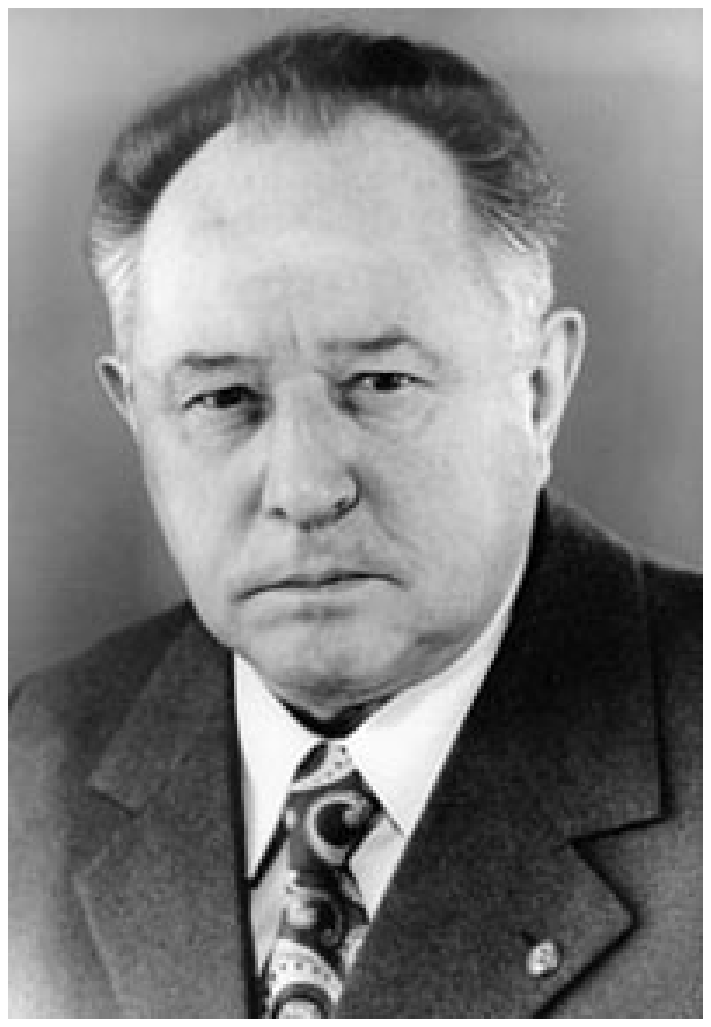,height=2cm} & 
\psfig{file=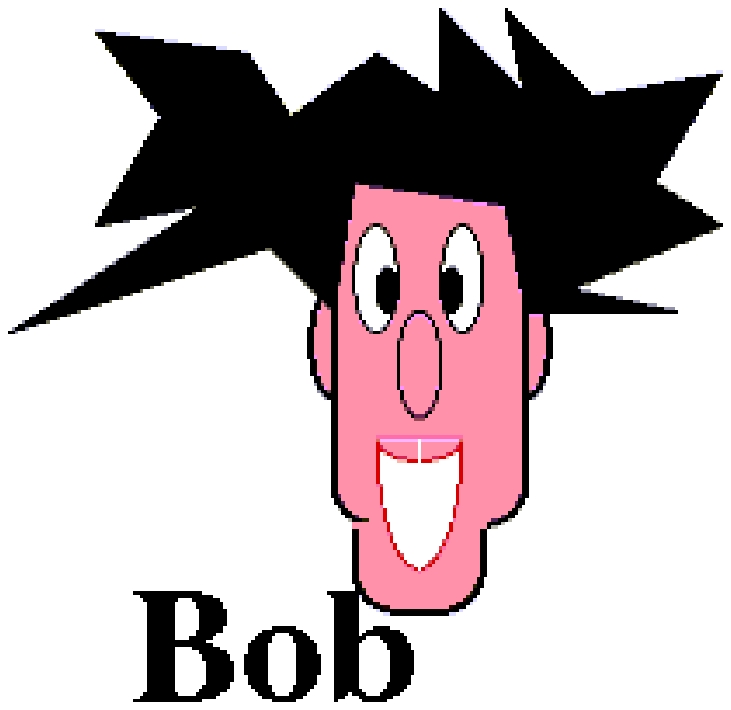,height=2cm} \\ \hline\hline
 & \multicolumn{3}{c||}
  {{\bf Generation of isomorphic graphs and a secret isomorphism}} \\ \hline
{\bf 1} & \parbox[t]{4cm}
  {chooses a large graph $G_1$, a random permutation $\pi$ on 
   $G_1$'s vertices, and computes the graph $G_2 = \pi(G_1)$; \\
   $(G_1, G_2)$ are public, $\pi$ is private} & & \\ \hline
 & \multicolumn{3}{c||}
  {{\bf Protocol}} \\ \hline
{\bf 2} &
\parbox[t]{4cm}
  {randomly chooses a permutation $\rho$ on $V(G_1)$ and a bit 
   $a \in \{1,2\}$,  computes $H = \rho(G_a)$} & & \\ \hline
{\bf 3} & & 
\mbox{\huge 
$\stackrel{\mbox{\normalsize $H$}}{\Rightarrow}$} & \\ \hline
{\bf 4} & & & 
 \parbox[t]{4cm}
  {chooses a bit $b \in \{1,2\}$ at random and wants to see an isomorphism
  between $G_b$ and $H$}
\\ \hline
{\bf 5} & & 
\mbox{\huge 
$\stackrel{\mbox{\normalsize $b$}}{\Leftarrow}$} & \\ \hline
{\bf 6} & \parbox[t]{4cm}
  {computes the permutation
\[
\sigma = \left\{
\begin{array}{ll}
\rho & \mbox{if $b = a$} \\
\rho \circ \pi & \mbox{if $1 = b \neq a = 2$} \\
\rho \circ \pi^{-1}  & \mbox{if $2 = b \neq a = 1$}
\end{array}
\right.
\]
satisfying $\sigma(G_b) = H$}  & & \\ \hline
{\bf 7} & & 
\mbox{\huge 
$\stackrel{\mbox{\normalsize $\sigma$}}{\Rightarrow}$} & \\ \hline
{\bf 8} & & & 
 \parbox[t]{4cm}
  {verifies that indeed
\[
\sigma(G_b) = H
\]
and accepts accordingly} \\ \hline\hline
\end{tabular}
\caption{The Goldreich-Micali-Wigderson zero-knowledge protocol 
for graph isomorphism.
\label{fig:goldreich-micali-wigderson}
}
\end{figure}

Since they cannot do better than guessing the bit~$b$, they can cheat with
probability at most~$\frac{1}{2}$.  Of course, they can always guess the
bit~$b$, which implies that their chance of cheating successfully is
exactly~$\frac{1}{2}$.  Hence, if Bob demands, say, $k$ independent rounds
of the protocol to be executed, he can make the cheating probability as small
as~$2^{-k}$, and thus is very likely to detect any cheater.  Note that after
only 20 rounds the odds of malicious Malice getting away with it undetected are
less than one to one million.  Hence, the protocol is correct.

\begin{figure}[!htb]
\centering
\begin{tabular}{||c||c|c|c||}
\hline\hline
\parbox[t]{.5cm}{\bf Step} & 
\psfig{file=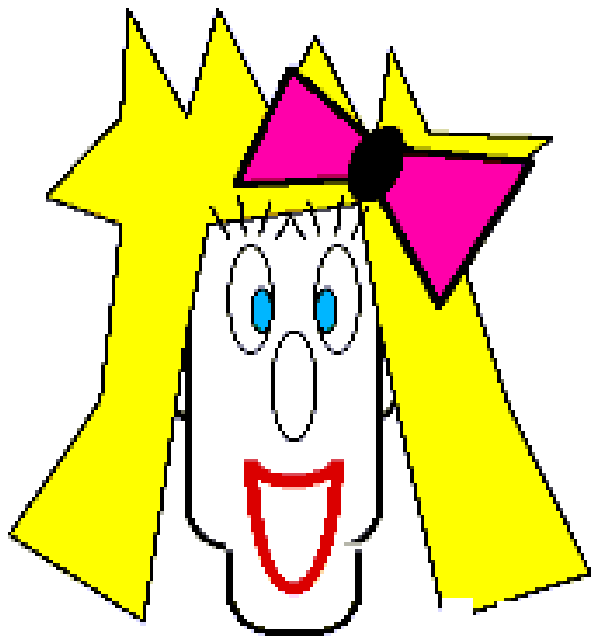,height=2cm} \hspace*{-1.2cm}  
\mbox{\bf\large Malice} & 
\psfig{file=mielke.ps,height=2cm} & 
\psfig{file=bob.ps,height=2cm} \\ \hline\hline
 & \multicolumn{3}{c||}
  {{\bf Simulated generation of isomorphic graphs}} \\ \hline
{\bf 1} & \parbox[t]{4cm}
  {knows the public pair $(G_1, G_2)$ of isomorphic graphs, does not know
   Alice's secret~$\pi$} & & \\ \hline
 & \multicolumn{3}{c||}
  {{\bf Simulated Protocol}} \\ \hline
{\bf 2} &
\parbox[t]{4cm}
  {randomly chooses a permutation $\rho$ on $V(G_1)$ and a bit 
   $a \in \{1,2\}$,  computes $H = \rho(G_a)$} & & \\ \hline
{\bf 3} & & 
\mbox{\huge 
$\stackrel{\mbox{\normalsize $H$}}{\Rightarrow}$} & \\ \hline
{\bf 4} & & & 
 \parbox[t]{4cm}
  {chooses a bit $b \in \{1,2\}$ at random and wants to see an isomorphism
  between $G_b$ and $H$}
\\ \hline
{\bf 5} & & 
\mbox{\huge 
$\stackrel{\mbox{\normalsize $b$}}{\Leftarrow}$} & \\ \hline
{\bf 6} &\parbox[t]{4cm}
  {if $b \neq a$ then $M$ deletes all messages transmitted in this round and
    repeats; \\
    if $b = a$ then $M$ sends $\sigma = \rho$}  & & \\ \hline
{\bf 7} & & 
\mbox{\huge 
$\stackrel{\mbox{\normalsize $\sigma$}}{\Rightarrow}$} & \\ \hline
{\bf 8} & & & 
 \parbox[t]{4cm}
  {$b = a$ implies that indeed
\[
\sigma(G_b) = H , 
\]
so Bob accepts ``Alice's'' identity} \\ \hline\hline
\end{tabular}
\caption{How to simulate the Goldreich-Micali-Wigderson protocol 
without knowing the secret~$\pi$.
\label{fig:goldreich-micali-wigderson-simulator}
}
\end{figure}

It remains to show that the protocol in
Figure~\ref{fig:goldreich-micali-wigderson} is zero-knowledge.
Figure~\ref{fig:goldreich-micali-wigderson-simulator} shows a simulated
protocol with Malice, who does not know the secret~$\pi$, replacing Alice.
The information communicated in one round of the protocol is given by a triple
of the form $(H, b, \sigma)$.  Whenever Malice chooses a bit $a$ with $a =
b$, she simply sends $\sigma = \rho$ and wins: Bob, or any independent
observer, will not detect that she in fact is Malice.  Otherwise, whenever $a
\neq b$, Malice fails.  However, that's no problem at all: She simply deletes
this round from the simulated protocol and repeats.  Thus, she can produce a
sequence of triples of the form $(H, b, \sigma)$ that is indistinguishable
from the corresponding sequence of triples in the original protocol between
Alice and Bob.  It follows that the Goldreich-Micali-Wigderson protocol is
zero-knowledge.
\end{proof}

\subsection{Fiat and Shamir's Zero-Knowledge Protocol}

Based on a similar protocol by Goldwasser, Micali and
Rackoff~\cite{gol-mic-rac:j:interactive-proof-systems},
Amos Fiat and Adi Shamir~\cite{fia-sha:c:fiat-shamir-zero-knowledge} proposed
a zero-knowledge protocol for a number-theoretical problem.  It is based on
the assumption that computing square roots in $\integers_{n}^{\ast}$ is
infeasible in practice.  Due to its properties, the Fiat-Shamir protocol is
particularly suitable for authentication of individuals in large computer
networks.  It is a public-key protocol, it is more efficient than other
public-key protocols such as the RSA algorithm, it can be implemented on a
chip card, and it is zero-knowledge.  These advantages resulted in a rapid
deployment of the protocol in practical applications.  The Fiat-Shamir
protocol is integrated in the ``Videocrypt'' Pay-TV
system~\cite{coh-has:p:controlling-access-broadcast}. 
The original Fiat-Shamir identification scheme has later been improved
by Feige, Fiat und
Shamir~\cite{fei-fia-sha:j:zero-knowledge-proof-of-identity}
to a zero-knowledge protocol in which not only the secret square roots
modulo~$n$ are not revealed, but also the information of whether or not
there {\em exists\/} a square root modulo~$n$ is not leaked.

The theory of zero-knowledge may also become important in future internet
technologies.  To prevent confusion, we note that Zero-Knowledge Systems,
Inc., a Montr\'{e}al-based company that was founded in 1997 and provides
products and services enabling users to protect their privacy on-line on the
world wide web, is not a commercial fielding of zero-knowledge
protocols~\cite{gol:perscomm:nov01}.

\begin{figure}[!htb]
\centering
\begin{tabular}{||c||c|c|c||}
\hline\hline
\parbox[t]{.5cm}{\bf Step} & 
\psfig{file=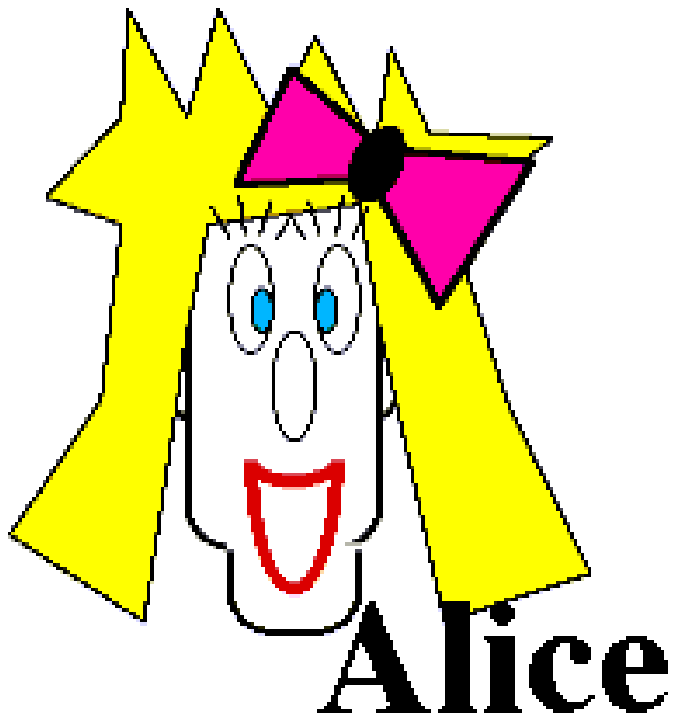,height=2cm} & 
\psfig{file=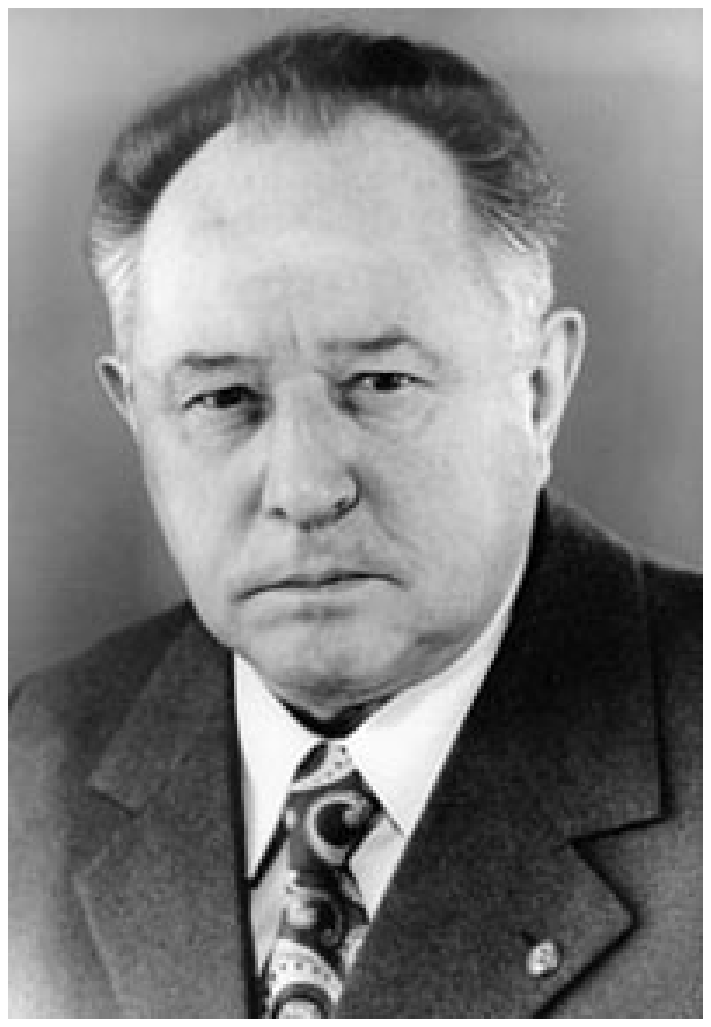,height=2cm} & 
\psfig{file=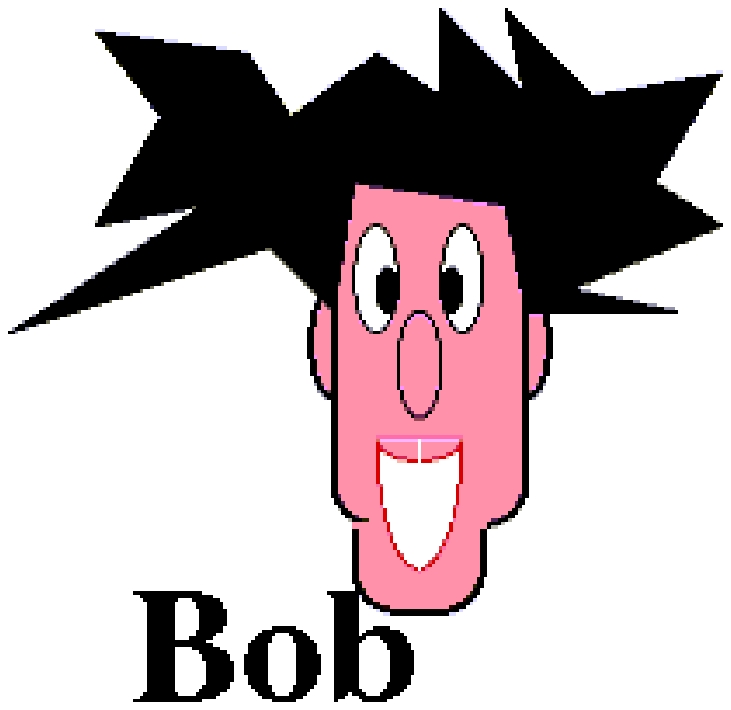,height=2cm} \\ \hline\hline
 & \multicolumn{3}{c||}
  {{\bf Key generation}} \\ \hline
{\bf 1} & \parbox[t]{4cm}
  {chooses two large primes $p$ and $q$ and a secret
   $s \in \integers_{n}^{\ast}$, $n = pq$,
   and computes $v = s^2 \mod n$; \\
   $p$, $q$, and $s$ are kept secret, \\
   whereas $n$ and $v$ are public} & & \\ \hline
 & \multicolumn{3}{c||}
  {{\bf Protocol}} \\ \hline
{\bf 2} &
\parbox[t]{4cm}
  {chooses $r \in \integers_{n}^{\ast}$ at random and computes $x = r^2 \mod n$} & & \\ \hline
{\bf 3} & & 
\mbox{\huge 
$\stackrel{\mbox{\normalsize $x$}}{\Rightarrow}$} & \\ \hline
{\bf 4} & & & 
 \parbox[t]{4cm}
  {chooses a bit $b \in \{0,1\}$ at random}
\\ \hline
{\bf 5} & & 
\mbox{\huge 
$\stackrel{\mbox{\normalsize $b$}}{\Leftarrow}$} & \\ \hline
{\bf 6} &\parbox[t]{4cm}
  {computes $y = r \cdot s^b \mod n$}  & & \\ \hline
{\bf 7} & & 
\mbox{\huge 
$\stackrel{\mbox{\normalsize $y$}}{\Rightarrow}$} & \\ \hline
{\bf 8} & & & 
 \parbox[t]{4cm}
  {verifies that indeed
\[
y^2 \equiv x \cdot v^b \mod n
\]
and accepts accordingly} \\ \hline\hline
\end{tabular}
\caption{The Fiat-Shamir zero-knowledge protocol.
\label{fig:fiat-shamir}
}
\end{figure}

\begin{figure}[!htb]
\centering
\begin{tabular}{||c||c|c|c||}
\hline\hline
\parbox[t]{.5cm}{\bf Step} & 
\psfig{file=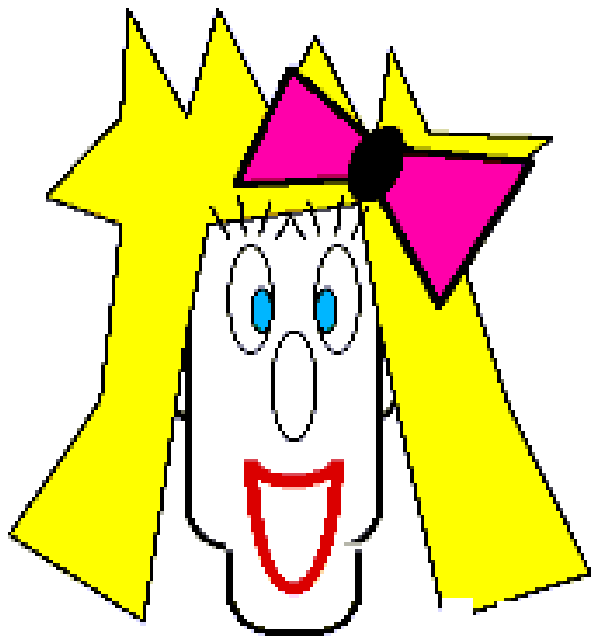,height=2cm} \hspace*{-1.2cm}  
\mbox{\bf\large Malice} & 
\psfig{file=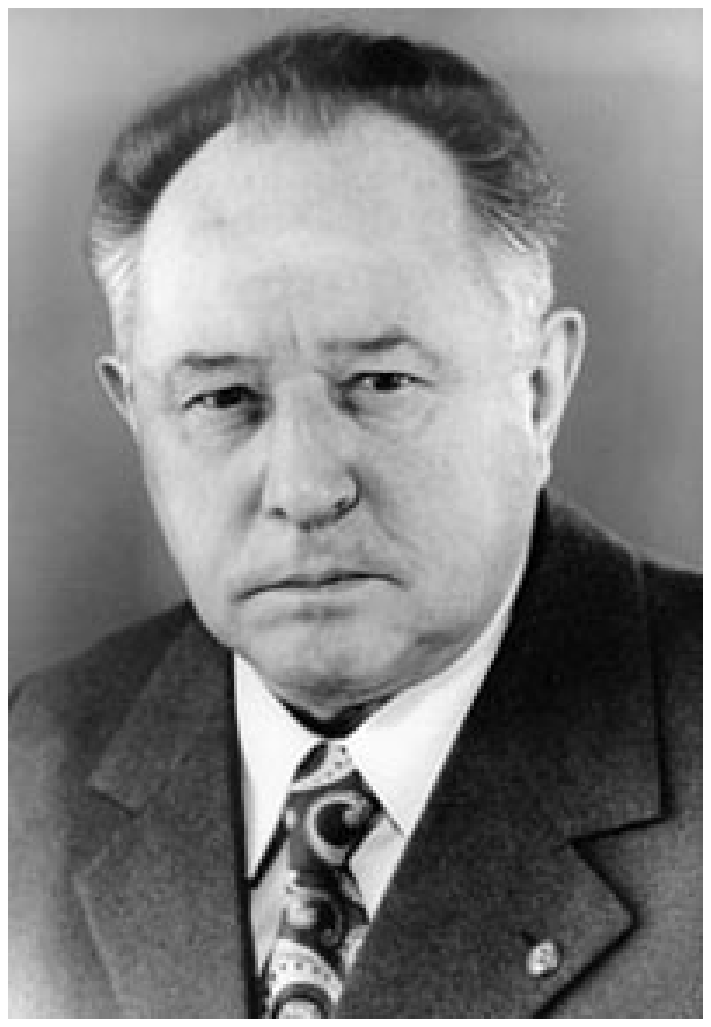,height=2cm} & 
\psfig{file=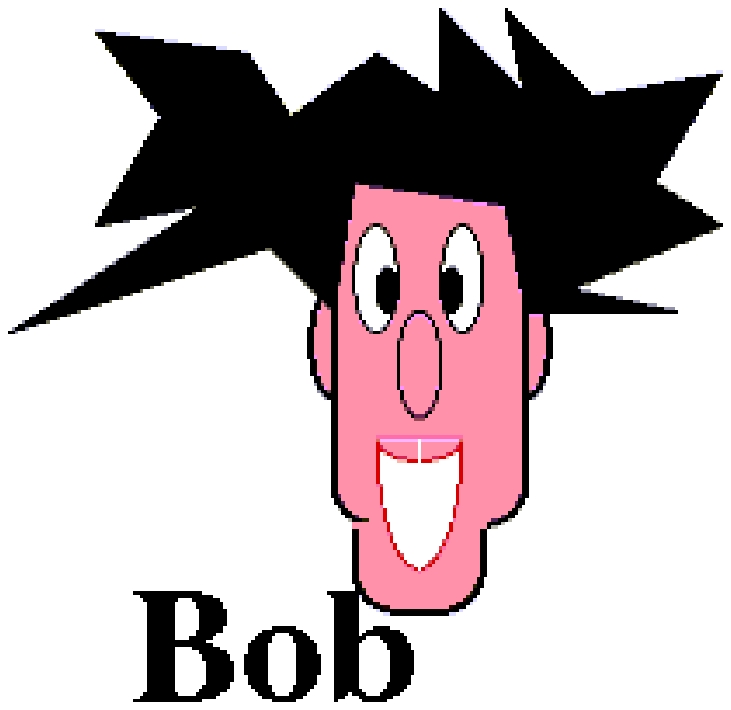,height=2cm} \\ \hline\hline
 & \multicolumn{3}{c||}
  {{\bf Simulated key generation}} \\ \hline
{\bf 1} & \parbox[t]{4cm}
  {knows the public $n = pq$ and \\ $v = s^2 \mod n$; \\
   does not know the private primes $p$ and $q$ and Alice's secret $s$}
 & & \\ \hline
 & \multicolumn{3}{c||}
  {{\bf Simulated Protocol}} \\ \hline
{\bf 2} &
\parbox[t]{4cm}
  {randomly chooses $r \in \integers_{n}^{\ast}$ and a bit $c \in \{0,1\}$, \\
   computes $x = r^2 \cdot v^{-c} \mod n$} & & \\ \hline
{\bf 3} & & 
\mbox{\huge 
$\stackrel{\mbox{\normalsize $x$}}{\Rightarrow}$} & \\ \hline
{\bf 4} & & & 
 \parbox[t]{4cm}
  {chooses a bit $b \in \{0,1\}$ at random}
\\ \hline
{\bf 5} & & 
\mbox{\huge 
$\stackrel{\mbox{\normalsize $b$}}{\Leftarrow}$} & \\ \hline
{\bf 6} &\parbox[t]{4cm}
  {if $b \neq c$ then $M$ deletes all messages transmitted in this round and
    repeats; \\
    if $b = c$ then $M$ sends $y = r$} & & \\ \hline
{\bf 7} & & 
\mbox{\huge 
$\stackrel{\mbox{\normalsize $y$}}{\Rightarrow}$} & \\ \hline
{\bf 8} & & & 
 \parbox[t]{4cm}
  {$b = c$ implies that indeed
\begin{eqnarray*}
y^2 & = & r^2 \ \; = \ \; r^2 v^{-c} v^{b} \\
    & \equiv & x \cdot v^b \mod n,
\end{eqnarray*}
so Bob accepts ``Alice's'' identity} \\ \hline\hline
\end{tabular}
\caption{How to simulate the Fiat-Shamir protocol without knowing 
         the secret~$s$.
\label{fig:fiat-shamir-simulator}
}
\end{figure}

\begin{theorem}
{\rm{}\cite{fia-sha:c:fiat-shamir-zero-knowledge}} \quad
\label{thm:fiat-shamir-zero-knowledge}
The Fiat-Shamir procedure given in Figure~\ref{fig:fiat-shamir}
is a zero-knowledge protocol.
\end{theorem}

\begin{proof} 
  Look at Figure~\ref{fig:fiat-shamir}.  The protocol is correct, since
  Alice knows the secret $s \in \integers_{n}^{\ast}$ that she has chosen, and
  thus she can compute~$y = r \cdot s^b$, where $b$ is the bit that Bob has
  chosen at random.  Hence, it holds in $\integers_{n}^{\ast}$ that
\[
y^2 \equiv (r \cdot s^b)^2 \equiv r^2 \cdot s^{2b} \equiv r^2 \cdot v^b \equiv
x \cdot v^b \mod n,
\]
so Bob accepts Alice's identity.

Suppose now that Erich or Malice want to cheat by pretending to be Alice.
They do not know her secret~$s$, nor do they know the primes $p$ and~$q$, but
they do know the public $n = pq$ and $v = s^2 \mod n$.  They want to convince
Bob that they know Alice's secret~$s$, the square root of $v$ modulo~$n$.  If,
by coincidence, Bob's bit $b$ equals zero then $y = r \cdot s^0 = r$ and they
win.  However, if $b = 1$, computing a $y$ that satisfies $y^2 \equiv x \cdot
v^b \mod n$ requires knowledge of the secret~$s$, assuming that computing
square roots modulo $n$ is hard.  Without knowing~$s$, if Malice or Erich were
able to compute the correct answer for both $b = 0$ and $b = 1$, say $y_b$
with $y_{b}^{2} \equiv x \cdot v^b \mod n$, they could efficiently compute
square roots modulo~$n$ as follows: $y_{0}^{2} \equiv x \mod n$ and $y_{1}^{2}
\equiv x \cdot v \mod n$ implies $(\frac{y_1}{y_0})^2 \equiv v \mod n$; hence,
$\frac{y_1}{y_0}$ is a square root of $v$ modulo~$n$.

It follows that they can cheat with probability at most~$\frac{1}{2}$.  Of
course, they can always guess the bit~$b$ in advance and prepare the answer
accordingly.  Choosing $x = r^2 \cdot v^{-b} \mod n$ and $y = r$ implies that
\begin{equation}
\label{eq:cheat-fiat-shamir}
y^2 \equiv r^2 \equiv r^2 \cdot v^{-b} \cdot v^{b} \equiv x \cdot v^{b} \mod n.
\end{equation}
Thus, Bob will not detect any irregularities and will accept.  Hence, their
chance to cheat successfully is exactly~$\frac{1}{2}$.  Again, if Bob
demands, say, $k$ independent rounds of the protocol to be executed, he can
make the cheating probability as small as desired and is very likely to
detect any cheater.

It remains to show that the Fiat-Shamir protocol in
Figure~\ref{fig:fiat-shamir} is zero-knowledge.
Figure~\ref{fig:fiat-shamir-simulator} shows a simulated protocol with Malice,
who does not know the secret~$s$, replacing Alice.  The information
communicated in one round of the protocol is given by a triple of the form
$(x, b, y)$.  In addition to the randomly chosen $r \in \integers_{n}^{\ast}$,
Malice guesses a bit $c \in \{0,1\}$ and computes $x = r^2 \cdot v^{-c} \mod
n$, which she sends to Bob.  Whenever $c$ happens to be equal to Bob's
bit~$b$, Malice simply sends $y = r$ and wins.  By an argument analogous to
Equation~(\ref{eq:cheat-fiat-shamir}) above, neither Bob nor any independent
observer will detect that she actually is Malice:
\[
y^2 \equiv r^2 \equiv r^2 \cdot v^{-c} \cdot v^{b} \equiv x \cdot v^{b} \mod n.
\]
Otherwise, whenever $c \neq b$, Malice fails.  However, that's no problem at
all: She simply deletes this round from the simulated protocol and repeats.
Thus, she can produce a sequence of triples of the form $(x, b, y)$ that is
indistinguishable from the corresponding sequence of triples in the original
protocol between Alice and Bob.  It follows that the Fiat-Shamir protocol is
zero-knowledge.
\end{proof}

We have chosen to give here the original Fiat-Shamir identification scheme as
presented in most books (see, e.g.,
\cite{gol:b:foundations,beu-sch-wol:b:kryptographie}).  Note, however, that
quite a number of modifications and improvements of the Fiat-Shamir protocol
have been proposed, including the ``zero-knowledge proof of knowledge''
protocol of Feige, Fiat und
Shamir~\cite{fei-fia-sha:j:zero-knowledge-proof-of-identity}.  We also note in
passing that we omitted many formal details in our arguments in this section.
A rigid formalism (see~\cite{gol:b:foundations}) is helpful in discussing many
subtleties that can arise in zero-knowledge protocols.  For example, looking
at Figure~\ref{fig:fiat-shamir}, Alice could be impersonated by anyone who
picks the value $r = 0$ without Bob detecting this fraud.  We refer to
Burmester and Desmedt~\cite{bur-des:j:remarks-soundness-of-proofs} for
appropriate modifications of the scheme.  Moreover, Burmester et
al.~\cite{bur-des-pip-wal:c:general-zero-knowledge,bur-des-bet:j:zero-knowledge-identification}
proposed efficient zero-knowledge protocols in a general algebraic setting.

\section{Strongly Noninvertible Associative One-Way Functions}
\label{sec:aowf}

Recall Rivest and Sherman's secret-key agreement protocol
(Figure~\ref{fig:rivest-sherman-secret-key}) and Rabi and Sherman's digital
signature protocol (Figure~\ref{fig:rabi-sherman-digital-signature}) presented
in Section~\ref{sec:riv-rab-she}.  Both of these protocols use a candidate for
a strongly noninvertible, associative one-way function.  Are these protocols
secure?  This question has two aspects: (1)~Are they secure under the
assumption that strongly noninvertible, associative one-way functions indeed
exist?  (2)~What evidence do we have for the existence of such functions?

The first question is an open problem.  Security here depends on
precisely how ``strong noninvertibility'' is defined, and in which model.
Traditional complexity theory is concerned with the worst-case model and has
identified a large number of problems that are hard in the worst case.
Cryptographic applications, however, require the more demanding average-case
model (see, e.g.,
\cite{gol:b:foundations,gol:b:modern-cryptography,lub:b:pseudorandom}) for
which much less is known.  As noted by Rabi and Sherman~\cite{rab-she:j:aowf},
no proof of security for the Rivest--Sherman and Rabi--Sherman protocols is
currently known, and even assuming the existence of associative one-way
functions that are strongly noninvertible in the weaker worst-case model would
not 
imply that the protocols are secure.  In that regard,
however, the Rivest--Sherman and Rabi--Sherman protocols are just like many
other protocols currently used in practical applications.  For example,
neither the Diffie--Hellman protocol nor the RSA protocol currently has a proof
of security.  There are merely heuristic, intuitive arguments about how to
avoid certain direct attacks.  The ``security'' of the Diffie--Hellman protocol
draws on the assumption that computing discrete logarithms is hard, and the
``security'' of the RSA protocol draws on the assumption that factoring large
integers is hard.  Breaking Diffie--Hellman is not even known to be as hard as
the discrete logarithm problem, and breaking RSA is not even known to be as
hard as the factoring problem.  In a similar vein, Rabi and
Sherman~\cite{rab-she:t-no-URL:aowf,rab-she:j:aowf} only give intuitive
arguments for the security of their protocols, explaining how to employ the
strong noninvertibility of associative one-way functions to preclude certain
direct attacks.

Turning to the second question raised above: What evidence do we have
that strongly noninvertible, associative one-way functions exist?  Assuming $\p
\neq \np$, we will show how to construct total, strongly noninvertible, 
commutative,\footnote{Commutativity is needed to extend the Rivest--Sherman 
and Rabi--Sherman protocols from two parties to $m > 2$ parties.  }
associative one-way functions~\cite{hem-rot:j:aowf}.  The question of whether
or not P equals NP is perhaps the most important question in theoretical
computer science.  It is widely believed that P differs from NP, although this
question has remained open for more than thirty years now.  For more
background on complexity theory, we refer
to the textbooks~\cite{bal-dia-gab:b:sctI:95,bov-cre:b:complexity,hem-ogi:b:companion,pap:b:complexity}.

\subsection{Definitions and Progress of Results}
\label{sec:definitions}

From now on, we adopt the worst-case notion of one-way functions that is due
to Grollmann and Selman~\cite{gro-sel:j:complexity-measures}, see also the
papers by Ko~\cite{ko:j:operators}, Berman~\cite{ber:thesis:iso}, and
Allender~\cite{all:thesis:invertible,all:coutdatedExceptForPUNCstuff:complexity-sparse},
and the surveys~\cite{sel:j:one-way,bey-hem-hom-rot:j:aowf-survey}.  
Recall that one-way functions are easy to compute but hard to invert.
To prevent the notion of noninvertibility from being trivialized, one-way
functions are required to be ``honest,'' i.e., to not shrink their inputs too
much.  Formal definitions of
various types of honesty can be found
in~\cite{gro-sel:j:complexity-measures,hem-rot-wec:c:easy-one-way-permutations,hem-rot:j:one-way,rot-hem:j:one-way,hem-pas-rot:c:strong-noninvertibility,hom:t:low-ambiguity-aowf,hom-tha:c:one-way-permutations}.

One-way functions are often considered to be one-argument functions.
Since the protocols from Section~\ref{sec:riv-rab-she} require two-argument
functions, the original definition is here tailored to the case of two-ary
functions.  Let $\rho : \nats \times \nats \rightarrow \nats$ be any two-ary
function; $\rho$ may be nontotal and it may be many-to-one.
We say that $\rho$ is {\em (polynomial-time) invertible\/} if there exists a
polynomial-time computable function $g$ such that for all $z \in
\image(\rho)$, it holds that $\rho(g(z)) = z$; otherwise, we call $\rho$
{\em not polynomial-time invertible}, or {\em noninvertible} for short. 
We say that $\rho$ is a {\em one-way function\/} if and only if $\rho$ is
honest, polynomial-time computable, and noninvertible.
One-argument one-way functions are well-known to exist if and only if $\p \neq
\np$; see, e.g.,~\cite{sel:j:one-way,bal-dia-gab:b:sctI:95}.  It is easy to
prove the analogous result for two-argument one-way functions,
see~\cite{hem-rot:j:aowf,rab-she:j:aowf}.  

We now define strong noninvertibility (strongness, for short).  
As with noninvertibility, strongness requires an appropriate notion of honesty
so as to not be trivial.  This notion is called ``s-honesty''
in~\cite{hem-pas-rot:c:strong-noninvertibility}, and since it is merely a
technical requirement, we omit a formal definition here.  Intuitively,
``s-honesty'' fits the notion of strong noninvertibility in that it is
measured not only in the length of the function value but also in the length
of the corresponding given argument.

\begin{definition}
  {\rm{}(see~\cite{rab-she:j:aowf,hem-rot:j:aowf})} \quad
\label{d:strong-oneway}
Let $\sigma : \nats \times \nats \rightarrow \nats$ be any two-ary function;
$\sigma$ may be nontotal and it may be many-to-one.
Let $\pair{\cdot, \cdot} : \nats \times \nats \rightarrow\, \nats$ be some
standard pairing function.  
\begin{enumerate}
\item We say that $\sigma$ is {\em (polynomial-time) invertible with respect
    to its first argument\/} if and only if there exists a polynomial-time
  computable function $g_1$ such that for all $z \in \image(\sigma)$ and for
  all $a$ and $b$ with $(a,b) \in \domain(\sigma)$ and $\sigma(a,b) = z$, it
  holds that $\sigma(a, g_1(\pair{a,z})) = z$.
  
\item We say that $\sigma$ is {\em (polynomial-time) invertible with respect
    to its second argument\/} if and only if there exists a polynomial-time
  computable funtion $g_2$ such that for all $z \in \image(\sigma)$ and for
  all $a$ and $b$ with $(a,b) \in \domain(\sigma)$ and $\sigma(a,b) = z$, it
  holds that $\sigma(g_2(\pair{b,z}), b) = z$.
  
\item We say that $\sigma$ is {\em strongly noninvertible\/} if and only if
  $\sigma$ is neither invertible with respect to its first argument nor
  invertible with respect to its second argument.
    
\item We say that $\sigma$ is a {\em strong one-way function\/} if and only if
  $\sigma$ is s-honest, polynomial-time computable, and strongly
  noninvertible.
\end{enumerate}
\end{definition}

Below, we define Rabi and Sherman's notion of associativity, which henceforth
will be called ``weak associativity.''

\begin{definition} 
{\rm{}\cite{rab-she:t-no-URL:aowf,rab-she:j:aowf}} \quad
\label{def:rs-associative}
A two-ary function $\sigma : \nats \times \nats \rightarrow\, \nats$ is said
to be {\em weakly associative\/} if and only if $\sigma(a, \sigma(b, c)) =
\sigma(\sigma(a, b), c)$ holds for all $a,b,c \in \nats$ for which each of
$(a,b)$, $(b,c)$, $(a, \sigma(b, c))$, and $(\sigma(a, b), c)$ belongs to the
domain of~$\sigma$.
\end{definition} 

Although this notion is suitable for total functions, weak associativity does
not adequately fit the nontotal function case.  More precisely, weak
associativity fails to preclude, for nontotal functions, equations from having
a defined value to the left, while being undefined to the right of their
equality sign.  Therefore, we present in Definition~\ref{def:associative}
below another notion of associativity for two-ary functions that is suitable
both for total and for nontotal two-ary functions.  This definition is due to
Hemaspaandra and Rothe~\cite{hem-rot:j:aowf} 
who note that the two notions of
associativity are provably distinct (see~Proposition\ref{prop:ass}), and this
distinction can be explained (see~\cite{hem-rot:j:aowf}) via Kleene's careful
discussion~\cite[pp.~327--328]{kle:b:metamathematics} of two distinct notions
of equality for partial functions in recursion theory: ``Weak equality''
between two partial functions explicitly allows ``specific, defined function
values being equal to undefined'' as long as the functions take the same
values on their joint domain.  In contrast, ``complete equality'' precludes
this unnatural behavior by additionally requiring that two given partial
functions be equal only if their domains coincide; i.e., whenever one is
undefined, so is the other.  Weak associativity from
Definition~\ref{def:rs-associative} is based on Kleene's weak equality between
partial functions, whereas associativity from Definition~\ref{def:associative}
is based on Kleene's complete equality.

\begin{definition} 
{\rm{}\cite{hem-rot:j:aowf}} \quad
\label{def:associative}
Let $\sigma : \nats \times \nats \rightarrow\, \nats$ be any two-ary function;
$\sigma$ may be nontotal.  Define $\nats_{\bot} \equalsdef \nats \cup
\{\bot\}$, and define an extension $\sigmabot : \nats_{\bot}
\times \nats_{\bot} \rightarrow\, \nats_{\bot}$ of $\sigma$ as follows:
\[
  \sigmabots{a,b} \equalsdef \left\{ 
\begin{array}{ll} \sigma(a,b)  &
  \mbox{if $a \neq \bot$ and $b \neq \bot$ and $(a,b) \in \mbox{\rm
  domain}(\sigma)$} \\ 
  \bot & \mbox{otherwise.}
\end{array} 
\right.
\]

We say that $\sigma$ is {\em associative\/} if and only if, for all $a,b,c \in
\nats$, it holds that
\begin{eqnarray*}
\sigmabots{\sigmabots{a,b},c} & = & \sigmabots{a,\sigmabots{b,c}}.
\end{eqnarray*}

We say that $\sigma$ is {\em commutative\/} if and only if, for all $a,b \in
\nats$, it holds that 
\begin{eqnarray*}
\sigmabots{a,b} & = & \sigmabots{b,a}.
\end{eqnarray*}
\end{definition}

The following proposition explores the relation between the two associativity
notions presented respectively in Definition~\ref{def:rs-associative} and in
Definition~\ref{def:associative}.  In particular, these are indeed
different notions.

\begin{proposition}
{\rm{}\cite{hem-rot:j:aowf}} \quad ~
\label{prop:ass}
\begin{enumerate}
\item Every associative two-ary function is weakly associative.
  
\item Every total two-ary function is associative exactly if it is weakly
  associative.
  
\item There exist two-ary functions that are weakly associative, yet not
  associative.
\end{enumerate}
\end{proposition}

Rabi and Sherman~\cite{rab-she:t-no-URL:aowf,rab-she:j:aowf} showed that $\p
\neq \np$ if and only if commutative, weakly associative one-way functions
exist.  However, they did not achieve strong noninvertibility.  They did not
achieve totality of their weakly associative one-way functions, although they
presented a construction that they claimed achieves totality of any weakly
associative one-way function.  Hemaspaandra and Rothe~\cite{hem-rot:j:aowf}
showed that Rabi and Sherman's claim is unlikely to be true: Any proof of this
claim would imply that $\np = \up$, which is considered to be unlikely.
Intuitively, the reason that Rabi and Sherman's construction is unlikely to
work is that the functions constructed
in~\cite{rab-she:t-no-URL:aowf,rab-she:j:aowf} are not associative in the
sense of Definition~\ref{def:associative}.  In contrast, the Rabi--Sherman
construction indeed is useful to achieve totality of the associative, strongly
noninvertible one-way functions constructed in~\cite{hem-rot:j:aowf}.

Thus, Rabi and Sherman~\cite{rab-she:t-no-URL:aowf,rab-she:j:aowf} left open
the question of whether there are plausible complexity-theoretic conditions
sufficient to ensure the existence of total, strongly noninvertible,
commutative, associative one-way functions.  They also asked whether such
functions could be {\em constructed\/} from any given one-way function.
Section~\ref{sec:creating} presents the answers to these questions.

\subsection{Creating Strongly Noninvertible, Total, Commutative, Associative
  One-Way Functions from Any One-Way Function}
\label{sec:creating}

Theorem~\ref{thm:aowf-equ} below is the main result of this section.  Since
$\p \neq \np$ is equivalent to the existence of one-way functions with no
additional properties required, the converse of the implication stated in
Theorem~\ref{thm:aowf-equ} is clearly also true.  However, we focus on
only the interesting implication directions in
Theorem~\ref{thm:aowf-equ} and in the upcoming Theorem~\ref{thm:c} and
Theorem~\ref{thm:d}.

\begin{theorem}
  {\rm{}\cite{hem-rot:j:aowf}}
  \quad
\label{thm:aowf-equ}
If $\p \neq \np$ then there exist total, strongly noninvertible, commutative,
associative one-way functions.
\end{theorem}

A detailed proof of Theorem~\ref{thm:aowf-equ} can be found
in~\cite{hem-rot:j:aowf}, see also the
survey~\cite{bey-hem-hom-rot:j:aowf-survey}.
Here, we briefly sketch the proof idea.  

Assume $\p \neq \np$.  Let $A$ be a set in $\np - \p$, and let $M$ be a fixed
NP machine accepting~$A$.  Let $x \in A$ be an input accepted by~$M$ in
time~$p(|x|)$, where $p$ is some polynomial.  A useful property of NP sets is
that they have polynomial-time checkable certificates.\footnote{Other common
  names for ``certificate'' are ``witness'' and ``proof'' and ``solution.''
}
That is, for each certificate $z$ for ``$x \in A$,'' it holds that: (a)~the
length of $z$ is polynomially bounded in the length of~$x$, and (b)~$z$
certifies membership of $x$ in $A$ in a way that can be verified
deterministically in polynomial time.  $\certificate{M}{x}$ denotes the set of
all certificates of $M$ on input~$x$.  Note that $\certificate{M}{x}$ is 
nonempty exactly if $x \in A$.

\begin{figure}[ht!]
\begin{center}
\psfig{figure=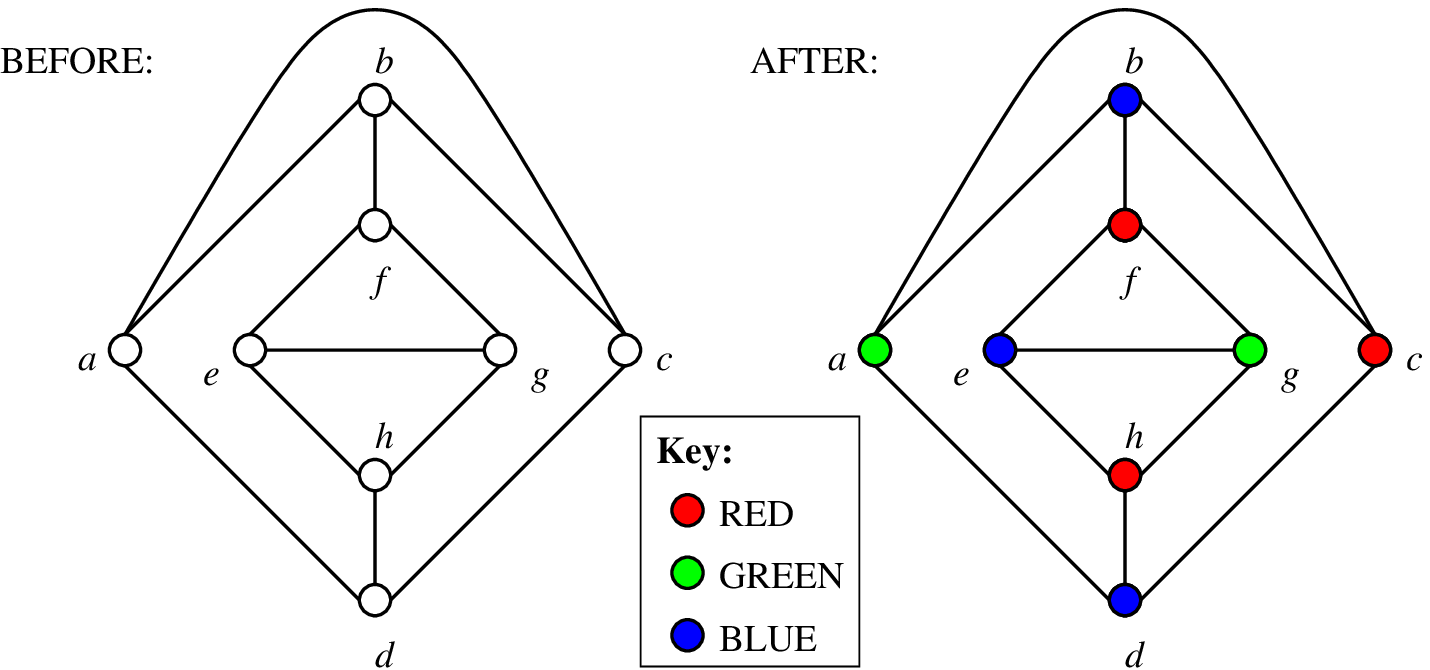,width=12cm}
\end{center}
\caption{\label{fig:dreif} The three-coloring $\psi$ of graph~$G$.}
\end{figure}

\begin{example}
  For concreteness, consider $\threecolor$, a well-known $\np$-complete
  problem that asks whether the vertices of a given graph can be colored with
  three colors such that no two adjacent vertices receive the same color.
  Such a coloring is called a legal three-coloring.  In other words, a legal
  three-coloring is a mapping $\psi$ from the vertex set of $G$ to the set of
  colors (\mbox{\rm RED}, \mbox{\rm GREEN}, \mbox{\rm BLUE}) such that the
  resulting color classes are independent sets.  Figure~\ref{fig:dreif} gives
  an example.
  
  The standard $\np$ machine for $\threecolor$ works as follows: Given a
  graph~$G$, nondeterministically guess a three-coloring $\psi$ of~$G$ (i.e.,
  a partition of the vertex set of $G$ into three color classes) and check
  deterministically whether $\psi$ is legal.
  
  Any legal three-coloring of~$G$ is a {\em certificate\/} for the
  three-colorability of $G$ (with respect to the above $\np$ machine).  For
  the specific graph from Figure~\ref{fig:dreif}, one certificate $\psi$ 
  is specified by the three color classes $\psi^{-1}(\mbox{\rm GREEN}) = \{ a
  , g \}$, $\psi^{-1}(\mbox{\rm RED}) = \{ c , f , h \}$, and
  $\psi^{-1}(\mbox{\rm BLUE}) = \{ b , d , e \}$.
  
  As is standard, graphs as well as three-colorings can be encoded as binary
  strings that represent nonnegative integers. 
\end{example}

Suppose that for each $x \in A$ and for each certificate $z$ for ``$x \in
A$,'' it holds that $|z| = p(|x|) > |x|$.
This is only a technical requirement that makes it easy to tell input strings
apart from their certificates.  For any integers $u, v, w \in \nats$, let
$\min(u,v)$ denote the minimum of $u$ and~$v$, and let $\min(u,v,w)$ denote
the minimum of $u$, $v$, and~$w$.  Define a two-ary function $\sigma : \nats
\times \nats \rightarrow\, \nats$ as follows:
\begin{itemize}
\item If $a = \pair{x,z_1}$ and $b = \pair{x,z_2}$ for some $x \in A$ with
  certificates $z_1,z_2 \in \certificate{M}{x}$ (where, possibly, $z_1 =
  z_2$), then define $\sigma(a,b) = \pair{x,\min(z_1,z_2)}$;
  
\item if there exists some $x \in A$ with certificate $z \in
  \certificate{M}{x}$ such that either $a = \pair{x,x}$ and $b = \pair{x,z}$,
  or $a = \pair{x,z}$ and $b = \pair{x,x}$, then define $\sigma(a,b) =
  \pair{x,x}$;

\item otherwise, $\sigma(a,b)$ is undefined.
\end{itemize}

What is the intuition behind the definition of~$\sigma$?  The number of
certificates contained in the arguments of $\sigma$ is decreased by one in a
way that ensures the associativity of~$\sigma$.  Moreover, $\sigma$ is
noninvertible, and it is also strongly noninvertible.  Why?  The intuition
here is that, regardless of whether none or either one of its arguments is
given in addition to $\sigma$'s function value, the inversion of~$\sigma$
requires information about the certificates for elements of~$A$.  However, our
assumption that $A \not\in \p$ guarantees that this information cannot
efficiently be extracted.

One can show that $\sigma$ is a commutative, associative one-way function that
is strongly noninvertible.  We will show associativity and strongness below.
Note that $\sigma$ is not a total function.  However, $\sigma$ can be extended
to a total function without losing any of its other properties already
established~\cite{hem-rot:j:aowf}.

We now show that $\sigma$ is strongly noninvertible.  For a contradiction,
suppose there is a polynomial-time computable inverter, $g_2$, for a fixed
second argument.  Hence, for each $w \in \image(\sigma)$ and for each second
argument $b$ for which there is an $a \in \nats$ with $\sigma(a,b) = w$, it
holds that
\[
\sigma(g_2(\pair{b,w}), b) = w.
\]
Then, contradicting our assumption that $A \not\in \p$, one could decide $A$
in polynomial time as follows:
\begin{quote} On input~$x$, compute $g_2 (\pair{\pair{x,x},\pair{x,x}})$,
  compute the integers $d$ and $e$ for which $\pair{d,e}$ equals $g_2
  (\pair{\pair{x,x},\pair{x,x}})$, and accept $x$ if and only if $d = x$ and
  $e \in \certificate{M}{x}$.
\end{quote}
Hence, $\sigma$ is not invertible with respect to its second argument.  An
analogous argument shows that $\sigma$ is not invertible with respect to its
first argument.  Thus, $\sigma$ is strongly noninvertible.

Next, we prove that $\sigma$ is associative.  Let $\sigmabot$ be
the total extension of $\sigma$ as in Definition~\ref{def:associative}.  Fix
any three elements of~$\nats$, say $a = \pair{a_1, a_2}$, $b = \pair{b_1,
  b_2}$, and $c = \pair{c_1, c_2}$.  To show that
\begin{eqnarray}
\label{equ:ass}
\sigmabots{\sigmabots{a, b}, c} & = & \sigmabots{a, \sigmabots{b, c}}
\end{eqnarray}
holds, distinguish two cases.

\begin{description}
\item[Case~1:] $a_1 = b_1 = c_1$ and $\{a_2, b_2, c_2 \} \seq \{a_1\} \cup
  \certificate{M}{a_1}$.
  
  Let $x, y \in \{a, b, c\}$ be any two fixed arguments of~$\sigma$.  As noted
  above, if $x$ and $y$ together contain $i$ certificates for ``$a_1\in A$,''
  where $i \in \{1, 2\}$, then $\sigma(x,y)$---and thus also
  $\sigmabots{x,y}$---contains exactly $\max\{0, i-1\}$
  certificates for ``$a_1\in A$.''~~In particular,
  $\sigmabots{x,y}$ preserves the minimum certificate if both $x$
  and $y$ contain a certificate for ``$a_1\in A$.''
  
  If exactly one of $x$ and $y$ contains a certificate for ``$a_1\in A$,''
  then $\sigmabots{x,y} = \pair{a_1, a_1}$.
  
  If none of $x$ and $y$ contains a certificate for ``$a_1\in A$,'' then
  $\sigma(x,y)$ is undefined, so $\sigmabots{x,y} = \bot$.
  
  Let $k\leq 3$ be a number telling us how many of $a_2$, $b_2$, and~$c_2$
  belong to $\certificate{M}{a_1}$.  For example, if $a_2 = b_2 = c_2 \in
  \certificate{M}{a_1}$ then $k = 3$.  Consequently:
\begin{itemize}
\item If $k \leq 1$ then both $\sigmabots{\sigmabots{a,b},c}$ and
  $\sigmabots{a,\sigmabots{b,c}}$ equals~$\bot$.
  
\item If $k=2$ then both $\sigmabots{\sigmabots{a,b},c}$ and
  $\sigmabots{a,\sigmabots{b,c}}$ equals~$\pair{a_1, a_1}$.
  
\item If $k=3$ then both $\sigmabots{\sigmabots{a,b},c}$ and
  $\sigmabots{a,\sigmabots{b,c}}$ equals~$\pair{a_1, \min(a_2, b_2, c_2)}$.
\end{itemize}
In each of these three cases, Equation~(\ref{equ:ass}) is satisfied.

\item[Case~2:] Suppose Case~1 is not true. 
  
  Then, either it holds that $a_1 \neq b_1$ or $a_1 \neq c_1$ or $b_1 \neq
  c_1$, or it holds that $a_1 = b_1 = c_1$ and $\{a_2, b_2, c_2 \}$ is not
  contained in $\{a_1\} \cup \certificate{M}{a_1}$.  By the definition
  of~$\sigma$, in both cases it follows that
\[
\begin{array}{rcccl}
\sigmabots{\sigmabots{a,b},c}
 &  = &  \bot & =  & 
\sigmabots{a,\sigmabots{b,c}} ,
\end{array}
\]
which satisfies Equation~(\ref{equ:ass}) and concludes the proof that $\sigma$
is associative.
\end{description}

Finally, we mention some related results of Chris
Homan~\cite{hom:t:low-ambiguity-aowf} who studied upper and lower bounds on
the ambiguity of associative one-way functions.  In particular, extending Rabi
and Sherman's~\cite{rab-she:j:aowf} result that no total, associative one-way
function is injective, he proved that no total, associative one-way function
can be constant-to-one.  He also showed that, under the plausible assumption
that $\p \neq \up$, there exist linear-to-one, total, strongly noninvertible,
associative one-way functions.  

On a slightly less related note, Homan and
Thakur~\cite{hom-tha:c:one-way-permutations} recently proved that one-way
permutations (i.e., one-way functions that are total, one-to-one, and onto)
exist if and only if $\p \neq \up \cap \coup$.  This result gives a
characterization of one-way permutations in terms of a complexity class
separation, and thus the ultimate answer to a question studied
in~\cite{gro-sel:j:complexity-measures,hem-rot-wec:c:easy-one-way-permutations,hem-rot:j:one-way,rot-hem:j:one-way}.

\subsection{If P $\neq$ NP then Some Strongly Noninvertible Functions are 
Invertible}
\label{sec:fct}

Is every strongly noninvertible function noninvertible?  Hemaspaandra,
Pasanen, and Rothe~\cite{hem-pas-rot:c:strong-noninvertibility} obtained the
surprising result that if $\p \neq \np$ then this is not necessarily the case.
This result shows that the term ``strong noninvertibility'' introduced
in~\cite{rab-she:t-no-URL:aowf,rab-she:j:aowf} actually is a misnomer, since
it seems to suggest that strong noninvertibility always implies
noninvertibility, which is not true.

\begin{theorem}
{\rm{}\cite{hem-pas-rot:c:strong-noninvertibility}} \quad
\label{thm:c}
If $\p \neq \np$ then there exists a total, honest two-ary function that is
strongly one-way but not a one-way function.
\end{theorem}

We give a brief sketch of the proof.  Assume $\p \neq \np$.  Then, there
exists a total two-ary one-way function, call it~$\rho$.  For any integer $n
\in \nats$, define the notation
\begin{eqnarray*}
\mbox{odd}(n) = 2n + 1 & \mbox{ and } & 
\mbox{even}(n) = 2n .
\end{eqnarray*}
Define a function $\sigma : \nats \times \nats \rightarrow
  \nats$ as follows.  Let $a, b \in \nats$ be any two arguments of~$\sigma$.
\begin{itemize}
\item If $a \neq 0 \neq b$, $a = \pair{x,y}$ is odd, and $b$ is even, then
  define $\sigma(a,b) = \mbox{even}(\rho(x,y))$.
  
\item If $a \neq 0 \neq b$, $a$ is even, and $b = \pair{x,y}$ is odd, then
  define $\sigma(a,b) = \mbox{even}(\rho(x,y))$.
  
\item If $a \neq 0 \neq b$, and $a$ is odd if and only if $b$ is odd, then
  define $\sigma(a,b) = \mbox{odd}(a+b)$.

\item If $a=0$ or $b=0$, then define $\sigma(a,b) = a+b$.
\end{itemize}

We claim that $\sigma$ is strongly noninvertible.
For a contradiction, suppose $\sigma$ were invertible with respect to its
first argument via an inverter, $g_1$.  By the definition of~$\sigma$,
for any $z \in \image(\rho)$ with $z \neq 0$, the function $g_1$ on input
$\pair{2,\mbox{even}(z)}$ yields an odd integer $b$ from which we can read the
pair $\pair{x,y}$ with $\rho(x,y) = z$.
Hence, using~$g_1$, one could invert $\rho$ in polynomial time, a
contradiction.  Thus, $\sigma$ is not invertible with respect to its first
argument.  Analogously, one can show that $\sigma$ is not invertible with
respect to its second argument.  So, $\sigma$ indeed is strongly
noninvertible.

But $\sigma$ is invertible!  By the fourth line in the definition of~$\sigma$,
every $z$ in the image of $\sigma$ has a preimage of the form $(0,z)$.
Thus, the function $g$ defined by $g(z) = (0, z)$ inverts $\sigma$ in
polynomial time.  Hence, $\sigma$ is not a one-way function.

Why don't we use a different notion of strongness that automatically implies
noninvertibility?  Here is an attempt to redefine the notion of strongness
accordingly, which yields a new notion that we will call ``overstrongness.''

\begin{definition}
{\rm{}\cite{hem-pas-rot:c:strong-noninvertibility}} \quad
\label{d:c-strong}
Let $\sigma : \nats \times \nats \rightarrow \nats$ be any two-ary function;
$\sigma$ may be nontotal and it may be many-to-one.  We say that $\sigma$ is
{\em overstrong\/} if and only if no polynomial-time computable function $f$
with $f : \{1,2\} \times \nats \times \nats \rightarrow \nats \times \nats$
satisfies that for each $i \in \{1,2\}$ and for each $z, a \in \nats$:
\[
((\exists b \in \nats) [(\sigma(a,b) = z \land i = 1) \lor
(\sigma(b,a) = z \land i = 2)]) \Lora \sigma(f(i,z,a)) = z .
\]
\end{definition}

Note that overstrongness implies both noninvertibility and strong
noninvertibility.  However, the problem with this new definition is that it
completely loses the core of why strongness precludes direct attacks on the
Rivest--Sherman and Rabi--Sherman protocols protocols.  To see why, look at
Figure~\ref{fig:rivest-sherman-secret-key} and
Figure~\ref{fig:rabi-sherman-digital-signature}, which give the protocols of
Rabi, Rivest, and Sherman.  In contrast to overstrongness, Rabi, Rivest, and
Sherman's original definition of strong noninvertibility (see
Definition~\ref{d:strong-oneway}) {\em respects the argument given}.  It is
this feature that precludes Erich from being able to compute Alice's secret
$x$ from the transmitted values $\sigma(x,y)$ and~$y$, which he knows.  In
short, overstrongness is {\em not well-motivated\/} by the protocols of Rabi,
Rivest, and Sherman.

We mention without proof some further results of Hemaspaandra, Pasanen, and
Rothe~\cite{hem-pas-rot:c:strong-noninvertibility}.  

\begin{theorem}
{\rm{}\cite{hem-pas-rot:c:strong-noninvertibility}} \quad
\label{thm:d}
\begin{enumerate}
\item If $\p \neq \np$ then there exists a total, honest, s-honest, two-ary
  overstrong function.  Consequently, if $\p \neq \np$ then there exists a
  total two-ary function that is both one-way and strongly one-way.

\item If $\p \neq \np$ then there exists a total, s-honest two-ary one-way
  function $\sigma$ such that $\sigma$ is invertible with respect to its first
  argument and $\sigma$ is invertible with respect to its second argument.
  
\item If $\p \neq \np$ then there exists a total, s-honest two-ary one-way
  function that is invertible with respect to either one of its arguments
  (thus, it is not strongly one-way), yet that is not invertible with respect
  to its other argument.
  
\item \label{thm:d4} If $\p \neq \np$ then there exists a total, honest,
  s-honest two-ary function that is noninvertible and strongly noninvertible
  but that is not overstrong.
\end{enumerate}
\end{theorem}

\begin{acks}
I am grateful to Pekka Orponen for inviting me to
be a lecturer of the 11th Jyv\"askyl\"a Summer School that was held in August,
2001, at the University of Jyv\"askyl\"a.  I thank Kari Pasanen for being a
great tutor of this tutorial, for carefully proofreading a preliminary draft
of this paper, and in particular for subletting his summer house on an island
of scenic Lake Keitele to me and my family during the summer school.  I am
grateful to Pekka and Kari for their hospitality, and I thank my 33 summer
school students from 16 countries for making this course so much fun and
pleasure.  I also thank Eric Allender, Godmar Back, Harald Baier, Lane
Hemaspaandra, Eike Kiltz, Alan Selman, Holger Spakowski, Gerd Wechsung, and
Peter Widmayer for their insightful advice and helpful comments and for their
interest in this paper.  Last but not least, I thank the anonymous ACM
Computing Surveys referees whose detailed comments very much helped to fix
errors in an earlier version and to improve the presentation, and the editor,
Paul Purdom, for his guidance during the editorial process.
\end{acks}

\bibliographystyle{alpha}

\bibliography{/home/inf1/rothe/BIGBIB/joergbib}

\begin{received}
Received Month Year;
revised Month Year; accepted Month Year
\end{received}

\end{document}